\documentclass[a4paper,12pt, leqno]{article}
\usepackage{amsmath}
\usepackage{amssymb, latexsym}
\usepackage{amsfonts}
\usepackage{makeidx}
\usepackage{ascmac}
\usepackage{amssymb}

\usepackage{color}

\date{}

\newtheorem{lem}{Lemma}[section]
\newtheorem{thm}{Theorem}[section]
\newtheorem{prop}{Proposition}[section]
\newtheorem{cor}{Corollary}[section]

\newcommand{\dbar}{d\!\!\!{\lower-0.6ex\hbox{$-$}}\!}
\newcommand{\dslash}{d\!\!\!{\lower-0.6ex\hbox{$-$}}}
\newcommand{\e}{\varepsilon}
\newcommand{\h}{\hbar}

\newcommand{\btt}{\lower-0.2ex\hbox{${\scriptscriptstyle{\bullet}}$}}
\newcommand{\ctt}{\lower-0.2ex\hbox{${\scriptscriptstyle{\circ}}$}}
\newcommand{\dtt}{\lower-0.2ex\hbox{${\scriptscriptstyle{\diamond}}$}}
\newcommand{\wt}{\lower-0.2ex\hbox{${\scriptstyle{\wedge}}$}}
\newcommand{\odt}{\lower-0.4ex\hbox{${\scriptscriptstyle{\odot}}$}}

\newcommand{\Cal}{\mathcal}

\begin{document}

\pagestyle{plain}

\title{Deformation Expression for Elements of Algebras (VII)\\
--Vacuum/Pseudo-vacuum Representations--}


\author{
     Hideki Omori\thanks{ Department of Mathematics,
             Faculty of Sciences and Technology,
        Tokyo University of Science, 2641, Noda, Chiba, 278-8510, Japan,
         email: omori@ma.noda.tus.ac.jp}
        \\Tokyo University of Science
\and  Yoshiaki Maeda\thanks{Department of Mathematics,
                Faculty of Science and Technology,
                Keio University, 3-14-1, Hiyoshi, Yokohama,223-8522, Japan,
                email: maeda@math.keio.ac.jp}
          \\Keio University
\and  Akira Yoshioka \thanks{Department of Mathematics,
          Faculty of Science, Tokyo University of Science,
         1-3, Kagurazaka, Tokyo, 102-8601, Japan,
         email: yoshioka@rs.kagu.tus.ac.jp}
           \\Tokyo University of Science}

\maketitle

\tableofcontents

\pagestyle{plain}



\setcounter{equation}{0}

\bigskip 

Thinking back the long history of physics, we see that 
the calculation used by physicists was nothing but the ordinary calculus. 
Another word, physicists have never wrote theories beyond the 
{\it basic axioms} of the calculus. 
This is not to declare of the victory of calculus or algebraic
topology. On the contrary, we are thinking that every theory of 
mathematical physics must suggest new frontier of ordinary calculus, 
which are never viewed by classical geometers. 

Weyl algebras or Heisenberg algebras are naturally involved in 
slightly extended systems of the algebra of ordinary calculus, and 
are supported by the classical notion of phase spaces 
on which the general mechanics are based.  
The theory of deformation quantizations gives a notion of 
quantization of ``phase space''. To explain its essence in brief we proposed 
in the previous note the notion of $\mu$-regulated algebra. 
 
Now, as it is a quantization of phase space, it does not give
Schr{\"o}dinger quantizations which are written by 
partial differential operators on ``configuration spaces''.

For instance, the classical motion described by the
Hamiltonian $H(q,p)$ is given by the equation on the phase space
$$
\frac{d}{dt}A_t(q,p){=}\{A_t, H\}
$$
by using a Poisson bracket. Its deformation quantization is
simply to write this in the form
$$
\frac{d}{dt}A_t(u,v){=}[A_t, H_*]
$$
by using the commutator bracket in a certain $\mu$-regulated
algebra. The solution is then given by using the $*$-exponential function 
${\rm{Ad}}(e_*^{tH_*})A_0$. But $e_*^{tH_*}$ and ${\rm{Ad}}(e_*^{tH_*})A_0$ are only 
elements of transcendentally extended $\mu$-regulated algebra. 
This does not give the solution of the Schr{\"o}dinger equation. 
We have to consider the (time independent) ``state vector'' $\psi$ and the 
expectation value $\langle\psi, A_t\psi\rangle$ 
to obtain the Schr{\"o}dinger picture. Here we have to note that 
the group $e_*^{tH_*}$ is often a covering group of ${\rm{Ad}}(e_*^{tH_*})$.

\medskip
Note that there is no mathematical criterion for the state
vectors, or the configuration space in the quantized phase space. 
We are vaguely thinking that state vectors are elements of a certain 
representation space and the configuration space is a place 
where one can draw differential geometrical pictures to objects.
 
In this series, we have introduced elements, called ``vacuums''   
 to consider the state vectors and 
the configuration spaces within the world of extended algebra 
of calculus with various expressions. We have found 
several strange elements, called {\it polar elements}, and an  
extended notions of vacuums, which were called 
{\it pseudo-vacuums} in \cite{ommy6}. These are not established
notions in mathematical physics, but we are thinking that 
these must propose new frontier for mathematical physics. 
We are thinking that vacuums and pseudo-vacuums are not unique, but 
the function algebra of the {\bf configuration spaces} must be    
an algebra similar to the {\bf Frobenius algebra} defined by vacuums. 

The point in this note is that to obtain classical pictures one 
has often to restrict the expression parameters, and there are 
two essentially different expression parameters.

\section{Preliminaries}\label{prelim}
First we recall the summary of the notions through the papers 
\cite{OMMY3}-\cite{ommy8} and former results together which will 
be used in this note. Throughout this note, we use notations as follows: 
\begin{equation}\label{Weyl}
\begin{aligned}
{\pmb u}&=(u_1,u_2,\cdots,u_{2m})=
(\tilde{\pmb u},\tilde{\pmb v}),\quad  
\tilde{\pmb u}=(\tilde{u}_1,\cdots,\tilde{u}_m),\,\,
\tilde{\pmb v}=(\tilde{v}_1,\cdots,\tilde{v}_m)\\
&{\pmb x}{=}(x_1,x_2,\cdots,x_{2m}){=}
(\tilde{\pmb x},\tilde{\pmb y}),\qquad {\pmb\xi}{=}
(\xi_1,\xi_2,\cdots,\xi_{2m})
{=}(\tilde{\pmb\xi},\tilde{\pmb\eta}).
\end{aligned}
\end{equation}
We also recall the Weyl algebra $W_{n}[\h]$, 
the associative algebra generated by 
$\tilde{\pmb u},\tilde{\pmb v}$ with the fundamental relations$$
[\tilde{u}_k,\tilde{v}_l]{=}-i\h\delta_{kl},\quad 
[\tilde{u}_k,\tilde{u}_l]{=}0{=}[\tilde{v}_k,\tilde{v}_l],\quad \h>0.
$$  
It is wellknown that this algebra is expressed by giving a product
formula on the space of polynomials.

\subsection{General product formulas and intertwiners}\label{PBW-theorem} 
Let ${\mathfrak S}_{\mathbb C}(n)$, $n{=}2m$, be 
the space of complex symmetric matrices. 
For a fixed $K{\in}{\mathfrak S}_{\mathbb C}(n)$ and the 
standard skew-symmetric matrix   
$J{=}\left[
\begin{smallmatrix} 
   0 & {-}I\\ 
   I & 0 
 \end{smallmatrix} 
\right]$ 
we set $\Lambda {=}K{+}J$ and define a product ${*}_{_{K}}$ on the space of polynomials   
${\mathbb C}[\pmb u]$ by the formula 
\begin{equation}
\label{eq:KK}
 f*_{_{K}}g=fe^{\frac{i\h}{2}
(\sum\overleftarrow{\partial_{u_i}}
{\Lambda}_{ij}\overrightarrow{\partial_{u_j}})}g
=\sum_{k}\frac{(i\h)^k}{k!2^k}
{\Lambda}_{i_1j_1}\!{\cdots}{\Lambda}_{i_kj_k}
\partial_{u_{i_1}}\!{\cdots}\partial_{u_{i_k}}f\,\,
\partial_{u_{j_1}}\!{\cdots}\partial_{u_{j_k}}g.   
\end{equation}
$({\mathbb C}[\pmb u],*_{_{K}})$ is known to be an associative algebra
isomorphic to the Weyl algebra $W_{n}[\h]$. 

\medskip
For another symmetric matrix $K'$, 
we have the following formula:
\begin{equation}
  \label{eq:Hochsch}
e^{\frac{i\h}{4}\sum{K}'_{ij}\partial_{u_i}\partial_{u_j}}
\Big(\big(e^{-\frac{i\h}{4}\sum{K}_{ij}
\partial_{u_i}\partial_{u_j}}f\big)
{*_{_K}}
\big(e^{-\frac{i\h}{4}\sum{K}_{ij}\partial_{u_i}\partial_{u_j}}g\big)\Big)
= f{*}_{_{K{+}K'}}g.
\end{equation}

The next one is proved immediately by the formula \eqref{eq:Hochsch}:
\begin{cor}
Let 
$I_0^{^K}(f)=
e^{\frac{i\h}{4}\sum{K}_{ij}\partial_{u_i}\partial_{u_j}}(f)$ 
and $I_{_K}^0(f)=
e^{-\frac{i\h}{4}\sum{K}_{ij}\partial_{u_i}\partial_{u_j}}(f)$. 
Then $I_0^{^K}$ is an isomorphism of 
$({\mathbb C}[\pmb u];{*}_{0})$ onto 
$({\mathbb C}[\pmb u];{*}_{_{K}})$.
\end{cor}
The isomorphism class is denoted by $(W_{2m}[\h], *)$ and 
called the Weyl algebra. The operator 
\begin{equation}
\label{intertwiner}
I_{_K}^{^{K'}}(f)=
\Big[\exp\Big(\frac{i\h}{4}\sum_{i,j}(K'_{ij}{-}K_{ij})
\partial_{u_i}\partial_{u_j}\Big)\Big](f) \,\,
(=I_{0}^{^{K'}}(I_{0}^{^{K}})^{-1}(f))  
\end{equation}
will be called the {\it intertwiner}.
Intertwiners do not change the algebraic structure $*$, 
but do change the expression of elements by the ordinary 
commutative structure.  

Let $H{\!o}l({\mathbb C}^n)$ be the space of all 
holomorphic functions on the complex $n$-plane ${\mathbb C}^n$ with 
the uniform convergence topology on each compact domain. 
$H{\!o}l({\mathbb C}^n)$ is a Fr{\'e}chet space defined by a countable family of seminorms. 
It is clear that the product $f{*_{_{K}}}g$ 
is defined if one of $f, g$ is 
a polynomial and another is a smooth function.

\begin{prop}\label{extholom} 
For every polynomial $p(\pmb u)\in{\mathbb C}[\pmb u]$, 
the left-multiplication 
$f\to p(\pmb u){*_{_{K}}}f$ and the right-multiplication 
$f\to f{*_{_{K}}}p(\pmb u)$ are both continuous 
linear mappings of $H\!ol({\mathbb C}^n)$ into itself. 

If two of $f, g, h$ are polynomials, then associativity 
$(f{*_{_{K}}}g){*_{_{K}}}h=
f{*_{_{K}}}(g{*_{_{K}}}h)$
holds.
\end{prop}

\subsection{Generic  expression parameters}
\label{Expinter}
Consider for instance an element
$u_1^2{*_{_K}}u_2{*_{_K}}u_1{*_{_K}}u^3_2$, which may be expressed
differently via commutation relations.  
Computing out this by the product formula \eqref{eq:KK} gives 
the way to express the element univalent way. In this sense, 
the $*_{_K}$-product formula is the $*_{_K}$-expression formula for 
elements of algebra.   
Note that according to the choice of $K=0, K_0, {-}K_0$, $I$,  
where 
$$
(0,\,\, K_0, {-}K_0, I)= 
\left(
 \left[
{\footnotesize
{\begin{matrix}
   0 & 0\\
   0 & 0
\end{matrix}}}
 \right],\,\,
\left[
{\footnotesize
{\begin{matrix}
   0 & I\\
   I & 0
 \end{matrix}}}
\right],\,\,
\left[
{\footnotesize
{\begin{matrix}
   0 &\!\!{-}I\\
   {-}I&\!\!0
 \end{matrix}}}\right],\,\,
\left[\footnotesize{
\begin{matrix}
   I&\!\!0\\
   0&\!\!I
 \end{matrix}}\right]
\right),
$$
\begin{tabular}{l|l} 
Choice of $K$ & (name of ordering)\\ \hline
 \medskip
$K=0$         & Weyl ordered expression\\ \hline 
 \medskip
$K_0=\left[
\begin{smallmatrix}
   0 & I\\
   I & 0
 \end{smallmatrix}
\right]$  & Normal ordered expression \\ \hline
 \medskip
$-K_0$    & Anti-normal ordered expression \\ \hline
\medskip
$\left[
\begin{smallmatrix}
   I& 0\\
   0 & I
 \end{smallmatrix}
\right]$  & Unit ordered expression \\ \hline
\medskip
General  $K$  & $K$-ordered expression
\end{tabular}
\hfill\parbox[c]{.4\linewidth}
{the product formulas \eqref{eq:KK} give the Weyl ordered 
expression and the normal ordered expression, 
the antinormal ordered expression respectively, but 
the unit ordered expression is not so familiar in physics.

For each ordered expression, the  product formulas are given  
respectively by the following formula:}
\begin{equation}\label{ppformula}
 \begin{aligned}[c]
f({\pmb u}){*{_{_0}}}g({\pmb u})=&
f\exp 
\frac{\h i}{2}\{\overleftarrow{\partial_{v}} 
     {\wedge}\overrightarrow{\partial_{u}}\}g,
     \quad{\text{(Moyal product formula)}}\\
f({\pmb u}){*{_{_{K_0}}}}g({\pmb u})=&
f\exp {\h i}\{\overleftarrow{\partial_{v}}\,\, 
       \overrightarrow{\partial_{u}}\}g,
       \qquad{\text{($\Psi$DO product formula)}}  \\
f({\pmb u}){*{_{_{{-}K_0}}}}g({\pmb u})=&
f\exp{-\h i}\{\overleftarrow{\partial_{u}}\,\, 
       \overrightarrow{\partial_{v}}\}g,
       \quad{\text{($\overline{\Psi}$DO product formula)}}   
 \end{aligned}
\end{equation} 
where 
$\overleftarrow{\partial_{v}}{\wedge}
\overrightarrow{\partial_{u}}
=\sum_i(\overleftarrow{\partial_{\tilde{v}_i}}
\overrightarrow{\partial_{\tilde{u}_i}}
-\overleftarrow{\partial_{\tilde{u}_i}}
\overrightarrow{\partial_{\tilde{v}_i}})$ and 
$\overleftarrow{\partial_{v}}\,\, 
       \overrightarrow{\partial_{u}}
=\sum_i\overleftarrow{\partial_{\tilde{v}_i}}\,\, 
       \overrightarrow{\partial_{\tilde{u}_i}}$.

\subsubsection{$V$-class of expression parameters}
The product formula for the unit ordered expression  
looks a bit complicated to write down, but there are many 
interesting phenomena which has never appeared in
Weyl-, normal-, or anti-normal-ordered expressions. 
We also give another family of expression parameters  
which will be often used in this note and involving 
$iI$ expression parameter as a special case. 

Set $n=2m$. Let $V$ be a real $n$-dimensional subspace of 
${\mathbb C}^n$ 
spanned by an $O(n)$ frame such that $JV{=}V$ and  ${\mathbb C}^n{=}V{\oplus}iV$. 

We denote $\pi_{re}$, $\pi_{im}$ the projections onto $V$, $iV$
respectively. As 
${\mathbb C}^n{=}
{\mathbb C}_{\tilde{\pmb x}}^{m}{\oplus}
{\mathbb C}_{\tilde{\pmb y}}^{m}$ and $V{\oplus}iV{=}{\mathbb C}^n$, we
see 
$$ 
{V}{=}V\cap{\mathbb C}_{\tilde{\pmb x}}^{m}{\oplus}V\!\cap{\mathbb C}_{\tilde{\pmb y}}^{m}, 
\quad J\big(V\cap{\mathbb C}_{\tilde{\pmb x}}^{m}\big){=}V\cap{\mathbb C}_{\tilde{\pmb y}}^{m}.
$$ 
We define a bilinear form $\langle\tilde{\pmb x},\tilde{\pmb
  x}'\rangle$ on $V$ 
 by reducing the canonical bilinear form to the subspace $V$. 
 Since $V$ is spanned by the $O(n)$ frame we see 
it is positive definite and 
$\langle\tilde{\pmb x}J,\,\tilde{\pmb x}'\rangle{=}-\langle\tilde{\pmb x},\,\tilde{\pmb x}'\!J\rangle$.

For a complex symmetric matrix
$K\in{\mathfrak S}_{\mathbb C}(n)$,  
$\pi_{re}K; V\to V$ and $\pi_{im}K; V\to iV$ are called the $V$-real
part and the $V$-imaginary part of $K$ respectively.
 
A class of expression parameters will be called a 
{\bf $V$-class}, denoted by ${\mathfrak H}_+(V)$, is a  
special class of expression parameters such that the real part of  
$iK$ is on $V$ is positive definite,  i.e. 
\begin{equation}\label{Siegel}
{\mathfrak H}_+(V)=
\{K;\frac{1}{\h}\langle  
\xi \pi_{re}(iK),\xi\rangle\geq
c_{K}|\xi|^2,  
\quad \forall\xi\in V\}.
\end{equation}
If $V{=}{\mathbb R}^n$, then $\pi_{re}(iK)$ is positive definite on 
${\mathbb R}^n$, but if $V{=}(i{\mathbb R})^n$, then 
$\pi_{re}(iK)$ is negative definite on ${\mathbb R}^n$. If 
$V{=}(\frac{1}{\sqrt{i}}\mathbb R)^{n}$, then $\pi_{re}K$ is 
positive definite on ${\mathbb R}^n$. 

\bigskip
Let $d{V}$ be the standard volume element on $V$. 
We denote by  
$\dbar{V}{=}( \frac{1}{\sqrt{2\pi\h}})^{n} d{V}$. 

\begin{lem}\label{Lemma1}
If $K{\in}{\mathfrak H}_+(V)$, then 
$\int_{V} e^{-\frac{1}{4\h}\langle\xi(iK),\xi\rangle}\dbar{V}_{\xi}{=}
\frac{2^m}{\sqrt{\det(iK})}$,
where $\sqrt{\det(iK})$ is determined without sign ambiguity in the
form 
$$
\sqrt{|\det(iK)|}\prod_{i=1}^m\sqrt{(1{+}ia_i)}, \quad
  \sqrt{|\det(iK)|}>0,\,\,{\rm{Re}}\sqrt{(1{+}ia_i)}>0. 
$$
If $K$ is fixed, this does not depend 
on the choice of $V$ whenever $K\in {\mathfrak H}_+(V)$.
\end{lem}

\noindent
{\bf Proof}\,\,\,We first fix $V$.  
Set $iK{=}R^2{+}iS$, where $R: V\to V$ is symmetric and  
positive definite on $V$  and
$S$ is any real linear mapping of $V\to V$. Let $\lambda_1, \cdots,\lambda_{2m}$be the eigenvalues of $R$. 
By a suitable $T\in SO(V)$, $iK$ is changed into 
$$
T(iK)T^{-1}{=}diag\{\lambda^2_1, \cdots,\lambda^2_{2m}\}{+}iTST^{-1}.
$$
Changing variables by setting $\eta_i{=}\lambda_i\xi_i$, the
integral turns out 
$$
\frac{1}{\det R}
\int_{\mathbb R^{2m}}e^{-\frac{1}{4\h}(\sum_{k=1}^{2m}\eta_k^2 
{+}i\sum_{kl}(R^{-1}SR^{-1})_{kl}\eta_k\eta_l)}\dbar{\pmb\eta},
$$
where $(R^{-1}SR^{-1})_{kl}$ is a real matrix.
Hence by a suitable $T'\in SO(2m)$, the integral becomes 
$$
\frac{1}{\det R}
\int_{\mathbb  R^{2m}}e^{-\frac{1}{4\h}\sum_{k=1}^{2m}(1{+}i\mu_k)\eta_k^2}\dbar{\pmb\eta}, 
\quad \mu_k\in{\mathbb R}.
$$
Note that Cauchy's integral theorem and rotation of the path of
integration gives 
$$
\frac{1}{\sqrt{2\pi\h}}\int_{\mathbb R}e^{-\frac{1}{4\h}(1{+}i\mu_k)\eta_k^2}d\eta{=}
\frac{\sqrt{2}}{\sqrt{(1{+}i\mu_k)}}, \quad 
{\rm{Re}}\sqrt{(1{+}i\mu_k)}>0. 
$$
It follows 
$$
\int_{V}
e^{-\frac{1}{4\h}\langle{\pmb\xi}(iK), {\pmb\xi}\rangle}\dbar{\pmb\xi}
{=}
{2^m}\frac{1}{\det R}\frac{1}{\sqrt{\det(I{+}R^{-1}(iS)R^{-1})}}=
\frac{2^m}{\sqrt{\det(iK)}},
$$
and $\sqrt{\det(iK)}{=}\sqrt{|\det(iK)|}\prod_{k=1}^m\sqrt{(1{+}i\mu_k)}$.
The result depends only on $K$, hence the integral gives the same
result for $K\in \mathfrak{H}_+(V)\cap\mathfrak{H}_+(V')$.
${}$\hfill$\Box$

\bigskip
In what follows we set the constant by  
\begin{equation}\label{impcnst}
C_{0}(K)=\frac{2^m}{\sqrt{\det iK}}{=}
\int_{V} e^{\frac{1}{4i\h}{\pmb\xi}{K}\,{}^t{\pmb\xi}}
\dbar{\pmb\xi}{=}
\int_{V} e^{-\frac{1}{4\h}{\pmb\xi}({iK})\,{}^t{\pmb\xi}}
\dbar{\pmb\xi},
\end{equation}
where $\dbar= \big(\frac{1}{\sqrt{2\pi\h}}\big)^{n/2}$. 
\subsubsection{Some remarks about Fourier transformations}
 
We denote by $\Cal S(\Bbb R^n)$ the space of all rapidly decreasing 
functions. 
First, express this space as the projective limit space 
of a family of  Hilbert spaces.
Taking the topological completion of $\Cal S(\Bbb R^n)$ 
by the norm topology defined by 
the weighted $C^k$-inner product $\langle\,\,,\,\,\rangle_k$. 
We denote it by $\Cal S^k(\Bbb R^n)$.
The Sobolev lemma gives that 
$\Cal S(\Bbb R^n)=\bigcap_k \Cal S^k(\Bbb R^n)$.  
Fourier transform is defined by 
$$
\frak F(f)(\pmb\xi)=\int_{\Bbb R^n}f(\pmb x)
e^{\frac{1}{i\h}\langle\pmb x,\pmb\xi\rangle}\dbar x, \quad
\text{where}\quad \dbar\pmb x=\left(\frac{1}{2\pi\h}\right)^{n/2} d{\pmb x}.  
$$ 
$\frak F(f)(\pmb\xi)$ is sometimes denoted by $\hat f(\pmb\xi)$.
Fourier transform is defined for $L^1$-functions at first, 
and extends in various ways. 
\begin{lem}
  \label{Planch}
If $f$, $f'$, $f''$ are continuous and summable, then  $\hat f(\xi)$ is 
summable, and it holds that $||f||_{L^2}=||\frak F(f)||_{L^2}$. 
\end{lem}
Fundamental properties of 
Fourier transform are all proved by integration by parts:
$$
\frak F(\partial^{\alpha}f)=(\frac{i}{\h}\xi)^{\alpha}\frak F(f), \qquad
\frak F((-\frac{i}{\h}x)^{\alpha}f)=\partial^{\alpha}\frak F(f).
$$
It is wellknown that the 
Fourier transform  $\frak F$ gives a topological isomorphism of 
$\Cal S(\Bbb R_x^n)$ onto $\Cal S(\Bbb R_{\xi}^n)$. 
Furthermore, it gives a topological isomorphism of 
$\Cal S^k(\Bbb R_x^n)$ onto $\Cal S^k(\Bbb R_{\xi}^n)$ for every 
$k$. 
Thus, $\frak F$ gives a topological isomorphism of the dual space 
$\Cal S'(\Bbb R_{\xi}^n)$ onto $\Cal S'(\Bbb R_x^n)$. 

Let $\Cal S^{-k}(\Bbb R_x^n)$ be the dual space of 
$\Cal S^{k}(\Bbb R_x^n)$. 
$\Cal S^{-k}(\Bbb R_x^n)$ is a Hilbert space. We see easily that 
$$
\Cal S^{-\infty}(\Bbb R_x^n)=
\bigcup_{k}\Cal S^{-k}(\Bbb R_x^n)
$$ 
with the inductive limit topology.
Elements of $\Cal S^{-\infty}(\Bbb R_x^n)$ 
are called {\bf tempered distributions}. 
If a tempered distribution $f$ is a function, that is, the value 
$f(x)$ is defined for every $x$, then 
$f$ is called a {\bf slowly increasing function}. 

For the convenience of notations, we denote 
$$
\delta(\pmb x{-}\pmb a){=}{\frak F}^{-1}(1){=}
\int_{\mathbb R^n}e^{-\frac{1}{i\h}\langle\pmb\xi,\,\,\pmb x{-}\pmb a\rangle}\dbar{\pmb\xi},
\quad
 \int_{\mathbb R^n}\delta(\pmb x{-}\pmb a)
e^{\frac{1}{i\h}\langle\pmb\xi,\,\,\pmb x\rangle}\dbar{\pmb x}{=}
e^{\frac{1}{i\h}\langle\pmb\xi,\,\,\pmb a\rangle}.
$$

\bigskip
Now let $(\mathcal B)$, called the space of all 
bounded derivatives class, be a little wider function space than $\mathcal S$
consisting of all smooth functions $f$ such that the differential 
$\partial^{\alpha}f$ is bounded on ${\mathbb R}^n$ for every $\alpha$.
The dual space $(\mathcal B)'$ is convenient to make the convolution product:
$$
f{\btt}g(\pmb x){=}
\int_{{\mathbb R}^n}{f}(\pmb y){g}({\pmb x}{-}{\pmb y})
\dbar\pmb y.
$$
\noindent
{\bf Remark}\,\, The convolution product is welldefined if one of them is a rapidly decreasing
distribution, where $f(\pmb x)$ is a rapidly decreasing
distribution, iff $f(\pmb x)(1{+}|\pmb\xi|^2)^k$ belongs to the dual
space of all bounded smooth functions on ${\mathbb R}^n$ 
for every integer $k$ (i.e. $(\mathcal B')$). 
 $(\mathcal B)$ forms a commutative algebra.
Hence $(\mathcal B')$ forms also a commutative algebra under the 
convolution product.

\bigskip
\noindent
{\bf Twisted convolution product $*_J$} 
\begin{equation}\label{twistconv}
f\,{*_J}\,g(\pmb\zeta){=}
\int_{{\mathbb R}^n}{f}(\pmb\xi){g}({\pmb\zeta}{-}{\pmb\xi})
e^{\frac{1}{2i\h}\langle{\pmb\xi}J,{\pmb\zeta}\rangle}\dbar\pmb\xi.
\end{equation}
As $|e^{\frac{1}{2i\h}\langle{\pmb\xi}J,{\pmb\zeta}\rangle}|{=}1$, the
twisted convolution product is welldefined under the same condition as above.

For instance setting $\delta_{\pmb a}(\pmb x){=}
\delta(\pmb x{-}\pmb a)$, we have 
${\mathcal F}(\delta_{\pmb a})(\pmb\xi):=\check{\delta}_{\pmb a}(\pmb\xi):=
e^{-\frac{1}{i\h}\langle\pmb\xi,\pmb a\rangle}$ 
and 
$$
\check{\delta}_{\pmb a}\,{*_J}\,\check{\delta}_{\pmb b}(\pmb\xi)
=\int_{{\mathbb R}^n}e^{-\frac{1}{i\h}\langle\pmb\zeta,{\pmb a}\rangle}
e^{\frac{1}{2i\h}\langle\pmb\zeta J,\pmb\xi\rangle}
e^{-\frac{1}{i\h} \langle(\pmb\xi{-}\pmb\zeta),{\pmb b}\rangle}
\dbar\pmb\zeta{=}\delta(\pmb b{-}\pmb a{-}\frac{1}{2}{\pmb\xi}J)
e^{-\frac{1}{i\h}\langle\pmb\xi,{\pmb b}\rangle}.
$$

\section{Star-exponential functions of linear forms}

Let $\langle{\pmb a},{\pmb u}\rangle$ be a linear function, 
Then, we have  
\begin{equation}
  \label{eq:intwin}
I_{_K}^{^{K'}}(e^{\frac{1}{i\h}\langle{\pmb a},{\pmb u}\rangle})
=e^{\frac{1}{4i\h}{\pmb a}(K'{-}K)\,{}^t\!{\pmb a}}
e^{\frac{1}{i\h}\langle{\pmb a},{\pmb u}\rangle}. 
\end{equation}
Hence, $\{e^{\frac{1}{4i\h}{\pmb a}K\,{}^t\!{\pmb a}}
e^{\frac{1}{i\h}\langle{\pmb a},{\pmb u}\rangle}; 
K\in{\mathfrak S}_{\mathbb C}(2m)\}$ is a family of mutually
isomorphic one parameter groups. 
We shall denote this family symbolically by 
$e_*^{\frac{1}{i\h}\langle{\pmb a},{\pmb u}\rangle}$, but 
it is often better to view this as an {\it element}.  
Its $K$-ordered expression will be  denoted by    
\begin{equation}\label{eq:tempexp}
:e_*^{\frac{1}{i\h}\langle{\pmb a},{\pmb u}\rangle}:_{_K}
=e^{\frac{1}{4i\h}\langle{\pmb a}K,{\pmb a}\rangle}
e^{\frac{1}{i\h}\langle{\pmb a},{\pmb u}\rangle}
=e^{\frac{1}{4i\h}\langle{\pmb a}K,{\pmb a}\rangle
{+}\frac{1}{i\h}\langle{\pmb a},{\pmb u}\rangle}
=  
e^{-\frac{1}{4\h}\langle{\pmb a}(iK),{\pmb a}\rangle
{+}\frac{1}{i\h}\langle{\pmb a},{\pmb u}\rangle}.
\end{equation}
Now, we see why the $V$-class of expression parameters and Lemma\,\ref{Lemma1} are crucial in this
note. In a $V$-class expression, ${:}e_*^{\frac{1}{i\h}\langle{\pmb\xi},{\pmb u}\rangle}{:}_{_K}$
is rapidly decreasing w.r.t. $\pmb\xi\in{V}$. Hence in a certain
$K$-expression, integrals 
$$
\int_{-\infty}^0e_*^{t\frac{1}{i\h}\langle{\pmb a},{\pmb u}\rangle}dt, \quad 
{-}\int_{0}^{\infty}e_*^{t\frac{1}{i\h}\langle{\pmb a},{\pmb u}\rangle}dt
$$  
give two different inverses 
$(\frac{1}{i\h}\langle{\pmb a},{\pmb u}\rangle)^{-1}_{\pm}$ 
such that  
$(\frac{1}{i\h}\langle{\pmb a},{\pmb u}\rangle)^{-1}_{+}
{-}(\frac{1}{i\h}\langle{\pmb a},{\pmb u}\rangle)^{-1}_{-}{=}
\int_{-\infty}^{\infty}e_*^{t\frac{1}{i\h}\langle{\pmb a},{\pmb u}\rangle}dt$.

\medskip
In general, for any holomorphic function $H_*$ of $\pmb u$, the $*$-exponential
function  $e_*^{tH_*}$ may be defined as the collection 
${:}e_*^{tH_*}{:}_{_K}$ of power series 
$\sum\frac{t^n}{n!}{:}H_*^n{:}_{_K}$ formally rearranged as power
series of $\h$. In the case $\h$ is not formal, the $*$-exponential function
$e_*^{tH_*}$ is not defined by a power series. Instead, these are considered as 
the real analytic solution of an evolution equation
\begin{equation}\label{starexpexp}
\frac{d}{dt}{:}f_t{:}_{_K}{=}{:}H_*{:}_{_K}{*_{_K}}{:}f_t{:}_{_K}, \quad {:}f_0{:}_{_K}{=}1. 
\end{equation} 
This is a differential equation, if $H_*$ is a polynomial, but the
solution may not exist in general. 
The solution (if exists uniquely) with initial data ${:}g_*{:}_{_K}$ will be denoted by 
${:}e_*^{tH_*}{*}g_*{:}_{_K}$. In particular, it is easy to have the exponential law
with a ordinary exponential functions 
$$
{:}e_*^{tH_*}e^{ta}{:}_{_K}{=}{:}e_*^{tH_*{+}ta}{:}_{_K}.
$$

\bigskip
Fundamental formulas for calculations are easily obtained by the
product formula: 
\begin{prop}\label{Fundcal}
For every expression parameter $K$ 
$$
{:}e_*^{\frac{1}{i\h}\langle{\pmb a},{\pmb u}\rangle}{*}
   e_*^{\frac{1}{i\h}\langle{\pmb b},{\pmb u}\rangle}{:}_{_K}
{=}
e^{\frac{1}{2i\h}\langle{\pmb a}J,{\pmb b}\rangle}
{:}e_*^{\frac{1}{i\h}\langle({\pmb a}{+}{\pmb b}),{\pmb u}\rangle}{:}_{_K}
{=}e^{\frac{1}{i\h}\langle{\pmb a}J,{\pmb b}\rangle}
{:}e_*^{\frac{1}{i\h}\langle{\pmb b},{\pmb u}\rangle}{*}
   e_*^{\langle\pmb a,\pmb u\rangle}{:}_{_K}.
$$
Hence this identity may be written without suffix
${:}\,\,{:}_{_K}$. On the other hand, the $*_{_K}$-product 
of ordinary exponential function and 
a holomorphic function $f({\pmb u})$ is  
\begin{equation*} 
e^{\frac{1}{i\h}\langle{\pmb a},{\pmb u}\rangle}{*_{_K}}f({\pmb u}){=}
e^{\frac{1}{i\h}\langle{\pmb a},{\pmb u}\rangle}
f({\pmb u}{+}\frac{1}{2}{\pmb a(K{+}J)}).
\end{equation*}
\end{prop}

Since the notations 
$\pmb u{=}(\tilde{\pmb u},\tilde{\pmb v}){=}(\tilde{\pmb  u},0){+}(0,\tilde{\pmb v}),\,\,
\pmb\xi{=}(\tilde{\pmb\xi},\tilde{\pmb\eta}){=}
(\tilde{\pmb\xi},0){+}(0,\tilde{\pmb\eta}),\,\, 
\pmb x{=}(\tilde{\pmb x},\tilde{\pmb y})
{=}(\tilde{\pmb  x},0){+}(0,\tilde{\pmb y})$, 
we see  
\begin{equation}\label{Fundcal2}
e_*^{\frac{1}{i\h}\langle{\pmb\xi},\,{\pmb u}{-}{\pmb x}\rangle}{=}
e^{-\frac{1}{2i\h}\langle{\tilde{\pmb\xi}}J,\,\tilde{\pmb\eta}\rangle}
e_*^{\frac{1}{i\h}\langle{\tilde{\pmb\xi}},\,\tilde{\pmb u}{-}\tilde{\pmb x}\rangle}{*}
e_*^{\frac{1}{i\h}\langle{\tilde{\pmb\eta}},\,\tilde{\pmb v}{-}\tilde{\pmb y}\rangle}{=}
e^{-\frac{1}{2i\h}\langle{\tilde{\pmb\eta}}J,\,\tilde{\pmb\xi}\rangle}
e_*^{\frac{1}{i\h}\langle{\tilde{\pmb\eta}},\,\tilde{\pmb v}{-}\tilde{\pmb y}\rangle}{*}
e_*^{\frac{1}{i\h}\langle{\tilde{\pmb\xi}},\,\tilde{\pmb u}{-}\tilde{\pmb x}\rangle},
\end{equation}
where we make attention to tricky notations 
$\langle{\tilde{\pmb\xi}}J,\tilde{\pmb\eta}\rangle{=}-\sum_{i=1}^m\tilde{\xi}_i\tilde{\eta}_i$ 
and $\langle{\tilde{\pmb\eta}}J,\tilde{\pmb\xi}\rangle{=}\sum_{i=1}^m\tilde{\eta}_i\tilde{\xi}_i$.

\bigskip
According to the notation ${\pmb\xi}{=}(\tilde{\pmb\xi}, \tilde{\pmb\eta})$ 
we denote the expression parameter $K$ by  
$$
K{=}
\begin{bmatrix}
K_{_{\btt\btt}}&K_{_{\btt\ctt}}\\
K_{_{\ctt\btt}}&K_{_{\ctt\ctt}}
\end{bmatrix}.
$$
If $K\in{\mathfrak H}_+(V)$, then $(iK_{_{\btt\btt}})_{re}$ 
and $(iK_{_{\ctt\ctt}})_{re}$ are positive definite on $V\cap{\mathbb C}^{m}_{\tilde{\pmb\xi}}$, 
$V\cap{\mathbb C}^{m}_{\tilde{\pmb\eta}}$, respectively.
Hence
$$
{:}e_*^{\frac{1}{i\h}\langle\tilde{\pmb\xi},\tilde{\pmb u}\rangle}{:}_{_K}{=}
e^{\frac{1}{4i\h}\langle\tilde{\pmb\xi}K_{_{\btt\btt}}, \tilde{\pmb\xi}\rangle}
e^{\frac{1}{i\h}\langle\tilde{\pmb\xi},\tilde{\pmb u}\rangle},\quad 
{:}e_*^{\frac{1}{i\h}\langle\tilde{\pmb\eta},\tilde{\pmb v}\rangle}{:}_{_K}{=}
e^{\frac{1}{4i\h}\langle\tilde{\pmb\eta}K_{_{\ctt\ctt}}, \tilde{\pmb\eta}\rangle}
e^{\frac{1}{i\h}\langle\tilde{\pmb\eta},\tilde{\pmb v}\rangle}
$$
are rapidly decreasing on 
$\tilde{\pmb\xi}{\in}V\cap{\mathbb C}^{m}_{\tilde{\pmb\xi}}$, 
$\tilde{\pmb\eta}{\in}V\cap{\mathbb C}^{m}_{\tilde{\pmb\eta}}$ respectively. 

\bigskip

The next theorem is fundamental and very useful: 
\begin{thm}\label{breakthru}
In a $V$-class expression $K (\in{\mathfrak H}_+(V))$,  
$$
{:}\int_{V}e_*^{\frac{1}{i\h}\langle\pmb\xi,{\pmb u}{-}{\pmb x}\rangle}\dbar{V_\xi}{:}_{_K}{=}
\frac{2^m}{\sqrt{\det iK}}
\,e^{{-}\frac{1}{\h}({\pmb u}{-}{\pmb x})\frac{1}{iK}\,{}^t({\pmb u}{-}{\pmb x})}
$$ 
holds without sign ambiguity and 
rapidly decreasing w.r.t. ${\pmb x}{\in}V$ with the order of
$e^{-\frac{1}{c_{_K}}|\pmb x|^2}$, and it is an entire function
w.r.t. $\pmb u$. 
Similarly,  
$$
{:}\int_{V\cap\,{\mathbb C}^{m}_{\tilde{\pmb\xi}}}
e_*^{\frac{1}{i\h}
\langle\tilde{\pmb\xi},\tilde{\pmb u}{-}\tilde{\pmb x}\rangle}\dbar{V_{\tilde{\pmb\xi}}}
{:}_{K_{_{\btt\btt}}}{=}
\frac{\sqrt{2}^m}{\sqrt{\det iK_{\ctt\ctt}}}
 \,e^{{-}\frac{1}{\h}(\tilde{\pmb u}{-}\tilde{\pmb x})
    \frac{1}{iK_{_{\btt\btt}}}\,{}^t(\tilde{\pmb u}{-}\tilde{\pmb x})}
$$ 
is defined without sign ambiguity and rapidly decreasing on 
$V\cap{\mathbb C}^{m}_{\tilde{\pmb x}}$ with the order of
$e^{-\frac{1}{c_{K_{_{\btt\btt}}}}|\tilde{\pmb x}|^2}$, and it is an 
entire function of $\tilde{\pmb u}$. 
\end{thm}

\noindent
{\bf Proof}\,\,\,We show first the first statement.
 Recall that   
${:}e_*^{\frac{1}{i\h}\langle{\pmb\xi, \,{\pmb u}{-}{\pmb x}\rangle}}{:}_{_K}{=}
e^{\frac{1}{4i\h}\langle{\pmb\xi}K,\,{\pmb\xi}\rangle}
e^{\frac{1}{i\h}\langle{\pmb\xi, \,{\pmb u}{-}{\pmb x}\rangle}}$, we see 
$$
\begin{aligned}
&\frac{1}{4i\h}\langle{\pmb\xi}K,\,{\pmb\xi}\rangle{+}
\frac{1}{i\h}\langle{\pmb\xi}, \,{\pmb u}{-}{\pmb x}\rangle{=}
{-}\frac{1}{4\h}\Big({\pmb\xi}iK\,{}^t\!{\pmb\xi}{+}
4i{\pmb\xi}{}^t\!({\pmb u}{-}{\pmb x})\Big)\\
&{=}
{-}\frac{1}{4\h}
\Big({\pmb\xi}{\sqrt{iK}}{+}2i({\pmb u}{-}{\pmb x})\frac{1}{\sqrt{iK}}\Big)
\,{}^t\!\Big({\pmb\xi}{\sqrt{iK}}{+}2i({\pmb u}{-}{\pmb x})\frac{1}{\sqrt{iK}}\Big)
{-}\frac{1}{\h}({\pmb u}{-}{\pmb x})\frac{1}{iK}\,{}^t({\pmb u}{-}{\pmb x}).
\end{aligned}
$$
Hence, the integral 
$$
\int_{V} e^{{-}\frac{1}{4\h}
\big({\pmb\xi}{\sqrt{iK}}{+}2i({\pmb u}{-}{\pmb x})\frac{1}{\sqrt{iK}}\big)
\,{}^t\!\big({\pmb\xi}{\sqrt{iK}}{+}2i({\pmb u}{-}{\pmb x})\frac{1}{\sqrt{iK}}\big)}
\dbar{V_{\pmb\xi}}
$$
converges, and is independent of ${\pmb u}{-}{\pmb x}$, for 
replacing $2i({\pmb u}{-}{\pmb x})\frac{1}{\sqrt{iK}}$ by ${\pmb\alpha}{\sqrt{iK}}$
$$
\frac{\partial}{\partial\alpha_i}
\int_{V} e^{{-}\frac{1}{4\h}
\big(({\pmb\xi}{+}{\pmb\alpha}){\sqrt{iK}}\big)
\,{}^t\!\big(({\pmb\xi}{+}{\pmb\alpha}){\sqrt{iK}}\big)}
\dbar{V_{\pmb\xi}}
{=}
\int_{V}\frac{\partial}{\partial\xi_i}
e^{{-}\frac{1}{4\h}
\big(({\pmb\xi}{+}{\pmb\alpha}){\sqrt{iK}}\big)
\,{}^t\!\big(({\pmb\xi}{+}{\pmb\alpha}){\sqrt{iK}}\big)}
\dbar{V_{\pmb\xi}}{=}0.
$$
By Lemma\,\ref{Lemma1}, we get 
$$
\int_{V}{:}e_*^{\frac{1}{i\h}
\langle{\pmb\xi, \,{\pmb u}{-}{\pmb x}\rangle}}{:}_{_K}\dbar{V_{\pmb\xi}}
{=}
\int_{V} e^{\frac{1}{4i\h}{\pmb\xi}{K}\,{}^t{\pmb\xi}}
\dbar{V_{\pmb\xi}}
\,\,e^{{-}\frac{1}{\h}({\pmb u}{-}{\pmb x})\frac{1}{iK}\,{}^t({\pmb u}{-}{\pmb x})}
{=}\frac{2^m}{\sqrt{\det iK}}
e^{{-}\frac{1}{\h}({\pmb u}{-}{\pmb x})\frac{1}{iK}\,{}^t({\pmb u}{-}{\pmb x})}.
$$  
Clearly, this  is rapidly decreasing w.r.t. $\pmb x\in V$.
The second one is similar.  \hfill $\Box$

\subsection{Star-delta functions  in $V$-class expressions}

Keeping the Fourier transform of $1$ in mind, we define 
$*$-{\it delta functions of full-variables} 
$\delta_*^{(V)}({\pmb u}{-}{\pmb x})$, and 
$*$-{\it delta functions of half-variables}  
$\delta_*^{(V)}(\tilde{\pmb u}{-}\tilde{\pmb x})$, 
$\delta_*^{(V)}(\tilde{\pmb v}{-}\tilde{\pmb y})$  
in a $V$-class expression as follows:
\begin{equation}
\begin{aligned}
&{:}\delta_*^{(V)}({\pmb u}{-}{\pmb x}){:}_{_K}{=}\!
\int_{V}
{:}e_*^{\frac{1}{i\h}
\langle\pmb\xi ,{\pmb u}{-}{\pmb x}\rangle}{:}_{_K}\dbar{V_{\pmb\xi}}
{=}\int_{V}
{:}e_*^{\frac{1}{i\h}
\langle\tilde{\pmb\xi},\tilde{\pmb u}{-}\tilde{\pmb x}\rangle{+}
\langle\tilde{\pmb\eta} ,\tilde{\pmb v}{-}\tilde{\pmb y}\rangle}
{:}_{_K}\dbar{V_{\tilde{\pmb\xi}}}\dbar{V_{\tilde{\pmb\eta}}},
\\
&{:}\delta_*^{(V)}(\tilde{\pmb u}{-}\tilde{\pmb x}){:}_{_K}{=}\!
\int_{V\cap\,{\mathbb C}^m_{\tilde{\pmb\xi}}}
{:}e_*^{\frac{1}{i\h}
\langle\tilde{\pmb\xi},\tilde{\pmb u}{-}\tilde{\pmb x}\rangle}{:}_{_K}
\dbar{V_{\tilde{\pmb\xi}}},
\quad
{:}\delta_*^{(V)}(\tilde{\pmb v}{-}\tilde{\pmb y}){:}_{_K}{=}\!
\int_{V\cap\,{\mathbb C}^m_{\tilde{\pmb\eta}}}
{:}e_*^{\frac{1}{i\h}
\langle\tilde{\pmb\eta},\tilde{\pmb v}{-}\tilde{\pmb  y}\rangle}{:}_{_K}
\dbar{V_{\tilde{\pmb\eta}}},
\end{aligned}
\end{equation}
where $\dbar{V_{\tilde{\pmb\xi}}}$, $\dbar{V_{\tilde{\pmb\eta}}}$ are
standard volume element of $V\cap\,{\mathbb C}^m_{\tilde{\pmb\xi}}$, 
$V\cap\,{\mathbb C}^m_{\tilde{\pmb\eta}}$ fractionalized by $(2\pi\h)^{m/2}$. 
By Theorem\,\ref{breakthru}, these are rapidly decreasing w.r.t. 
$\tilde{\pmb x},\tilde{\pmb y}$ and entire functions w.r.t. 
$\tilde{\pmb u},\tilde{\pmb v}$ respectively.

\medskip
As 
$\frac{1}{i\h}\tilde{u}_i{*}e_*^{\frac{1}{i\h}\langle\tilde{\pmb\xi},\tilde{\pmb u}\rangle}
{=}\partial_{\tilde{\xi}_i}
e_*^{\frac{1}{i\h}\langle\tilde{\pmb\xi},\tilde{\pmb u}\rangle}$
we have the property similar to a usual $delta$ function:
$$
\frac{1}{i\h}(\tilde{u}_i{-}\tilde{x}_i){*}\delta_{*}^{(V)}(\tilde{\pmb u}{-}\tilde{\pmb x})
{=}0{=}\delta_{*}^{(V)}(\tilde{\pmb u}{-}\tilde{\pmb x})
{*}\frac{1}{i\h}(\tilde{u}_i{-}\tilde{x}_i).
$$
As changing the variables shows that  
$\dbar{V_{\tilde{\pmb x}}}{=}e^{i\theta}\dbar{\tilde{\pmb x}'}$,
$\tilde{\pmb x}'{\in}{\mathbb R}^n$, and  there is a linear transformation $T$ 
$$
\int_{V\cap\,{\mathbb C}^m_{\tilde{\pmb\xi}}}
e^{\frac{1}{i\h}\langle\tilde{\pmb\xi},\tilde{\pmb x}\rangle}\dbar{V_{\tilde{\pmb x}}}{=}
e^{i\theta}\int_{{\mathbb R}^m}
e^{\frac{1}{i\h}\langle\tilde{\pmb\xi}T,\tilde{\pmb x}'\rangle}\dbar{\tilde{\pmb x}'}
{=}e^{i\theta}\delta(\tilde{\pmb\xi}T).
$$
For the convenience of computations we define 
\begin{equation}
\delta^{(V)}(\tilde{\pmb\xi}){=}e^{i\theta}\delta(\tilde{\pmb\xi}T).
\end{equation}

Hence
\begin{equation}
\int_{V\cap\,{\mathbb C}^m_{\tilde{\pmb\xi}}}
{:}\delta_*^{(V)}(\tilde{\pmb u}{-}\tilde{\pmb x}){:}_{_K}\dbar{V_{\tilde{\pmb x}}}{=}1,\quad
\int_{V\cap\,{\mathbb C}^m_{\tilde{\pmb\eta}}}
{:}\delta_*^{(V)}(\tilde{\pmb v}{-}\tilde{\pmb y}){:}_K
\dbar{V_{\tilde{\pmb y}}}{=}1.
\end{equation}

The $*$-product of $*$-delta functions is given as follows:   
\begin{equation}\label{proddeltatilde}
{:}\delta_*^{(V)}(\tilde{\pmb u}{-}\tilde{\pmb x}){*}
\delta^{(V)}_*(\tilde{\pmb u}{-}\tilde{\pmb x}'){:}_{_K}{=}
\int_{V}
\delta^{(V)}(\tilde{\pmb x}{-}\tilde{\pmb x}')
{:}e_*^{\frac{1}{i\h}\langle\tilde{\pmb\zeta}, \tilde{\pmb u}{-}\tilde{\pmb x}'\rangle}{:}_{_K}
\dbar{V_{\tilde{\pmb\zeta}}}
=\delta^{(V)}(\tilde{\pmb x}{-}\tilde{\pmb x}')
{:}\delta^{(V)}_*(\tilde{\pmb u}{-}\tilde{\pmb x}'){:_{_K}}.
\end{equation}
 We have also 
\begin{equation}\label{proddeltatilde2}
{:}\delta^{(V)}_*(\tilde{\pmb v}{-}\tilde{\pmb y}){*}
\delta^{(V)}_*(\tilde{\pmb v}{-}\tilde{\pmb y}'){:}_{_K}{=}
\delta^{(V)}(\tilde{\pmb y}{-}\tilde{\pmb y}')
{:}\delta^{(V)}_*(\tilde{\pmb v}{-}\tilde{\pmb y}'){:}_{_K}.
\end{equation}
\bigskip 
For a tempered distribution $h(\tilde{\pmb x})$ on
$V$, we define the inverse Fourier transform by 
$$
\check{h}^{(V)}(\tilde{\pmb\xi}){=}\int_{V}h(\tilde{\pmb x})
e^{-\frac{1}{i\h}\langle\tilde{\pmb\xi},\,\tilde{\pmb x}\rangle}\dbar{V_{\tilde{\pmb x}}},\quad 
\tilde{\pmb\xi}\in V\cap\,{\mathbb C}^m_{\tilde{\pmb\xi}}.
$$
Putting the reciprocity
$h(\tilde{\pmb x}'{-}\tilde{\pmb x}){=}
\int_{V}\check{h}^{(V)}(\tilde{\pmb\xi})
e^{\frac{1}{i\h}\langle\tilde{\pmb\xi},\,\tilde{\pmb x}'{-}\tilde{\pmb x}\rangle}\dbar{V_{\tilde{\pmb\xi}}}$
in mind, we define for $K{\in}{\mathfrak H}_+(V)$, a $*$-function 
\begin{equation}\label{StarFunc00}
{:}h_*^{(V)}(\tilde{\pmb u}{-}\tilde{\pmb x}){:}_K{=}
\int_{V}\check{h}^{(V)}(\tilde{\pmb\xi})
{:}e_*^{\frac{1}{i\h}
\langle\tilde{\pmb\xi},\,\tilde{\pmb u}{-}\tilde{\pmb x}\rangle} {:}_K
\dbar{V_{\tilde{\pmb\xi}}}{=}
\int_{V}\int_{V}{h}(\tilde{\pmb x}')
{:}e_*^{\frac{1}{i\h}\langle\tilde{\pmb\xi},\,
\tilde{\pmb u}{-}\tilde{\pmb x}'{-}\tilde{\pmb x}\rangle}{:}_K
\dbar{V_{\tilde{\pmb x}'}}\dbar{V_{\tilde{\pmb\xi}}}.
\end{equation}
As ${:}e_*^{\frac{1}{i\h}\langle\tilde{\pmb\xi},\,\tilde{\pmb u}\rangle}{:}_K$
is rapidly decreasing under $V$-class expressions,
one may write this by 
$$
{:}h^{(V)}_*(\tilde{\pmb u}{-}\tilde{\pmb x}){:}_{_K}{=}
\int_{V}h(\tilde{\pmb x})
{:}\delta^{(V)}_*(\tilde{\pmb u}{-}\tilde{\pmb x}'{-}\tilde{\pmb x}){:}_{_K}\dbar{V_{\tilde{\pmb x}'}}.
$$
By Theorem\,\ref{breakthru}
${:}h^{(V)}_*(\tilde{\pmb u}{-}\tilde{\pmb x}){:}_{_K}$ is rapidly decreasing 
w.r.t. $\pmb x{\in}V$ in a $V$-class expression, and an entire function w.r.t. $\tilde{\pmb u}$.

\bigskip
Let $\check{f}(\tilde{\pmb\xi})$, $\check{g}(\tilde{\pmb\xi})$ be
tempered distributions on $V$. Suppose the 
convolution product is welldefined as a tempered distribution. Then, 
$$
{:}f^{(V)}_*(\tilde{\pmb u}){*}g^{(V)}_*(\tilde{\pmb u}){:}_{_K}{=}
\int_{V}\Big(\int_{V}
\check{f}(\tilde{\pmb\xi})\check{g}(\tilde{\pmb\zeta}{-}\tilde{\pmb\xi})
\dbar\tilde{\pmb\xi}\Big)
{:}e_*^{\frac{1}{i\h}\langle\tilde{\pmb\zeta}, \tilde{\pmb u}{-}\tilde{\pmb x}
\rangle}{:}_{_K}
\dbar{V_{\tilde{\pmb x}}}\dbar{V_{\tilde{\pmb\zeta}}}{=}
\int_{V}(gh)(\tilde{\pmb x})
{:}\delta^{(V)}(\tilde{\pmb u}{-}\tilde{\pmb x}){:}_{_K}\dbar{V_{\tilde{\pmb x}}}
$$
by noting that 
${:}e_*^{\frac{1}{i\h}\langle\tilde{\pmb\zeta}, \tilde{\pmb u}{-}\tilde{\pmb x}\rangle}{:}_{_K}$
is rapidly decreasing in $\tilde{\pmb\zeta}$ under any $V$-class expression parameters. 

The basic properties of these $*$-functions are 
$$
f_*^{(V)}(\tilde{\pmb u}){*}g_*^{(V)}(\tilde{\pmb u}){=}
\int_{V}(fg)(\tilde{\pmb x})
\delta^{(V)}_*(\tilde{\pmb u}{-}\tilde{\pmb x})\dbar{V_{\tilde{\pmb x}}},
\quad 
[\tilde{v}_i, h^{(V)}_*(\tilde{\pmb u})]{=}
\int_{V}i\h\partial_{\tilde{x}_i}{h}(\tilde{\pmb x})
\delta^{(V)}_*(\tilde{\pmb u}{-}\tilde{\pmb x})\dbar{V_{\tilde{\pmb x}}}.
$$

%

\bigskip
By setting ${\pmb x}{=}
(\tilde{\pmb x}, \tilde{\pmb y})$, 
the above arguments may be applied to the case of
``full-variables.'' 
\begin{prop}
If $f(\tilde{\pmb y})$ is a tempered distribution  on
$V\cap\,{\mathbb C}^m_{\tilde{\pmb\eta}}$, then 
$\int_{V}f(\tilde{\pmb y})
{:}\delta^{(V)}_*(\pmb u-\pmb x){:}_{_K}\dbar{V_{\pmb x}}$
is a half-variable $*$-function  
$$
\int_{V}f(\tilde{\pmb y}){:}\delta^{(V)}_*(\pmb u-\pmb x){:}_{_K}\dbar{V_{\pmb x}}{=}
\int_{V\cap\,{\mathbb C}^m_{\tilde{\pmb y}}}
f(\tilde{\pmb y}) 
{:}\delta_*(\tilde{\pmb v}-\tilde{\pmb y}){:}_{_K} 
\dbar{V_{\tilde{\pmb y}}}
{=}{:}f^{(V)}_*(\tilde{\pmb v}){:}_{_K}.
$$
\end{prop}

\noindent
{\bf Proof}\,\, Note that 
${\pmb x}{=}(\tilde{\pmb x}, \tilde{\pmb y}),\quad 
{\pmb u}{=}(\tilde{\pmb u}, \tilde{\pmb v}),\quad 
\dbar{V_{\pmb x}}=\dbar{V_{\tilde{\pmb x}}}\,\dbar{V_{\tilde{\pmb y}}}, \quad 
\dbar{V_{\pmb\xi}}=\dbar{V_{\tilde{\pmb\xi}}}\,\,\dbar{V_{\tilde{\pmb\eta}}}$
and \eqref{Fundcal2} gives 
$$ 
\iint{:}\delta^{(V)}_*(\pmb u-\pmb x){:}_{_K}\dbar{V_{\tilde{\pmb\xi}}}\,\dbar{V_{\tilde{\pmb x}}}{=}
\iint e^{-\frac{1}{2i\h}\langle\tilde{\pmb\xi}J,\tilde{\pmb\eta\rangle}}
{:}e_*^{\frac{1}{i\h}\langle\tilde{\pmb\xi},\tilde{\pmb u}{-}\tilde{\pmb x}\rangle}{*}
e_*^{\frac{1}{i\h}\langle\tilde{\pmb\eta},\tilde{\pmb v}{-}\tilde{\pmb y}\rangle}{:}_{_K}
\dbar{V_{\tilde{\pmb\xi}}}\,\dbar{V_{\tilde{\pmb x}}}
{=}{:}e_*^{\frac{1}{i\h}\langle\tilde{\pmb\eta},\tilde{\pmb v}{-}\tilde{\pmb y}\rangle}{:}_{_K},
$$
for integrating by $\tilde{\pmb x}$ first gives   
$\int e^{-\frac{1}{i\h}\langle\tilde{\pmb\xi},\tilde{\pmb x}\rangle}\dbar{V_{\tilde{\pmb x}}}
{=}\delta^{(V)}(\tilde{\pmb\xi})$.   
Plugging this  we obtain the result. ${}$\hfill $\Box$

\bigskip
Now, let $\rho(\tilde{\pmb y})$ be a tempered distribution on 
$V\cap\,{\mathbb C}^m_{\tilde{\pmb\eta}}$ or a continuous function on 
$V\cap\,{\mathbb C}^m_{\tilde{\pmb\eta}}$ such
that $e^{it\rho(\tilde{\pmb y})}$ is the growth order 
$e^{c|\tilde{\pmb y}|^{\alpha}}$, $\alpha{<}2$. 
Then, we have a remarkable result as follows: 
\begin{prop}\label{apple}
 In  $V$-class expressions, 
${:}e_*^{it\rho_*(\tilde{\pmb v})}{:}_{_K}{=}
\int_{V}e^{it\rho(\tilde{\pmb y})}{:}\delta_*(\pmb u{-}\pmb x){:}_{_K}\dbar{V_{\pmb x}}$
is same to  
$$
{:}e_*^{it\rho_*(\tilde{\pmb v})}{:}_{_K}{=}
\int_{V\cap\,{\mathbb C}^m_{\tilde{\pmb y}}}e^{it\rho(\tilde{\pmb y})}
{:}\delta^{(V)}_*(\tilde{\pmb v}{-}\tilde{\pmb y}){:}_{_K} 
\dbar{V_{\tilde{\pmb y}}}
$$
and it is welldefined one parameter group w.r.t. $t$ and an entire
function w.r.t. $\tilde{\pmb v}$.
\end{prop}

\noindent
{\bf Proof}\,\,
It is enough to prove the group property. 
For this, we show that this satisfies the equation 
$$
\frac{d}{dt}{:}e_*^{it\rho_*(\tilde{\pmb v})}{:}_{_K}
{=}i{:}\rho_*(\tilde{\pmb  v}){*}
\int_{{\mathbb R}^{m}}e^{it\rho(\tilde y)}
\delta^{(V)}_*(\tilde{\pmb v}{-}\tilde{\pmb y})\dbar{V_{\tilde{\pmb y}}}{:}_{_K}.
$$
Note that 
$\frac{d}{dt}{:}e_*^{it\rho_*(\tilde{\pmb v})}{:}_{_K}{=}
\int_{V\cap\,{\mathbb C}^m_{\tilde{\pmb y}}}i\rho(\tilde{\pmb y})
e^{it\rho(\tilde{\pmb y})}
{:}\delta_*(\tilde{\pmb v}{-}\tilde{\pmb y}){:}_{_K}
\dbar{V_{\tilde{\pmb y}}}$, 
and  
$i{:}\rho_*(\tilde{\pmb v}){:}_{_K}
{=}\int_{V\cap\,{\mathbb C}^m_{\tilde{\pmb y}}} 
i\rho(\tilde{\pmb y}'){:}\delta_*(\tilde{\pmb v}{-}\tilde{\pmb y}'){:}_{_K}
\dbar{V_{\tilde{\pmb y'}}}$. 
Hence applying \eqref{proddeltatilde2}, we have  
$$
\iint i\rho(\tilde{\pmb y}')e^{it\rho(\tilde{\pmb y})}
{:}\delta^{(V)}_*(\tilde{\pmb v}-\tilde{\pmb y}'){*}\delta^{(V)}_*(\tilde{\pmb v}-\tilde{\pmb y}){:}_{_K}
\dbar{V_{\tilde{\pmb y'}}}\dbar{V_{\tilde{\pmb y}}}
{=}
\int i\rho(\tilde{\pmb y})e^{it\rho(\tilde{\pmb y})}
{:}\delta^{(V)}_*(\tilde{\pmb v}-\tilde{\pmb y}){:}_{_K}
\dbar{V_{\tilde{\pmb y}}}. 
$$
${ }$\hfill$\Box$
 
\bigskip
Note also that $e_*^{it\rho_*(\tilde{\pmb v})}$ may not be real
analytic in $t$. But the above Theorem shows that for any polynomial
$p(\tilde{\pmb y})$, the  $*$-exponential function 
$e_*^{itp_*(\tilde{\pmb y})}$ is welldefined in any $V$-class
expression as a real one parameter group. However, we have seen in 
\cite{OMMY3} that such one parameter group must have singularities
in the complex domain. Indeed, if $\deg p(\tilde{\pmb y})\geq 3$ then
the radius of convergence of the series 
$\sum_k \frac{z^k}{k!}{:}p_*(\tilde{\pmb v})^k{:}_{_K}$ is $0$
(cf.\cite{OMMY3}, Proposition 1.2).

\medskip
Proposition\,\ref{apple} can be applied to the hybrid case.
Let $\rho(\tilde{\pmb x},\tilde{\pmb y})$ be a tempered distribution on 
$V\cap\,{\mathbb C}^m_{\tilde{\pmb y}}$ w.r.t. $\tilde{\pmb y}$, 
or a continuous function on $V\cap\,{\mathbb C}^m_{\tilde{\pmb y}}$ such
that $e^{it\rho(\tilde{\pmb x},\tilde{\pmb y})}$ is growth order 
$e^{c|\tilde{\pmb y}|^{\alpha}}$, $\alpha{<}2$ w.r.t. $\tilde{\pmb y}$
at each fixed $\tilde{\pmb x}$.

\begin{prop}\label{apple0}
Under a V-class expression, the $*$-exponential function 
$$
e_*^{it\rho_*(\tilde{\pmb x},\tilde{\pmb v})}
{=}
\int_{V\cap\,{\mathbb C}^m_{\tilde{\pmb y}}}e^{it\rho(\tilde{\pmb x},\tilde{\pmb y})}
\delta_*(\tilde{\pmb v}{-}\tilde{\pmb x})
\dbar{V_{\tilde{\pmb x}}}
$$
is welldefined one parameter group w.r.t. $t$ and an entire
function w.r.t. $\tilde{\pmb v}$.
\end{prop}

%

\bigskip
For $K{\in}{\mathfrak H}_{+}(V)$, we have defined 
$\delta^{(V)}_*(\tilde{\pmb v})$, $\delta^{(V)}_*(\tilde{\pmb u})$
separately. By setting ${\pmb u}{=}(\tilde{\pmb u},\tilde{\pmb v})$, 
${\pmb\xi}{=}(\tilde{\pmb\xi},\tilde{\pmb\eta})$, $\pmb
x{=}(\tilde{\pmb x},\tilde{\pmb y})$, 
the $K$-ordered expression of the $*$-product 
$\delta^{(V)}_*(\tilde{\pmb v}{-}\tilde{\pmb y})
{*}\delta^{(V)}_*(\tilde{\pmb u}{-}\tilde{\pmb x})$ is computed as
follows:
$$
{:}\delta^{(V)}_*(\tilde{\pmb v}{-}\tilde{\pmb y})
{*}\delta^{(V)}_*(\tilde{\pmb u}{-}\tilde{\pmb x}){:}_{_K}{=}
\iint e^{\frac{1}{2i\h}\langle\tilde{\pmb\eta}J,\,\tilde{\pmb\xi}\rangle}
{:}e_*^{\frac{1}{i\h}\langle{\pmb\xi},\,{\pmb u}{-}{\pmb x}\rangle}{:}_{_K}
\dbar{V_{\tilde{\pmb\xi}}}\dbar{V_{\tilde{\pmb\eta}}}{=}
\int_{V}e^{-\frac{1}{4\h}\langle{\pmb\xi}i(K{+}C),\,{\pmb\xi}\rangle}
e^{\frac{1}{i\h}\langle{\pmb\xi},\,{\pmb u}{-}{\pmb x}\rangle}\dbar{V_{\pmb\xi}},
$$
where
$\langle\tilde{\pmb\eta}J,\,\tilde{\pmb\xi}\rangle{=}\sum_i\tilde{\eta}_i\tilde{\xi}_i$, \,
$C=
{\footnotesize{
\begin{bmatrix}
0&I\\
I&0
\end{bmatrix}}}$.
So, suppose 
$K{+}
{\footnotesize{
\begin{bmatrix}
0&I\\
I&0
\end{bmatrix}}}$ is non-singular. Then
\begin{equation}\label{formvac}
{:}\delta^{(V)}_*(\tilde{\pmb v}{-}\tilde{\pmb y})
{*}\delta^{(V)}_*(\tilde{\pmb u}{-}\tilde{\pmb x}){:}_{_K}{=}
\frac{2^m}{\sqrt{\det(i(K{+}C))}}
e^{-\frac{1}{\h}\langle(\pmb u{-}\pmb x)\frac{1}{i(K{+}C)},\pmb u{-}\pmb x\rangle},\quad
{\pmb u}{=}(\tilde{\pmb u},\tilde{\pmb v}), \,\,{\pmb x}{=}(\tilde{\pmb x},\tilde{\pmb y}) .
\end{equation}
Note that the r.h.s. makes sense whenever $K{+}C$ is non-singular, but
this has $\pm$ sign ambiguity in $\sqrt{\det(i(K{+}C))}$. In
\S\,\ref{VacVac}, it will be shown that 
${:}\delta^{(V)}_*(\tilde{\pmb v}{-}\tilde{\pmb y})
{*}\delta^{(V)}_*(\tilde{\pmb u}{-}\tilde{\pmb x}){:}_{_K}$ is an
idempotent element, and this relates what we called the vacuum.

\subsection{Several properties of ${*}$-functions of full-variables}

On the other hand, we have defined the $*$-{\it delta function} (of full
variables) as follows:
\begin{equation}\label{fulldelta}
{:}\delta^{(V)}_*({\pmb u}{-}{\pmb x}){:}_{_K}{=}\!\int_{V}
{:}e_*^{\frac{1}{i\h}\langle{\pmb\xi},{\pmb u}{-}{\pmb x}\rangle}
{:}_{_K}\dbar{V_{\pmb\xi}},\quad {\pmb x}{\in} V.
\end{equation}
We note here that the exponential law gives 
$$ 
e_*^{\frac{1}{i\h}\langle{\pmb\xi},{\pmb u}{-}{\pmb x}\rangle}{=}
e_*^{\frac{1}{i\h}\langle{\pmb\xi},{\pmb u}\rangle}
e^{-\frac{1}{i\h}\langle{\pmb\xi},{\pmb x}\rangle},\quad
{\text{and}}\quad 
\int_{V} e_*^{\frac{1}{i\h}\langle{\pmb\xi},{\pmb u}{-}{\pmb x}\rangle}
\dbar{V_{\pmb x}}{=}\delta^{(V)}(\pmb\xi)e_*^{\frac{1}{i\h}\langle{\pmb\xi},{\pmb u}\rangle}.
$$ 
It follows  
\begin{equation}
\int_{V}\delta^{(V)}_*({\pmb u}{-}{\pmb x})\dbar{\pmb x}{=}1. 
\end{equation}

%
%

\medskip
Let $f({\pmb x})$ be a tempered distribution on
$V$ and let 
$\check{f}^{(V)}(\pmb\xi){=}\int_{V}f({\pmb x})
e^{-\frac{1}{i\h}\langle{\pmb\xi},\,{\pmb x}\rangle}\dbar{V_{\pmb  x}}$
be the inverse Fourier transform. 
Noting the wellknown reciprocity formula 
$f({\pmb x}){=}
\int_{V}\check{f}^{(V)}({\pmb\xi})
e^{\frac{1}{i\h}\langle{\pmb\xi},\,{\pmb x}\rangle}\dbar{\pmb\xi}$, 
we define $*$-function corresponding to $f(\pmb x)$ as 
\begin{equation}\label{StarFunc}
{:}f_*^{(V)}({\pmb u}{-}{\pmb x}){:}_{_K}{=}
\int_{V}\check{f}^{(V)}({\pmb\xi})
{:}e_*^{\frac{1}{i\h}\langle{\pmb\xi},\,{\pmb u}{-}{\pmb x}\rangle}{:}_{_K}
\dbar{V_{\pmb x}}
{=}
\int_{V}\int_{V}{f}({\pmb x}')
 e^{-\frac{1}{i\h}\langle{\pmb\xi},\,{\pmb x}'\rangle}
{:}e_*^{\frac{1}{i\h}\langle{\pmb\xi},\,{\pmb u}{-}{\pmb x}\rangle}{:}_{_K}
\dbar{V_{\pmb x'}}\dbar{V_{\pmb\xi}}.
\end{equation}
The Weyl ordered ($K{=}0$) expression of 
$\delta_*^{(V)}(\pmb u{-}\pmb x)$ is given by 
${:}\delta_*^{(V)}(\pmb u{-}\pmb x){:}_{0}{=}\delta^{(V)}(\pmb u{-}\pmb x)$. 
Thus we see 
\begin{equation}\label{WeylWeyl}
{:}f^{(V)}_*(\pmb u){:}_0{=}\int f(\pmb x)\delta^{(V)}(\pmb u{-}\pmb x)\dbar{\pmb x}{=}f(\pmb u).
\end{equation}
\begin{prop}\label{$*$-funcWeyl}
The inverse of the correspondence $f(\pmb x)\to f^{(V)}_*(\pmb u)$ is
given by its Weyl ordered expression and replacement of $\pmb u$ by
$\pmb x$. 
\end{prop}

\medskip
Applying the exponential law in the r.h.s. gives 
\begin{equation}\label{Vclass}
{:}f_*^{(V)}({\pmb u}{-}{\pmb x}){:}_{_K}{=}
\int_{V}\int_{V}{f}({\pmb x}')
{:}e_*^{\frac{1}{i\h}\langle{\pmb\xi},\,{\pmb u}{-}{\pmb x}'{-}{\pmb x}\rangle}{:}_{_K}
\dbar{V_{\pmb x'}}\dbar{V_{\pmb\xi}}{=}
\int_{V}{f}({\pmb x}')
{:}\delta_*^{(V)}({\pmb u}{-}{\pmb x}'{-}{\pmb x}){:}_{_K}\dbar{V_{\pmb x'}}.
\end{equation}

\medskip
The next theorem is the main tool to extend the class of 
$*$-functions via Fourier transform.
\begin{thm}
  \label{funddefm}
For every tempered distribution $f({\pmb x})$ on $V$, the $V$-
class expression \eqref{Vclass} of the integral 
$\int_{V}f(\pmb x')\delta^{(V)}_*({\pmb u}{-}\pmb x'{-}\pmb x)\dbar{V_{\pmb x'}}$ 
is rapidly decreasing w.r.t. $\pmb x$ and an entire function of ${\pmb u}$. 
In particular we see 

$$
{:}\delta^{(V)}_*({\pmb u}{-}\pmb a){:}_{_K}
=\int_{V}\delta^{(V)}({\pmb x}{-}\pmb a)
{:}\delta^{(V)}_*({\pmb u}{-}\pmb x){:}_{_K} \dbar{V_{\pmb x}}.
$$
Moreover, even if $f(x)$ is $e^{|x|^{\alpha}}$\!\!-growth on $V$ with 
$0{<}\alpha{<}2$, the integral 
$\int_{V}\!f(\pmb x){:}\delta^{(V)}_*({\pmb u}{-}\pmb x){:}_{_K}\dbar{V_{\pmb{x}}}$ 
is well-defined to give an entire function w.r.t. $\pmb u$.
\end{thm} 

\noindent
{\bf Proof}\,\,The $K$-expression is 
$\int_{V}\int_{V}f(\pmb x)
e^{-\frac{1}{4\h}\langle{\pmb\xi}(iK),{\pmb\xi}\rangle
  {+}\frac{1}{i\h}\langle{\pmb\xi},{\pmb u}{-}{\pmb x}\rangle}
\dbar{V_{\pmb\xi}}\dbar{V_{\pmb x}}$. 
By restricting ${\pmb u}$ to an arbitrary compact subset of 
$\mathbb C^{2m}$, Theorem\,\ref{breakthru} of [24] gives that  
$\int_{V}e^{-\frac{1}{4\h}\langle{\pmb\xi}(iK),{\pmb\xi}\rangle
  {+}\frac{1}{i\h}\langle{\pmb\xi},{\pmb u}{-}{\pmb x}\rangle}$ is rapidly decreasing 
w.r.t. $\pmb x{\in}V$ with the growth order $e^{-c_{_K}|\pmb x|^2}$. 
Hence, the integral  
$\int_{V}f({\pmb x})\delta_*({\pmb u}{-}{\pmb x})\dbar{V_{\pmb x}}$ 
exists for every tempered distribution $f(\pmb x)$ on $V$. 
Since the complex differentiation $\partial_{u_i}$ does not suffer 
the convergence, we see that this is holomorphic w.r.t. $\pmb u$.  \hfill $\Box$

\bigskip
\noindent
{\bf Warning}\, 
${:}\partial_{\xi_i}e_*^{\frac{1}{i\h}\langle{\pmb\xi},{\pmb u}\rangle}{:}_{_K}{=}
\partial_{\xi_i}{:}e_*^{\frac{1}{i\h}\langle{\pmb\xi},{\pmb u}\rangle}{:}_{_K}
=\partial_{\xi_i}\big(e^{\frac{1}{4i\h}\langle{\pmb\xi}K,{\pmb\xi}\rangle
{+}\frac{1}{i\h}\langle{\pmb\xi},{\pmb u}\rangle}\big).
$
But this is {\it not} ${:}\frac{1}{i\h}u_i{*}e_*^{\frac{1}{i\h}\langle{\pmb\xi},{\pmb u}\rangle}{:}_{_K}$. 
It is very easy to make a confusion. We have in fact
$$
{:}\frac{1}{i\h}u_i{*}e_*^{\frac{1}{i\h}\langle{\pmb\xi},{\pmb u}\rangle}{:}_{_K}
{=}
e^{\frac{1}{4i\h}\langle{\pmb\xi}K,{\pmb\xi}\rangle}
\frac{1}{i\h}u_i{*_{_K}}e^{\frac{1}{i\h}\langle{\pmb\xi},{\pmb u}\rangle}{=}
 e^{\frac{1}{4i\h}\langle{\pmb\xi}K,{\pmb\xi}\rangle}
\Big(\frac{1}{i\h}u_i
{+}\frac{1}{2}\sum_{j}\big((K{+}J)_{ij}{+}(K{+}J)_{ji}\big)\xi_j\Big)
e^{\frac{1}{i\h}\langle{\pmb\xi},{\pmb u}\rangle},
$$
$$
{:}e_*^{\frac{1}{i\h}\langle{\pmb\xi},{\pmb u}\rangle}{*}\frac{1}{i\h}u_i{:}_{_K}
{=}e^{\frac{1}{4i\h}\langle{\pmb\xi}K,{\pmb\xi}\rangle}
\frac{1}{i\h}u_i{*_{_K}}e^{\frac{1}{i\h}\langle{\pmb\xi},{\pmb u}\rangle}{=}
 e^{\frac{1}{4i\h}\langle{\pmb\xi}K,{\pmb\xi}\rangle}
\Big(\frac{1}{i\h}u_i
{+}\frac{1}{2}\sum_{j}\big((K{-}J)_{ij}{+}(K{-}J)_{ji}\big)\xi_j\Big)
e^{\frac{1}{i\h}\langle{\pmb\xi},{\pmb u}\rangle}.
$$
Summing up to eliminate the terms involving $J$, we see 
\begin{equation}\label{usepolar}
{:}\frac{1}{i\h}u_i{*}e_*^{\frac{1}{i\h}\langle{\pmb\xi},{\pmb u}\rangle}{+}
e_*^{\frac{1}{i\h}\langle{\pmb\xi},{\pmb u}\rangle}{*}\frac{1}{i\h}u_i{:}_{_K}{=}
\partial_{\xi_i}{:}e_*^{\frac{1}{i\h}\langle{\pmb\xi},{\pmb u}\rangle}{:}_{_K}. 
\end{equation}

Thus, we see that 
\begin{prop}\label{anticomgen}
In a $V$-class expression, $\delta_*^{(V)}(\pmb u)$ anti-commutes with
every generator i.e. 
\begin{equation*}
{:}\frac{1}{i\h}u_i{*}\delta^{(V)}_{*}(\pmb u)
{+}\delta_{*}^{(V)}(\pmb u){*}\frac{1}{i\h}u_i{:}_{_K}{=}
\int_{V}\partial_{\xi_i}
{:}e_*^{\frac{1}{i\h}\langle{\pmb\xi},{\pmb u}\rangle}{:}_{_K}d\pmb\xi=0,
\quad i=1\sim 2m.
\end{equation*}
In particular $\delta^{(V)}_{*}(\pmb u){*}f^{(V)}_*(\pmb u){=}
f^{(V)}_*(-\pmb u){*}\delta^{(V)}_{*}(\pmb u)$.
\end{prop}

\noindent
{\bf Proof}\,\,The proof is given also by using \eqref{Fundcal2} and
the integration by parts. For the second identity\\
$\delta^{(V)}_{*}(\pmb u){*}f^{(V)}_*(\pmb u){=}
f^{(V)}_*(-\pmb u){*}\delta^{(V)}_{*}(\pmb u)$, note first that 
$e_*^{\frac{1}{i\h}\langle{\pmb\xi}, 
\pmb u\rangle}{*}\delta^{(V)}_{*}(\pmb u){=}
\delta^{(V)}_{*}(\pmb u)
e_*^{{-}\frac{1}{i\h}\langle{\pmb\xi}, 
\pmb u\rangle}$, 
and hence 
$$
e_*^{\frac{1}{i\h}\langle{\pmb\xi}, 
\pmb u{-}{\pmb x}\rangle}{*}\delta^{(V)}_{*}(\pmb u){=}
\delta^{(V)}_{*}(\pmb u){*}e_*^{-\frac{1}{i\h}\langle{\pmb\xi}, 
\pmb u{+}{\pmb x}\rangle}.
$$
Since 
$f_*^{(V)}(\pmb u)=\int_{V}f(x)\delta_*^{(V)}(\pmb u{-}\pmb x)\dbar{V_{\pmb x}}$,
the changing variables in the integration gives the result.  
 \hfill $\Box$

\bigskip
\noindent
{\bf Adjoint actions.}\,\,
On the other hand, it is easy to see that in every $*_{_K}$-product  
$$
{\rm{ad}}(\frac{1}{i\h}\langle{\pmb a},{\pmb u}\rangle)(u_1,u_2,\cdots,u_{2m}){=}(a_1,a_2,\cdots,a_{2m})J.
$$
It follows 
$e^{{\rm{ad}}(\frac{1}{i\h}\langle{\pmb a},{\pmb u}\rangle)}(u_1,u_2,\cdots,u_{2m}){=}
(u_1,u_2,\cdots,u_{2m}){+}(a_1,a_2,\cdots,a_{2m})J$, 
and   
$$
{\rm{Ad}}(e_*^{\frac{1}{i\h}\langle{\pmb a},{\pmb u}\rangle})(u_1,u_2,\cdots,u_{2m})
{=}(u_1,u_2,\cdots,u_{2m}){+}(a_1,a_2,\cdots,a_{2m})J.
$$
Since ${\rm{Ad}}(e_*^{\frac{1}{i\h}\langle{\pmb a},{\pmb u}\rangle})$
gives a $*$-isomorphism, we have 
\begin{equation}\label{afjoint}
{:}e_*^{\frac{1}{i\h}\langle{\pmb a},{\pmb u}\rangle}
{*}p_*({\pmb u}){*}e_*^{-\frac{1}{i\h}\langle{\pmb a},{\pmb u}\rangle}{:}_{_K}
{=}{:}p_*({\pmb u}{+}{\pmb a}J){:}_{_K},\quad 
{:}e_*^{\frac{1}{i\h}\langle{\pmb a},{\pmb u}\rangle}{*}
e_*^{\frac{1}{i\h}\langle{\pmb\xi},{\pmb u}\rangle}{*}
e_*^{-\frac{1}{i\h}\langle{\pmb a},{\pmb u}\rangle}{:}_{_K}{=}
{:}e_*^{\frac{1}{i\h}\langle{\pmb\xi},{\pmb u}{+}{\pmb a}J\rangle}{:}_{_K}.
\end{equation}
The identity \eqref{afjoint} extends easily to 
\begin{equation}\label{afjointdelta}
\begin{aligned}
&{:}e_*^{\frac{1}{i\h}\langle{\pmb a},{\pmb u}\rangle}
{*}\delta^{(V)}_*({\pmb u}{-}{\pmb x}){*}e_*^{-\frac{1}{i\h}\langle{\pmb a},{\pmb u}\rangle}{:}_{_K}
{=}{:}\delta^{(V)}_*({\pmb u}{-}{\pmb x}{+}{\pmb a}J){:}_{_K},\\
{:}e_*^{\frac{1}{i\h}\langle{\pmb a},{\pmb u}\rangle}
{*}\delta^{(V)}_*(\tilde{\pmb v}{-}\tilde{\pmb y})&{*}e_*^{-\frac{1}{i\h}\langle{\pmb a},{\pmb u}\rangle}{:}_{_K}
{=}{:}\delta^{(V)}_*(\tilde{\pmb v}{-}\tilde{\pmb y}{-}\tilde{\pmb a}){:}_{_K},\quad  
{:}e_*^{\frac{1}{i\h}\langle{\pmb a},{\pmb u}\rangle}
{*}\delta^{(V)}_*(\tilde{\pmb u}{-}\tilde{\pmb x}){*}e_*^{-\frac{1}{i\h}\langle{\pmb a},{\pmb u}\rangle}{:}_{_K}
{=}{:}\delta^{(V)}_*(\tilde{\pmb u}{-}\tilde{\pmb x}{+}\tilde{\pmb b}){:}_{_K},
\end{aligned}
\end{equation}
where ${\pmb a}{=}(\tilde{\pmb a},\tilde{\pmb b})$ and 
${\pmb a}J{=}(\tilde{\pmb b}, {-}\tilde{\pmb a})$.

\medskip
A $*$-delta function $\delta^{(V)}_*({\pmb u})$ of full variables has
a peculiar property. 
The first equality of \eqref{afjointdelta} gives that
$f_t({\pmb u}){=}
{\rm{Ad}}(e_*^{\frac{t}{i\h}\langle{\pmb a},{\pmb u}\rangle})\delta^{(V)}_*({\pmb u})$
satisfies the Heisenberg equation 
$$
\frac{d}{dt}f_t({\pmb u}){=}
\frac{1}{i\h}[\langle{\pmb a},{\pmb u}\rangle, f_t({\pmb u})],\quad 
f_0({\pmb u}){=}\delta^{(V)}_*({\pmb u}).
$$
On the other hand, since $\delta^{(V)}_*({\pmb u})$ anti-commutes with
generators, $f_t({\pmb u})$ satisfies the evolution equation
$$
\frac{d}{dt}f_t({\pmb u}){=}
\frac{2}{i\h}\langle{\pmb a},{\pmb u}\rangle{*}f_t({\pmb u}),\quad 
f_0({\pmb u}){=}\delta^{(V)}_*({\pmb u}).
$$
Although $f^{(V)}_*({\pmb u}{-}\pmb x)$ is defined similarly to 
$f^{(V)}_*(\tilde{\pmb u}{-}\tilde{\pmb  x})$, the nature of those are 
very different. 
As it will be seen below,
${:}\delta^{(V)}_*({\pmb u}{-}\pmb x){*}\delta^{(V)}_*({\pmb u}{-}\pmb x){:}_{_K}{=}2^{2m}$, while 
${:}\delta^{(V)}_*(\tilde{\pmb u}{-}\tilde{\pmb  x'}){*}
\delta^{(V)}_*(\tilde{\pmb u}{-}\tilde{\pmb  x}){:}_{_K}
{=}{:}\delta^{(V)}(\tilde{\pmb x}'{-}\tilde{\pmb x})
\delta^{(V)}_*(\tilde{\pmb  u}{-}\tilde{\pmb  x}){:}_{_K}$.

\medskip
For tempered distribution $f(\pmb x)$, $g(\pmb x)$ on $V$, 
let $\check{f}(\pmb\xi)$, $\check{g}(\pmb\eta)$ be their Fourier transforms, 
and suppose the convolution product is welldefined as a tempered distribution  
$\check{f}{\btt}\check{g}(\pmb\zeta){=}
\int_{V}\check{f}(\pmb\eta)\check{g}({\pmb\zeta}{-}{\pmb\eta})\dbar{V_{\pmb\eta}}
{=}\check{g}{\btt}\check{f}(\pmb\zeta)$. Then as $V$ is assumed $JV{=}V$,  
the twisted convolution product $\check{f}{*_J}\check{g}(\pmb\zeta)$
(cf.,\eqref{twistconv}) is also welldefined. We denote 
\begin{equation}
(f{{\btt}_J}g)(\pmb x){=}
\int_{V}\check{f}{*_J}\check{g}(\pmb\xi)
e^{\frac{1}{i\h}\langle\pmb\xi, \pmb x\rangle}\dbar{V_{\pmb\xi}},\quad
{\pmb x}{\in}V.
\end{equation}
Hence we have 
\begin{equation}\label{notdelta}
{:}f^{(V)}_*({\pmb u}){*}g^{(V)}_*(\pmb u){:}_{_K}
{=}\int_{V}\int_{V}
\check{f}(\pmb\xi)\check{g}(\pmb\zeta{-}\pmb\xi)
e^{\frac{1}{2i\h}\langle{\pmb\xi}J,{\pmb\zeta}\rangle}
{:}e_*^{\frac{1}{i\h}\langle\pmb\zeta, \pmb u\rangle}{:}_{_K}
\dbar\pmb\xi\dbar\pmb\zeta
=\int_{V}\check{f}{*_J}\check{g}(\pmb\xi)
{:}e_*^{\frac{1}{i\h}\langle{\pmb\xi},{\pmb u}\rangle}{:}_{_K}\dbar{V_{\pmb\xi}}.
\end{equation}

If $K\in {\mathfrak H}_+(V)$, then
$$
{:}f_*^{(V)}({\pmb u}){*}g_*^{(V)}({\pmb u}){:}_{_K}{=}
\int_{V}
(f{{\btt}_J}g)(\pmb x){:}\delta^{(V)}_*(\pmb u{-}\pmb x)\dbar{V_{\pmb x}}{:}_{_K}.
$$
For instance setting $\delta_{\pmb a}^{(V)}(\pmb x){=}
\delta^{(V)}(\pmb x{-}\pmb a)$, we have 
$\check{\delta}^{(V)}_{\pmb a}(\pmb\xi)=e^{-\frac{1}{i\h}\langle\pmb\xi,\pmb a\rangle}$ 
and 
$$
\begin{aligned}
(\delta^{(V)}_{\pmb a}{{\btt}_J}\delta^{(V)}_{\pmb b})(\pmb x)&{=}
\int_{V}\int_{V}
e^{-\frac{1}{i\h}\langle\pmb\zeta,{\pmb a}\rangle}
e^{\frac{1}{2i\h}\langle\pmb\zeta J,\pmb\xi\rangle}
e^{-\frac{1}{i\h} \langle(\pmb\xi{-}\pmb\zeta),{\pmb b}\rangle}
e^{\frac{1}{i\h}\langle\pmb\xi, \pmb x\rangle}\dbar{V_{\pmb\zeta}}\dbar{V_{\pmb\xi}}\\
&{=}\int_{V}\Big(\int_{V}
e^{-\frac{1}{i\h}\langle\pmb\zeta,{\pmb a}{-}{\pmb b}{+}\frac{1}{2}\pmb\xi J\rangle}\dbar{V_{\pmb\zeta}}\Big)
e^{\frac{1}{i\h}\langle\pmb\xi, \pmb x{-}\pmb b\rangle}\dbar{V_{\pmb\xi}}
{=}\int_{V}\delta^{(V)}({\pmb a}{-}{\pmb b}{+}\frac{1}{2}\pmb\xi J)
e^{\frac{1}{i\h}\langle\pmb\xi, \pmb x{-}\pmb b\rangle}\dbar{V_{\pmb\xi}}\\
&{=}2^{2m}e^{-\frac{2}{i\h}\langle(\pmb a{-}\pmb b)J, \pmb x{-}\pmb b\rangle}
{=}2^{2m}e^{2\frac{1}{i\h}\langle{\pmb a}J,{\pmb b}\rangle} 
e^{-2\frac{1}{i\h}\langle({\pmb a}{-}{\pmb b})J,{\pmb x}\rangle}
 \end{aligned}
$$
by setting $\frac{1}{2}\pmb\xi{=}\pmb\xi'$ in the delta function. 
It follows 
\begin{equation}\label{Nice00}
(\delta^{(V)}_{\pmb a}{{\btt}_J}\delta^{(V)}_{\pmb b})(\pmb x){=}
2^{2m}e^{2\frac{1}{i\h}\langle\pmb aJ,{\pmb b}\rangle}
e^{-2\frac{1}{i\h}\langle({\pmb a}{-}{\pmb b})J,{\pmb x}\rangle},\quad 
(\delta^{(V)}_{\pmb a}{{\btt}_J}\delta^{(V)}_{\pmb a})(\pmb x){=}2^{2m}.
\end{equation}

By \eqref{Nice00}, we have a remarkable formula
\begin{equation}\label{2jyou}
\begin{aligned}
{:}\delta^{(V)}_*&({\pmb u}{-}{\pmb a}){*}\delta^{(V)}_*({\pmb u}{-}\pmb b){:}_{_K}{=}
\int_{V}
(\delta^{(V)}_{\pmb a}{{\btt}_J}
\delta^{(V)}_{\pmb b})(\pmb x){:}\delta_*(\pmb u{-}\pmb x){:}_{_K}\dbar{V_{\pmb x}}\\
&{=}
\int_{V}
e^{2\frac{1}{i\h}\langle{\pmb a}J,{\pmb b}\rangle}
e^{-2\frac{1}{i\h}\langle(\pmb a{-}\pmb b)J,{\pmb x}\rangle}
{:}\delta^{(V)}_*(\pmb u{-}\pmb x){:}_{_K}\dbar{V_{\pmb x}}\\
&{=}2^{2m}e^{2\frac{1}{i\h}\langle\pmb aJ,{\pmb b}\rangle}
{:}e_*^{-2\frac{1}{i\h}\langle(\pmb a{-}\pmb b)J,{\pmb u}\rangle}{:}_{_K}
{=}2^{2m}{:}e_*^{-2\frac{1}{i\h}\langle(\pmb a{-}\pmb b)J,\,{\pmb u}{-}\pmb b\rangle}{:}_{_K}.
\end{aligned}
\end{equation}
In particular, we have 
\begin{equation}\label{proddelta2}
{:}\delta^{(V)}_*({\pmb u}{-}{\pmb a}){*}\delta^{(V)}_*({\pmb u}{-}\pmb a){:}_{_K}{=}2^{2m}.
\end{equation}

\medskip

This is obtained also by another direct calculation as follows:
$$
\begin{aligned}
{:}\delta^{(V)}_*({\pmb u}{-}\pmb a){*}\delta^{(V)}_*({\pmb u}{-}\pmb b){:}_{_K}
&{=}\int_{V}
{:}e_*^{\frac{1}{i\h}\langle{\pmb\xi},\,{\pmb u}{-}{\pmb a}\rangle}{:}_{_K}
\dbar{V_{\pmb\xi}}{*_{_K}}
\int_{V}
{:}e_*^{\frac{1}{i\h}\langle{\pmb\eta},\,{\pmb u}{-}{\pmb b}\rangle}{:}_{_K}
\dbar{V_{\pmb\eta}}\\
&{=}
\int_{V}\int_{V}
{:}e_*^{\frac{1}{i\h}\langle{\pmb\xi},\,{\pmb u}{-}{\pmb a}\rangle}{*}
e_*^{\frac{1}{i\h}\langle{\pmb\eta},\,{\pmb u}{-}{\pmb b}\rangle}{:}_{_K}
\dbar{V_{\pmb\xi}}\dbar{V_{\pmb\eta}}
\end{aligned}
$$
By the exponential law and \eqref{Fundcal2}, we have 
$$
\begin{aligned}
{:}\delta^{(V)}_*({\pmb u}{-}\pmb a){*}
\delta^{(V)}_*({\pmb u}{-}\pmb b){:}_{_K}&{=}
\int_{V}\int_{V}
e^{\frac{1}{2i\h}\langle{\pmb\xi}J,\,{\pmb\eta}\rangle}
{:}e_*^{\frac{1}{i\h}\langle{\pmb\xi}{+}{\pmb\eta},\,{\pmb u}\rangle}{:}_{_K}
e^{-\frac{1}{i\h}\langle{\pmb\xi},{\pmb a}\rangle}
e^{-\frac{1}{i\h}\langle{\pmb\eta},{\pmb b}\rangle}
\dbar{V_{\pmb\xi}}\dbar{V_{\pmb\eta}}\\
&{=}
\int_{V}\int_{V}
e^{\frac{1}{2i\h}\langle{\pmb\xi}J,\,{\pmb\zeta}{-}{\pmb\xi}\rangle}
e^{\frac{1}{4i\h}e^{\langle{\pmb\zeta}K,\,{\pmb\zeta}\rangle}}
e^{\frac{1}{i\h}\langle{\pmb\zeta},\,{\pmb u}\rangle}
e^{-\frac{1}{i\h}\langle{\pmb\xi},{\pmb a}\rangle}
e^{-\frac{1}{i\h}\langle{\pmb\zeta}{-}{\pmb\xi},{\pmb b}\rangle}
\dbar{V_{\pmb\xi}}\dbar{V_{\pmb\zeta}}\\
&{=}\int_{V}
e^{-\frac{1}{i\h}\langle{\pmb\xi},\,{\pmb a}{-}{\pmb b}\rangle}
\Big(\int_{V}
e^{-\frac{1}{2ih}\langle{\pmb\zeta}J,\,{\pmb\xi}\rangle}
e^{\frac{1}{4i\h}\langle{\pmb\zeta}K,\,{\pmb\zeta}\rangle}
e^{\frac{1}{i\h}\langle{\pmb\zeta},\,{\pmb u}{-}\pmb b\rangle}
\dbar\pmb\zeta\Big)\dbar{V_{\pmb\xi}}.
\end{aligned}
$$
Noting that  
$\langle{\pmb\zeta}J,\,{\pmb\xi}\rangle{=}-\langle{\pmb\zeta},\,{\pmb\xi}J\rangle$, we have 
$$
\begin{aligned}
\frac{1}{i\h}\langle{\pmb\zeta}K,&\,{\pmb\zeta}\rangle{+}
 \frac{2}{i\h}\langle{\pmb\xi}J,\,{\pmb\zeta}\rangle {+}
 \frac{4}{i\h}\langle{\pmb\zeta},\,{\pmb u}{-}{\pmb b}\rangle\\
&{=}{-}\frac{1}{\h}\Big({\pmb\zeta}{\sqrt{iK}}{-}
({\pmb\xi}J{-}2({\pmb u}{-}{\pmb b}))\frac{i}{\sqrt{iK}}\Big)
     \,{}^t\!\Big({\pmb\zeta}{\sqrt{iK}}{-}
({\pmb\xi}J{-}2({\pmb u}{-}{\pmb b}))\frac{i}{\sqrt{iK}}\Big)\\
&\qquad\qquad{-}\frac{1}{\h}(({\pmb\xi}J{-}2({\pmb u}{-}{\pmb b}))\frac{i}{\sqrt{iK}})
\,{}^t(({\pmb\xi}J{-}2({\pmb u}{-}{\pmb b}))\frac{i}{\sqrt{iK}}).
\end{aligned}
$$
By the same reasoning as in Theorem\,\ref{breakthru} and 
 by setting $C_0(K){=}\frac{2^m}{\sqrt{\det(iK)}}$, we have 
$$
{:}\delta^{(V)}_*({\pmb u}{-}\pmb a){*}\delta^{(V)}_*({\pmb u}{-}\pmb b){:}_{_K}{=}
C_{0}(K)\int_{V}
e^{-\frac{1}{4\h}(({\pmb\xi}J{-}2({\pmb u}{-}{\pmb b}))\frac{i}{\sqrt{iK}})
\,{}^t(({\pmb\xi}J{-}2({\pmb u}{-}{\pmb b}))\frac{i}{\sqrt{iK}}){-}
\frac{1}{i\h}\pmb\xi\,{}^t({\pmb a}{-}{\pmb b})}
\dbar{V_{\pmb\xi}}.
$$
Noting that 
$$
\begin{aligned}
-\frac{1}{4\h}&\{(({\pmb\xi}J{-}2({\pmb u}{-}{\pmb a}))\frac{i}{\sqrt{iK}})
    \,{}^t(({\pmb\xi}J{-}2({\pmb u}{-}{\pmb a}))\frac{i}{\sqrt{iK}}){-}
      4i\pmb\xi\,{}^t({\pmb a}{-}{\pmb b})\}\\
&{=}
-\frac{1}{4\h}\Big(({\pmb\xi}J{-}2({\pmb u}{-}{\pmb a}))
  \frac{i}{\sqrt{iK}}{-}({\pmb a}{-}{\pmb b})\sqrt{iK}\Big)
\,{}^t\Big(({\pmb\xi}J{-}2({\pmb u}{-}{\pmb a}))
  \frac{i}{\sqrt{iK}}{-}({\pmb a}{-}{\pmb b})\sqrt{iK}\Big)\\
&\qquad\qquad-\frac{1}{4\h}({\pmb a}{-}\pmb b)4(iK)\,{}^t({\pmb  a}{-}\pmb b)
{+}2\frac{1}{i\h}({\pmb a}{-}\pmb b)J\,{}^t({\pmb u}{-}\pmb b), 
\end{aligned}
$$
and $J\frac{1}{iK}J$ is positive definite as $JV{=}V$, we have by
\eqref{2jyou} that 
\begin{equation}
{:}\delta^{(V)}_*({\pmb u}{-}\pmb a){*}\delta^{(V)}_*({\pmb u}{-}\pmb b){:}_{_K}{=}
C_{0}(K)C_{0}(J\frac{1}{K}J)
e^{-\frac{1}{\h}({\pmb a}{-}\pmb b)(iK)\,{}^t({\pmb a}{-}\pmb b)}
e^{-\frac{2}{i\h}\langle({\pmb a}{-}\pmb b)J,\,{\pmb u}{-}\pmb b\rangle}.
\end{equation}

\bigskip
For a fixed $V$, we define $V_k$ by
$V_k{=}V\cap\{s\tilde{u}_k{+}t\tilde{v}_k; (s,t)\in{\mathbb C}^2\}$
and by $\dbar{V_k}$ we denote the volume form on $V_k$ divided by ${2\pi\h}$.
Define as follows and note that this is not a hybrid $*$-delta
function but a {\it partial} $*$-delta function:
$$
{:}\delta^{(V)}_*(\tilde{u}_k,\tilde{v}_k){:}_{_K}{=}
\int_{V_k}{:}e_*^{\frac{1}{i\h}(s\tilde{u}_k{+}t\tilde{v}_k)}{:}_{_K}\dbar{V_k},\quad 
K\in {\mathfrak H}_+(V).
$$
By Theorem\,\ref{breakthru}, the $K$-ordered expression is 
$$
{:}\delta^{(V)}_*(\tilde{u}_k,\tilde{v}_k){:}_{_K}{=}
\frac{2}{\sqrt{\det(iK(k))}}
e^{-\frac{1}{\h}(\tilde{u}_k,\tilde{v}_k)\frac{1}{iK(k)}\,{}^t\!(\tilde{u}_k,\tilde{v}_k))},
\quad {\text{where}}\,\,K(k){=}
\footnotesize
{\begin{bmatrix}
K_{k,k}& K_{k,m{+}k}\\
K_{m+k,k}&K_{m{+}k,m{+}k}
\end{bmatrix}
}.
$$   
As $[\tilde{u}_i,{\tilde{v}_j}]{=}0$ for $i\not=j$, we see 
${:}\delta^{(V)}_*(\tilde{u}_i,\tilde{v}_i){*}
\delta^{(V)}_*(\tilde{u}_j,\tilde{v}_j){:}_K{=}
{:}\delta^{(V)}_*(\tilde{u}_j,\tilde{v}_j){*}
\delta^{(V)}_*(\tilde{u}_i,\tilde{v}_i){:}_K$ and 
$$
\delta^{(V)}_*({\pmb u}){=}
\delta^{(V)}_*(\tilde{\pmb u},\tilde{\pmb v}){=}
\delta^{(V)}_*(\tilde{u}_1,\tilde{v}_1){*}
\delta^{(V)}_*(\tilde{u}_2,\tilde{v}_2){*}\cdots{*}
\delta^{(V)}_*(\tilde{u}_m,\tilde{v}_m).
$$
For simplicity, we denote 
$\delta_*(k){=}\delta^{(V)}_*(\tilde{u}_k,\tilde{v}_k)$ in what
follows. Then we see  
$$
\delta_*(k)^2
{=}4,\quad \delta_*(k){*}\delta_*(l){=}
\delta_*(l){*}\delta_*(k)
$$
$$
\tilde{u}_k{*}\delta_*(k)=
-\delta_*(k){*}\tilde{u}_k,\quad
\tilde{v}_k{*}\delta_*(k)=-\delta_*(k){*}\tilde{v}_k,
\quad 
\tilde{u}_l{*}\delta_*(k)=
\delta_*(k){*}\tilde{u}_l \quad(k\not=l).
$$
In the next section, we see that every
$\delta_*(\tilde{u}_i,\tilde{v}_i)$ is on a compact one parameter subgroup of 
a $*$-exponential function of a quadratic form. 
Moreover, Theorem\,\ref{Prop3.4} in \S\,\ref{subsecpolar} below shows that 
$\frac{1}{2}\delta_*(k)$ is one of $\pm{\e}_{00}(k)$, $\pm i{\e}_{00}(k)$.

\section{Star-exponential functions of quadratic forms}

In the previous section  we have treated elements obtained by 
the integrations of $*$-exponential functions of linear forms. 
In this section, we give some interesting relations between these 
elements and the $*$-exponential functions of quadratic forms.

Let $H{=}\sum_{k=1}^m\frac{1}{i\h}{\tilde x}_k{\tilde y}_k$. Then, by
\eqref{WeylWeyl}, we have 
\begin{equation}
\int_{\mathbb R^{2m}}(\sum_{k=1}^m
\frac{1}{i\h}{\tilde x}_k{\tilde y}_k)\delta_*({\pmb u}{-}\pmb x)\dbar{\pmb x}{=}
\sum_{k=1}^m\frac{1}{i\h}{\tilde u}_k{\ctt}{\tilde v}_k,
\end{equation}
where $a{\ctt}b=\frac{1}{2}(a{*}b{+}b{*}a)$. We denote 
$H_*=\sum_{k=1}^m\frac{1}{i\h}{\tilde u}_k{\ctt}{\tilde v}_k$.
The $*$-exponential functions of ``half-variables'' can be managed by
the method mentioned in the previous sections, but 
$\int e^{itH}\delta_*({\pmb u}{-}\pmb x)\dbar{\pmb x}$ does not form a group.
The $*$-exponential functions of quadratic forms of full variables are defined by
the real analytic solutions of evolution equations \eqref{starexpexp}. 
Indeed, this was the main tool in the previous notes \cite{OMMY5}, \cite{ommy6}.

\subsection{Summary of blurred Lie groups }

The space of quadratic forms is isomorphic to the 
Lie algebra of $Sp(m,\mathbb C)$, i.e. 
$$
\{\langle{\pmb u}A,{\pmb u}\rangle; A\in{\mathfrak S}(2m)\}
\cong {\mathfrak{sp}}(m,{\mathbb C})=
\{\alpha; \alpha J{+}J\,{}^t\!\alpha=0\}
$$ 
as Lie algebras under the commutator bracket product. 
Let $E_{2m}{=}
\{\langle{\pmb\xi},{\pmb u}\rangle;
{\pmb\xi}{\in}{\mathbb C}^{2m}\}$. 
For every quadratic form 
$\frac{1}{2i\h}\langle{\pmb u}A,{\pmb u}\rangle_*$, 
${\rm{ad}}(\frac{1}{2i\h}\langle{\pmb u}A,{\pmb u}\rangle_*)$ 
is welldefined as a linear mapping independent of expression parameters:  
$$
{\rm{ad}}(\frac{1}{2i\h}\langle{\pmb u}A,{\pmb u}\rangle_*):
E_{2m}\to E_{2m},\quad 
{\rm{ad}}(\frac{1}{2i\h}\langle{\pmb u}A,{\pmb u}\rangle_*):
H\!ol({\mathbb C}^{2m})\to H\!ol({\mathbb C}^{2m})
$$ 
It is easy to see that
${\text{ad}}(\langle{\pmb u}
(\frac{1}{2i\h}\alpha J),{\pmb u}\rangle)
{=}{-}\alpha{\in}{\mathfrak{sp}}(m,{\mathbb C})$.

\medskip
For every $\alpha{\in}{\mathfrak{sp}}(m,{\mathbb C})$, 
the $K$-ordered expression of the $*$-exponential
function is given as follows:  
\begin{equation}\label{fundamental}
\begin{aligned}
{:}e_*^{\frac{t}{i\h}
\langle{\pmb u}(\alpha J), {\pmb u}\rangle_*}{:}_{_K}
{=}
\frac{2^m}{\sqrt{\det(I{-}{\kappa}{+}e^{-2t\alpha}(I{+}\kappa))}}
e^{\frac{1}{i\h}\langle{\pmb u}
\frac{1}{I{-}{\kappa}{+}e^{-2t\alpha}(I{+}\kappa)}
(I{-}e^{-2t\alpha})J,{\pmb u}\rangle}
\end{aligned}
\end{equation}
where $\kappa{=}J\!K$. 
As \eqref{fundamental} has double branched singular points in generic
$K$, we have to use two sheets 
by setting {\it slits} in the complex plane  
to treat ${:}e_*^{tH_*}{:}_{_K}$ univalent way.  Because of these
singularities, $*$-products of these $*$-exponential functions of
quadratic forms are defined only for some open subsets depending on 
expression parameters. Thus, expression parameters may be viewed as   
``local coordinate system'' for that object, and intertwiners are  
``changing coordinate''.  However in general intertwiners applied 
to $*$-exponential functions {\bf do not satisfy the cocycle conditions}.
Hence 
${:}e_*^{\frac{t}{i\h}
\langle{\pmb u}(\alpha J), {\pmb u}\rangle_*}{:}_{_K}$ 
do not form a group, but by considering generic $K$ all together, 
these generate an object $Sp^{(\frac{1}{2})}_{\mathbb C}(m)$, 
called the blurred Lie group which looks like a double covering group
of $Sp(m,{\mathbb C})$ which is known to be simply connected. 

\medskip
In spite of this, $Sp^{(\frac{1}{2})}_{\mathbb C}(m)$ can contain 
several genuine groups, when intertwiners on the object $G$ 
satisfies the cocycle conditions. Some of them are given in 
\cite{OMMY5}. Here we give another comment which may be used later. 

\begin{prop}{\label{twocover}}
If the object $G$ is expressed by the collection 
of ${:}U_1{:}_{_{K_1}}$, ${:}U_2{:}_{_{K_2}}$ with the intertwiner 
$I_{_{K_1}}^{^{K_2}}$, then the joint object 
$({:}U_1{:}_{_{K_1}},{:}U_2{:}_{_{K_2}})$ gives a 
a point set picture to the object $G$.
\end{prop}

\medskip
Note that $J{\in}{\mathfrak{sp}}(m,{\mathbb C})$ and also  
$J{\in}{Sp}(m,{\mathbb C}){=}
\{g\in G\!L(2m,{\mathbb C}); gJ\,{}^t\!g{=}J\}$.
For every $g{\in}Sp(m,{\mathbb C})$, 
$\tilde J=gJg^{-1}$ is both an element of Lie algebra and 
a group element satisfying 
$$
\tilde J^2{=}{-}I, \quad 
e^{t\tilde J}{=}\cos tI{+}(\sin t)\tilde J.
$$ 
Setting $\alpha=\tilde J$ in \eqref{fundamental} and noting 
$\alpha J=gJg^{-1}J={-}g\,{}^t\!g$, we  have in \cite{OMMY5} in generic $K$-ordered
expression that 
\begin{equation}\label{tildeKKK}
\begin{aligned}
{:}e_*^{\frac{t}{i\h}
\langle{\pmb u}g, 
{\pmb u}g\rangle_*}{:}_{_{K}}
&{=}
\frac{1}{\sqrt{\det(\cos t I{-}(\sin t){}^t\!gKg)}}
e^{\frac{1}{i\h}\langle{\pmb u}g 
\frac{\sin t}{\cos tI-\sin t \,{}^t\!g Kg},{\pmb u}g\rangle}.
\end{aligned}
\end{equation}

\medskip
Now, one may assume in generic ordered expressions, 
${}^t\!g Kg$ has disjoint $2m$ simple eigenvalues. 
Considering the diagonalization of ${}^t\!g Kg$  
in \eqref{tildeKKK}, we easily see that 
\begin{lem}\label{singularpts00}
In a generic $($open dense$)$ ordered expression, 
the singular points distributed $\pi$-periodically along $2m$ lines parallel 
to the real axis, 
%
%
and the singular points are all simple double 
branched singular points.
Moreover, ${:}e_*^{\frac{t}{i\h}
\langle{\pmb u}g, 
{\pmb u}g\rangle_*}{:}_{_{K}}$ 
is rapidly decreasing along lines parallel to 
the pure imaginary axis of the growth order $e^{-|t|^m}$, where $2m=n$. 
\end{lem}

\subsubsection{Polar elements}\label{subsecpolar}
For any fixed $g{\in}Sp(m,{\mathbb C})$, the behavior of
$*$-exponential function $e_*^{t\frac{1}{2i\h}\langle{\pmb u}g, {\pmb u}g\rangle_*}$
is very strange. At $t{=}0$, the initial direction $\frac{1}{2i\h}\langle{\pmb u}g, {\pmb u}g\rangle_*$
depends on $g{\in}Sp(m,{\mathbb C})$. 
At $t{=}\pi$, we have 
${:}e_*^{\frac{\pi}{i\h}\langle{\pmb u}g,{\pmb u}g\rangle_*}{:}_{_{K}}{=}\sqrt{1}{=}\pm 1$.
As $Sp(m,{\mathbb C})$ is connected, this looks determined without
$\pm$ ambiguity. If $K{=}0$, then \eqref{tildeKKK} is 
$$
{:}e_*^{\frac{t}{i\h}
\langle{\pmb u}g, 
{\pmb u}g\rangle_*}{:}_0{=}
\frac{1}{(\cos t)^m}e^{\frac{1}{i\h}\langle{\pmb u}g 
\frac{\sin t}{\cos tI},{\pmb u}g\rangle},\quad 
{:}e_*^{\frac{\pi}{i\h}
\langle{\pmb u}g, 
{\pmb u}g\rangle_*}{:}_0{=}(-1)^m
$$ 
without sign ambiguity by requesting $1$ at the initial point $t=0$. 
However choosing a suitable $K$, there is a case where 
${:}e_*^{\frac{\pi}{i\h}
\langle{\pmb u}g, 
{\pmb u}g\rangle_*}{:}_{_K}{=}1$.

On the other hand, 
${:}e_*^{\pi\frac{1}{2i\h}\langle{\pmb u}g, {\pmb u}g\rangle_*}{:}_{_K}$ 
also looks to be a focal point.  
\begin{lem}\label{madamada} 
Denoting by $[0{\sim}a]$ a path starting from the origin $0$ ending at
the point $a$
avoiding singular points, but evaluated at $a$, we have 
\begin{equation}\label{defpolar22}
{:}e_*^{[0{\sim}\pi]\frac{1}{2i\h}
\langle{\pmb u}g, {\pmb u}g\rangle_*}{:}_{_{K}}
=
\frac{1}{\sqrt{\det{K}}}
e^{{-}\frac{1}{\h}\langle{\pmb u}\frac{1}{iK},{\pmb u}\rangle}.
\end{equation}
The sign of $\sqrt{\det{K}}$ is determined by the sheet on which the 
end point of the path $[0{\sim}\pi]$ is sitting. 
\end{lem}

At a first glance, as  $Sp(m,{\mathbb C})$ is connected, this looks to be determined without $\pm$ ambiguity. 
Note however that for every $g{\in}Sp(m,{\mathbb C})$ 
there is an opposite $k{\in}Sp(m,{\mathbb C})$ such that 
$-\langle{\pmb u}g,{\pmb u}g\rangle_*{=}
\langle{\pmb u}k,{\pmb u}k\rangle_*$. 
This is shown for instance 
\begin{equation}\label{sp-inverse}
g
{\footnotesize
{\begin{bmatrix}
iI&0\\
0&{-i}I
\end{bmatrix}
\begin{bmatrix}
iI&0\\
0&{-i}I
\end{bmatrix}}}
\,{}^t\!g= -g\,{}^t\!g. 
\end{equation}
Thus, we have 
$$
{:}e_*^{\pi\frac{1}{2i\h}\langle{\pmb u}g, {\pmb u}g\rangle}{:}_{_K}{=}
\frac{1}{\sqrt{\det{K}}}
e^{{-}\frac{1}{\h}\langle{\pmb u}\frac{1}{iK},{\pmb u}\rangle}{=}
{:}e_*^{-\pi\frac{1}{2i\h}\langle{\pmb u}k, {\pmb u}k\rangle}{:}_{_K}.
$$
Considering a path $g(t)$ from $g$ to $k$ in $Sp(m,{\mathbb C})$
together with line segment $st$, $0{\leq}s{\leq}1$, for each $t$.
We have also
$$
{:}e_*^{\pi\frac{1}{2i\h}\langle{\pmb u}g, {\pmb u}g\rangle}{:}_{_K}{=}
\frac{1}{\sqrt{\det{K}}}
e^{\frac{1}{\h}\langle{\pmb u}\frac{1}{iK},{\pmb u}\rangle}{=}
{:}e_*^{\pi\frac{1}{2i\h}\langle{\pmb u}k, {\pmb u}k\rangle}{:}_{_K}.
$$
Thus, if ${:}e_*^{2\pi\frac{1}{2i\h}\langle{\pmb u}g, {\pmb u}g\rangle}{:}_{_K}{=}{-1}$,
it looks to cause a contradiction. The reason of this strange phenomenon is that 
a singular point appears on the path $g(st)$ from $g$ to $k$.
One cannot set the path $[0{\sim}\pi]$ continuously in $g\in Sp(m,{\mathbb C})$,
the sign changes discontinuously at some $g$.
In fact, the sign depends on the path $e_*^{t\frac{1}{2i\h}\langle{\pmb u}g, {\pmb u}g\rangle_*}$
of the $*$-exponential function. Two paths are equivalent only when 
these can continuously change each other by keeping the parity of 
crossing slits invariant. 
To distinguish the sign, we use the notation   
\begin{equation}\label{strictnotion}
{:}{\e}_{00}[g]{:}_{_K}=
{:}e_*^{[0{\to}\pi]
\frac{1}{2i\h}\langle{\pmb u}g, {\pmb u}g\rangle_*}{:}_{_K}=
\frac{1}{\sqrt{\det(\cos([0{\to}1]{\pi})I{-}(\sin([0{\to}1]{\pi}){}^t\!gKg)}}
e^{-\frac{1}{i\h}\langle{\pmb u}\frac{1}{K}, {\pmb u}\rangle},
\end{equation}
where $[0{\to}a]$ is the path along the straight line segment. 
Note that ${:}{\e}_{00}[g]{:}_{_K}$  
may not be defined at some $g$, when a singular point appears in 
the interval $(0,\pi]$. ${\e}_{00}[g]$ is not 
continuous w.r.t.\! $g$. The sign changes discontinuously at some $g$. 
For a generic $K$, there is $g{\in}Sp(m,{\mathbb C})$ such that 
${}^t\!gKg$ is a real diagonal matrix. Hence 
${:}e_*^{t\frac{1}{2i\h}\langle{\pmb u}g, {\pmb u}g\rangle_*}{:}_{_K}$
has a singular point. One can make a detour of a singular point 
by a slight change of path, but a double branched singular point  
forces to change the sheets.

\medskip
For every $k$, ${\e}_{00}(k){=}
e_*^{\frac{\pi i}{i\h}{\tilde u}_k{\ctt}{\tilde v}_k}$ is 
called a {\bf polar element}. 
It is interesting that polar element ${\e}_{00}(k)$ behaves just like a
scalar, but it behaves various ways. Sometimes, it behaves as if it
were $-1$, and sometimes it looks as if $i$ depending
on $K$. We call such elements $q$-scalars. 
But, to treat this as a univalent element, 
we have to distinguish more strictly, or it is better to treat 
${\e}_{00}(k)$ as two-valued elements. Although the additive structure 
is difficult to control, the multiplicative structure is treated
within such a multi-valued structure by calculations
such as $\sqrt{a}\sqrt{b}{=}\sqrt{ab}$.

Polar elements are obtained also by the formula
\eqref{genericparam00} in the case $m{=}1$ 
at $t=\pm\frac{\pi}{2}$. 
Set $t=\pi i$, $\frac{1}{2}\pi i$ in \eqref{genericparam00}. Then 
$$
{:}e_*^{\pi i\frac{1}{i\h}2\tilde{u}{\ctt}\tilde{v}}{:}_{_K}
{=}\frac{2}{\sqrt{4}},\quad 
{:}e_*^{\pi i\frac{1}{2i\h}2\tilde{u}{\ctt}\tilde{v}}{:}_{_K}
{=}
\frac{2}{\sqrt{4(\delta\delta'{-}c^2)}}
\,\,e^{\frac{1}{i\h}
\frac{2i}{4(\delta\delta'{-}c^2)}
\big((2i)(\delta'\tilde{u}^2{+}\delta\tilde{v}^2){+}2c\tilde{u}\tilde{v}\big)}.
$$
By the bumping identity we have already seen in the previous note 
remarkable properties of polar elements:  
$$
{\e}_{00}(k){*}\tilde{u}_{l}{=}(-1)^{\delta_{kl}}\tilde{u}_l{*}{\e}_{00}(k),\quad 
{\e}_{00}(k){*}\tilde{v}_{l}{=}(-1)^{\delta_{kl}}\tilde{v}_l{*}{\e}_{00}(k),\quad 
(k=1{\sim}m).
$$ 
On the other hand, Proposition\,\ref{anticomgen} shows that  
in a $V$-class expression $\delta_*(\pmb u)$ anti-commutes with
every generator. Now, comparing with the $K$-ordered expression in Theorem\,\ref{breakthru} of
$\delta_*(\pmb u)$, we see 
\begin{thm}\label{Prop3.4}
A polar element ${:}e_*^{[0\sim\pi]\frac{1}{2i\h}
\langle{\pmb u}g, {\pmb u}g\rangle_*}{:}_{_{K}}$ is one of 
$\pm 2^{-m}\delta_*(\pmb u)$, $\pm i2^{-m}\delta_*(\pmb u)$. The
constant is determined by $K$ and the path $[0\sim\pi]$ and the parity of
times of crossing slits.
\end{thm}

Note that ${:}e_*^{[0\sim\pi]\frac{1}{2i\h}
\langle{\pmb u}g, {\pmb u}g\rangle_*}{:}_{_{K}}$ is defined in generic
ordered expression, while $\delta_*(\pmb u)$ is defined in 
$V$-class expressions without sign ambiguity.   
However, $2^{-m}\delta_*(\pmb u)$ belongs to various $*$-one parameter 
subgroups given in the form 
$e^{i\,\omega\, t}e_*^{t\frac{1}{i\h}\langle{\pmb u}g, {\pmb u}g\rangle_*}$ such that  
$e^{\pi\,\omega i}e_*^{\pi\frac{1}{i\h}\langle{\pmb u}g, {\pmb u}g\rangle_*}{=}1$. 
We define  the {\bf total polar element} by 
$$
{\e}_{00}(L){=}{\e}_{00}(1){*}{\e}_{00}(2){*}\cdots{*}{\e}_{00}(m).
$$

\medskip
In the case $m=1$, setting 
$g={\footnotesize{
\begin{bmatrix}
a&b\\
c&d
\end{bmatrix}}}\in S\!L(2,{\mathbb C})$, the quadratic form $\langle{\pmb u}g,{\pmb u}g\rangle_*$ is
$$
(\tilde{u},\tilde{v})
\begin{bmatrix}
a^2{+}b^2& ac{+}bd\\
ac{+}bd& c^2{+}d^2
\end{bmatrix}
\begin{bmatrix}
\tilde{u}\\\tilde{v}
\end{bmatrix}{=}
(a^2{+}b^2)\tilde{u}^2{+}(c^2{+}d^2)\tilde{v}^2{+}2(ac{+}bd)\tilde{u}{\ctt}\tilde{v}.
$$
In particular this covers the quadratic forms given by the Lie algebra
${\mathfrak{su}}(2)$ of $SU(2)$.   Let ${\mathfrak{su}}_1(2)$ be the 
space of all traceless skew-hermitian matrices
with determinant $1$. Thus by identifying ${\mathfrak{su}}_1(2)J$ with  
a space of bilinear forms, we set 
\begin{equation}\label{x,y,rho}
\begin{bmatrix}
a^2{+}b^2& ac{+}bd\\
ac{+}bd& c^2{+}d^2
\end{bmatrix}{=}
\begin{bmatrix}
x{+}iy& i\rho\\
i\rho & x{-}iy
\end{bmatrix}, \quad x, y, \rho\in {\mathbb R},\quad
x{+}iy{=}\sqrt{1{-}\rho^2}\,e^{i\theta}, \quad |\rho|< 1. 
\end{equation}
Then we easily see 
$$
\begin{bmatrix}
a& b\\
c& d
\end{bmatrix}{=}
\begin{bmatrix}
\sqrt[4]{1{-}\rho^2}\,e^{i\theta/2}\cosh\xi&\sqrt[4]{1{-}\rho^2}\,ie^{i\theta/2}\sinh\xi \\
\sqrt[4]{1{-}\rho^2}\,ie^{-i\theta/2}\sinh\eta&\sqrt[4]{1{-}\rho^2}\,e^{-i\theta/2}\cosh\eta\\ 
\end{bmatrix}, 
\quad \cosh(\xi{+}\eta){=}\frac{1}{\sqrt{1{-}\rho^2}},\,\,
\sinh(\xi{+}\eta){=}\frac{\rho}{\sqrt{1{-}\rho^2}},
$$
where $\xi, \eta$ are real variables. If $\rho{=}\pm 1$, we set 
$$
\begin{bmatrix}
a& b\\
c& d
\end{bmatrix}{=}
\frac{1}{\sqrt{2}}
\begin{bmatrix}
1& i\\
i& 1
\end{bmatrix}.
$$
Denote by $S'$ the subset 
\begin{equation}\label{setS'}
S'{=}\{g\in S\!L(2,{\mathbb C}); 
\frac{1}{i\h}\langle{\pmb u}g,{\pmb u}g\rangle_*\in {\mathfrak{su}}_1(2)J\}.
\end{equation}

To control the sign ambiguity of \eqref{tildeKKK}, we have to prepare two sheets and have
to fix slits. However, note that the way of setting slits is not unique.
Let us consider the case $m{=}1$ and a quadratic 
form $2\tilde{u}{\ctt}\tilde{v}$ by setting  
$g{=}\frac{1}{\sqrt{2}}
{\footnotesize{
\begin{bmatrix}
1&i\\
i&1
\end{bmatrix}
}}$.
We take a general expression parameter  
$K{=}
\footnotesize
{\begin{bmatrix}
\delta&c\\
c&\delta'
\end{bmatrix}}$. 
By setting 
$\Delta{=}e^t{+}e^{-t}{-}c(e^t{-}e^{-t})$, the $K$-ordered expression
of $e_*^{t\frac{1}{i\h}2\tilde{u}{\ctt}\tilde{v}}$ is given in \cite{OMMY5} by 
\begin{equation}\label{genericparam00}
{:}e_*^{t\frac{1}{i\h}2\tilde{u}{\ctt}\tilde{v}}{:}_{_{K}}{=}
\frac{2}{\sqrt{\Delta^2{-}(e^t{-}e^{-t})^2\delta\delta'}}
\,\,e^{\frac{1}{i\h}
\frac{e^t-e^{-t}}{\Delta^2{-}(e^t{-}e^{-t})^2\delta\delta'}
\big((e^t-e^{-t})(\delta'\tilde{u}^2{+}\delta\tilde{v}^2)
{+}2\Delta \tilde{u}\tilde{v}\big)}.
\end{equation}

\subsection{Some special classes of expression parameters}
As ${\e}_{00}^2{=}\pm 1$, polar elements have double valued
nature, but ${\e}_{00}$ is contained in various one parameter 
subgroups. Moreover as the periodicity depends also on 
the expression parameter, polar elements must be considered together with that 
one parameter subgroup. 
If ${\e}_{00}$ is given as $e_*^{\pi iH_*}$, then  
there are $K$, $K'$
such that ${:}e_*^{2\pi iH_*}{:}_{_K}{=}1$, 
${:}e_*^{2\pi iH_*}{:}_{_{K'}}{=}-1$, and 
if ${:}e_*^{2\pi iH_*}{:}_{_{K'}}{=}-1$, then
the argument in the previous subsection shows
there must be a quadratic form $H'_*$ such that 
${:}e_*^{2\pi iH'_*}{:}_{_{K'}}{=}1$.

\subsubsection{A special class ${\mathfrak K}_{re}$}\label{anothersp}
We found in \cite{OMMY5} there is a 
special class ${\mathfrak K}_{re}$ of expression parameters: 
\begin{prop}\label{nicenice22}
If 
$K_{re}{=}
\footnotesize{
\begin{bmatrix}
\rho&ic'\\
ic'&\rho
\end{bmatrix}}$ with $c'{=}c{+}i\theta, \,\, c, \rho {\in}{\mathbb
R}$,\,\,$|\theta|$ is small, satisfies 
$|\frac{\rho{+}ic'{+}1}{\rho{+}ic'{-}1}|{\not=}1$, 
then $K_{re}$ ordered expressions of $*$-exponential 
functions  
$$
e_*^{\frac{t}{i\h}\langle{ug}, ug\rangle_*}; \quad \forall g{\in}S'
\quad ({\text{cf}}.\eqref{setS'}),
$$
in particular 
$$
e_*^{\frac{it}{i\h}2\tilde{u}{\ctt}\tilde{v}}, \quad 
e_*^{\frac{t}{i\h}(\tilde{u}^2+\tilde{v}^2)},\quad
e_*^{\frac{it}{i\h}(\tilde{u}^2-\tilde{v}^2)}, 
$$
have no singular point on the real axis and $\pi$-periodic, 
but each of them has singular points sitting $\pi$-periodically along 
two lines parallel to the real axis on both upper and  
lower half plane.

Hence, the polar element ${\e}_{00}$ may be written in the  
$K_{re}$-expression by 
$$
{:}{\e}_{00}{:}_{_{K_{re}}}=
{:}e_*^{\frac{\pi i}{i\h}\tilde{u}{\ctt}\tilde{v}}{:}_{_{K_{re}}}=
{:}e_*^{\frac{\pi i}{2i\h}(\tilde{u}^2{-}\tilde{v}^2)}{:}_{_{K_{re}}}=
{:}e_*^{-\frac{\pi}{2i\h}(\tilde{u}^2+\tilde{v}^2)}{:}_{_{K_{re}}}{=}
{:}e_*^{\frac{\pi}{2i\h}\langle{ug}, ug\rangle_*}{:}_{_{K_{re}}}; \,\, \forall g{\in}S'.
$$
We have ${\e}_{00}^2=1$, but ${\e}_{00}$ is not a scalar, as this
anti-commutes with generators. 
Moreover ${\e}_{00}$ has three square roots 
$$
e_1=e_*^{\frac{\pi i}{2i\h}\tilde{u}{\ctt}\tilde{v}}, \quad   
e_2=e_*^{\frac{\pi}{4i\h}(\tilde{u}^2{+}\tilde{v}^2)},\quad 
e_3=e_*^{\frac{\pi i}{4i\h}(\tilde{u}^2-\tilde{v}^2)} 
$$
such that $e_i^2={\e}_{00}$. Furthermore, we see 
$e_1{*}e_2=e_3$ in the $K_{re}$-ordered expression. 
\end{prop}

Generally, adjoint relations of quadratic forms give 
the following master relations (cf.\cite{ommy6}) for elements of square roots 
of the polar element.
\begin{lem}\label{master}
Let $H_*$ be a quadratic form with the discriminant $1$. 
Then,  
$e^{\pi i{\rm{ad}}(H_*)}_**e_j{=}e_j^{-1}$. 
This implies that $e_i{*}e_j{*}e_i^{-1}{=}e_j^{-1}$. 
These relations  hold without sign ambiguity. 
\end{lem}
By this master relation, we have in general
\begin{equation}\label{master22}
e_i{*}e_j=e_j^{-1}{*}e_i
={\e}_{00}{*}e_j{*}e_i.
\end{equation}
By the identity $e_3=e_1{*}e_2$, we 
have 
$$
e_2{*}e_3=
e_2{*}e_1{*}e_2=
e_2{*}e_2^{-1}{*}e_1=e_1. 
$$  
Similarly, 
$$
e_3{*}e_1=
e_3{*}e_2{*}e_3=
e_3{*}e_3^{-1}{*}e_2=e_2.
$$
The bumping identity gives the interesting commutation relations:
$$
\begin{aligned}
\tilde{u}{*}e_1&{=}-ie_1{*}\tilde{u},\quad
\tilde{u}{*}e_2{=}{-}e_2{*}\tilde{v},\quad
\tilde{u}{*}e_3{=}{i}e_3{*}\tilde{v},\\  
\tilde{v}{*}e_1&{=}ie_1{*}\tilde{v},\quad
\tilde{v}{*}e_2{=}e_2{*}\tilde{u},\quad
\tilde{v}{*}e_3{=}{i}e_3{*}\tilde{u}.
\end{aligned}
$$
The polar element anti-commutes with generators, and  we have

\noindent
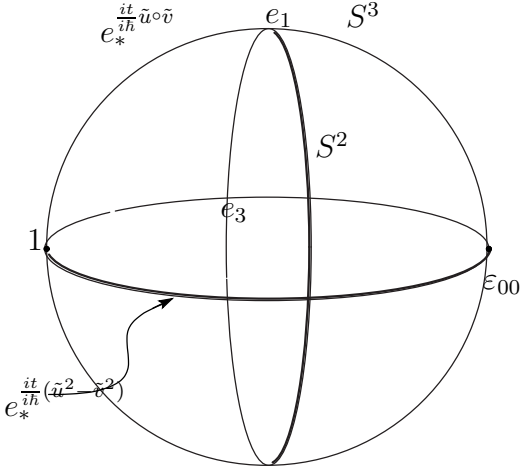
\unitlength 0.07in
\begin{picture}( 25.4000, 25.0000)(1.0000,-38.6000)
%
\special{pn 8}%
\special{ar 1430 1420 1140 1140  0.0069929 6.2831853}%
%
\special{pn 8}%
\special{ar 1430 1430 1150 270  3.9617389 6.2831853}%
\special{ar 1430 1430 1150 270  0.0000000 3.9269908}%
%
\special{pn 8}%
\special{ar 1440 1420 220 1140  3.0019416 6.2831853}%
\special{ar 1440 1420 220 1140  0.0000000 2.9932069}%
%
\special{pn 13}%
\special{ar 1440 1420 1150 270  0.0678955 2.9948295}%
%
\special{pn 13}%
\special{ar 1440 1420 210 1130  4.8542860 6.2831853}%
\special{ar 1440 1420 210 1130  0.0000000 1.4743226}%
%
\special{pn 13}%
\special{sh 1}%
\special{ar 290 1430 10 10 0  6.28318530717959E+0000}%
\special{sh 1}%
\special{ar 290 1430 10 10 0  6.28318530717959E+0000}%
%
\special{pn 13}%
\special{sh 1}%
\special{ar 2580 1430 10 10 0  6.28318530717959E+0000}%
\special{sh 1}%
\special{ar 2580 1430 10 10 0  6.28318530717959E+0000}%
%
\special{pn 8}%
\special{pa 300 2190}%
\special{pa 334 2190}%
\special{pa 368 2190}%
\special{pa 406 2188}%
\special{pa 444 2186}%
\special{pa 482 2182}%
\special{pa 520 2178}%
\special{pa 556 2170}%
\special{pa 590 2160}%
\special{pa 622 2148}%
\special{pa 650 2134}%
\special{pa 674 2118}%
\special{pa 694 2096}%
\special{pa 706 2072}%
\special{pa 714 2044}%
\special{pa 714 2014}%
\special{pa 712 1980}%
\special{pa 708 1948}%
\special{pa 706 1914}%
\special{pa 706 1882}%
\special{pa 712 1852}%
\special{pa 726 1826}%
\special{pa 744 1804}%
\special{pa 768 1784}%
\special{pa 794 1764}%
\special{pa 826 1748}%
\special{pa 858 1732}%
\special{pa 892 1718}%
\special{pa 910 1710}%
\special{sp}%
%
\special{pn 8}%
\special{pa 840 1740}%
\special{pa 940 1690}%
\special{fp}%
\special{sh 1}%
\special{pa 940 1690}%
\special{pa 872 1702}%
\special{pa 892 1714}%
\special{pa 890 1738}%
\special{pa 940 1690}%
\special{fp}%
\put(1.0000,-33.2000){\makebox(0,0)[lb]{$e_*^{\frac{it}{i\h}(\tilde{u}^2{-}\tilde{v}^2)}$}}%
\put(2.7000,-20.5000){\makebox(0,0)[lb]{$1$}}%
\put(36.4000,-23.8000){\makebox(0,0)[lb]{${\e}_{00}$}}%
\put(24.0000,-13.6000){\makebox(0,0)[lb]{$S^2$}}%
\put(8.1000,-5.3000){\makebox(0,0)[lb]{$e_*^{\frac{it}{i\h}\tilde{u}{\ctt}\tilde{v}}$}}%
\put(20.3000,-3.9000){\makebox(0,0)[lb]{$e_1$}}%
\put(26.3000,-3.9000){\makebox(0,0)[lb]{$S^3$}}%
\put(17.0000,-18.4000){\makebox(0,0)[lb]{$e_3$}}%
\end{picture}
\hfill
\parbox[b]{.55\linewidth}{
\begin{thm}\label{surprise0}
In the $K_{re}$-ordered expression such that 
$|\frac{1{+}\rho{+}ic'}{1{-}\rho{-}ic'}|{\not=}1$,  
$\{{\e}_{00}, e_1, e_2, e_3\}$ 
generates an algebra ${\mathcal A}$ in which   
two idempotent elements $\frac{1}{2}(1{+}{\e}_{00})$,  
$\frac{1}{2}(1{-}{\e}_{00})$ exist such that 
$$
1=\frac{1}{2}(1{+}{\e}_{00}){+}\frac{1}{2}(1-{\e}_{00}), \quad 
\frac{1}{2}(1{+}{\e}_{00}){*}\frac{1}{2}(1-{\e}_{00})=0.
$$
\end{thm}

The subalgebra 
$\frac{1}{2}(1{-}{\e}_{00}){*}{\mathcal A}$ is naturally isomorphic to 
the complexification ${\mathbb C}{\otimes}{\mathbb H}$ 
of the quaternion field $\mathbb H$ such that by denoting $\hat
1=\frac{1}{2}(1{-}{\e}_{00})$,  
 $\hat{e}_i=\frac{1}{2}(1{-}{\e}_{00}){*}{e}_i$:} 
$$
\hat{\e}_{00}=\frac{1}{2}(1{-}{\e}_{00}){*}{\e}_{00}=-\hat 1, \quad
\hat{e}_i^2=-\hat 1, \quad  
\hat{e}_i{*}\hat{e}_j={-}\hat 1{*}\hat{e}_j{*}\hat{e}_i,\quad 1\leq i,j\leq 3.
$$
On the other hand, the subalgebra 
$\frac{1}{2}(1{+}{\e}_{00}){*}{\mathcal A}$ is the group ring over
${\mathbb C}$ of the Klein's four group. 
Obviously, $\frac{1}{2}(1{-}{\e}_{00}){*}{\mathcal A}$ and $\frac{1}{2}(1{+}{\e}_{00}){*}{\mathcal A}$
are not isomorphic.  

\medskip
Viewing ${\e}_{00}$ as the representative of ${\e}_{00}(k)$, we define
for each $k$ square roots $e_1(k), e_2(k), e_3(k)$ of ${\e}_{00}(k)$,
and denote by ${\mathcal A}(k)$ the algebra generated by 
${\e}_{00}(k), e_1(k), e_2(k), e_3(k)$.
Then 
$$
e_{i_1}^{\e_1}(1){*}e^{\e_2}_{i_2}(2){*}\cdots{*}e^{\e_m}_{i_m}(m); \quad {\e}_i{=}0,1,2,3
$$
form an $m$-tensor algebra 
${\mathcal A}(1){\otimes}{\mathcal A}(2){\otimes}\cdots{\otimes}{\mathcal A}(m)$.

\subsubsection{Another special class ${\mathfrak K}_s$} 
On the other hand, 
 we have shown in \cite{OMMY4} the following: 
\begin{prop}\label{Cliffemarging}
There is a small class ${\mathfrak K}_s$ (called another special class) of expression parameters such that 
polar elements ${\e}_{00}(1),\,{\e}_{00}(2),\cdots,{\e}_{00}(m)$ 
given by ${\e}_{00}(k){=}e_*^{\frac{\pi i}{i\h}\tilde{u}_k{\ctt}\tilde{v}_k}$
form a Clifford algebra ${\mathcal C}(m)$ of $m$ generators such that  
${\e}_{00}(k)^2{=}-1$ for every $k$. 
\end{prop}

The shape of matrices in ${\mathfrak K}_s$ is given mainly as follows:
$$
K_s{=}
\begin{bmatrix}
i\rho I_m& (c)\\
(c)&i\rho I_m
\end{bmatrix},\quad \rho, \,\,c\,\,{\in}{\mathbb R},\quad \rho>0, \quad
(c){=}m{\times}m\,\, \text{matrix, all entries are }c  
$$
This is an expression parameter such that ${\rm{Re}}(iK_s)$ is negative
definite. 

The proof of Proposition\,\ref{Cliffemarging} is based on the fact that   
$
{:}e_*^{\frac{it}{i\h}(u_k{\ctt}v_k{+}u_{\ell}{\ctt}v_{\ell})}{*_{_{K_s}}}$
has singular points on the open intervals $(0,\pi)$ and $(\pi,2\pi)$,
but no other singular point in    
$$
e_*^{\frac{1}{i\h}(is u_k{\ctt}v_k{+}it u_{\ell}{\ctt}v_{\ell})},\quad
(s,t)\in [0,\pi]{\times}[0,\pi].
$$

For simplicity of notations we denote in what follows ${\e}_{00}(k)$ in the special
class expression $K_s$ by ${\e}_k$, i.e. 
${:}{\e}_k{:}_{_{K_s}}{=}{:}{\e}_{00}(k){:}_{_{K_s}}$.
The next formula is easy to see 
$$
e_*^{t({\e}_k\cos\theta{+}{\e}_{\ell}\sin\theta)}
{=}\cos t{+}({\e}_k\cos\theta{+}{\e}_{\ell}\sin\theta)\sin t.
$$ 
That is, ${\e}_k, {\e}_{\ell}$ are 
``infinitesimal operators'', although these are $*$-exponentiated 
elements.

\begin{prop}\label{GrassCliff}
A system of polar elements $\{{\e}_1,{\e}_2,\cdots,{\e}_m\}$ forms a Clifford algebra in
a ${\mathfrak K}_s$-ordered expression such that 
${\e}_i{*}{\e}_j{+}{\e}_j{*}{\e}_i{=}-2\delta_{ij}$.
If $m{=}2d$ then 
$$
\xi_k{=}\frac{1}{2}({\e}_{2k{-}1}{+}i{\e}_{2k}),\quad
\eta_k{=}\frac{1}{2}({\e}_{2k{-}1}{-}i{\e}_{2k}),\quad k=1\sim d,  
$$
satisfy in $+$ bracket notations  
$$
\{\xi_k,\xi_{\ell}\}{=}0{=}\{\eta_k,\eta_{\ell}\},\quad \{\xi_k,\eta_{\ell}\}{=}-\delta_{k\ell}.
$$
Hence, $\xi_1,\cdots,\xi_d$,\,$\eta_1,\cdots,\eta_d$ respectively generate
the Grassmann algebra 
$\Lambda_d(\xi)$,\, $\Lambda_d(\eta)$. 
\end{prop}

In the later section, we will show that a Grassmann algebra of $m$
generators appears more naturally by making a vacuum representation.

\bigskip
\noindent
{\bf Remark 1}\,\,In the previous subsection we have shown the relation that the polar element 
${\e}_k$ defined above is one of $\pm 2^{-1}\delta_*(\tilde{u}_k,\tilde{v}_k)$, 
$\pm 2^{-1}i\delta_*(\tilde{u}_k,\tilde{v}_k)$. However, note this is a very anomalous
phenomenon. In spite that the commutativity 
$\delta_*(\tilde{u}_k,\tilde{v}_k){*}\delta_*(\tilde{u}_l,\tilde{v}_l){=}
\delta_*(\tilde{u}_l,\tilde{v}_l){*}\delta_*(\tilde{u}_k,\tilde{v}_k)$
is easily checked in a ${\mathfrak H}_+(\mathbb R^n)$- 
class expression,
Proposition\,\ref{Cliffemarging} insists that 
${\e}_k{*}{\e}_{l}{=}{-}{\e}_{l}{*}{\e}_k$. Indeed, a polar element
${\e}_k$ should be defined together with path in a $*$-exponential function
from the origin. The 
equality above means only the endpoint of the path is one of  
$\pm 2^{-1}\delta_*(\tilde{u}_k,\tilde{v}_k)$, 
$\pm 2^{-1}i\delta_*(\tilde{u}_k,\tilde{v}_k)$. In this sense, polar
elements are {\it not } elements of $H\!ol(\mathbb C^{2m})$, but
elements of a certain space which is one step up the degree of 
fine identifications. This does not seem to be a simple homotopical phenomena.

\subsubsection{A class ${\mathfrak K}_{im}$ } 
Note now that all $\e_{k}$ in this subsection is defined by 
$e_*^{\frac{\pi i}{i\h}u_k{\ctt}v_k}$, but a polar element is 
a member of various one parameter subgroups of $*$-exponential 
functions of quadratic forms. 
Since ${:}{\e}^2_k{:}_{_{K_s}}{=}-1$, the argument in \S\,\ref{subsecpolar}  
shows that there must exist a quadratic form $H_*$ such that 
${:}e_*^{\frac{\pi i}{i\h}H_*}{:}_{_{K_s}}{=}{:}{\e}_k{:}_{_{K_s}}$,
but ${:}e_*^{\frac{2\pi i}{i\h}H_*}{:}_{_{K_s}}{=}1$. Hence, it seems  
to be better to think 
${:}e_*^{\frac{\pi i}{i\h}H_*}{:}_{_{K_s}}
{\not=}{:}{\e}_k{:}_{_{K_s}}$, but how the properties 
of square roots $e_*^{\frac{\pi i}{2i\h}H_*}$, 
$e_*^{\frac{\pi i}{2i\h}u_k{\ctt}v_k}$ are explained. 

In the previous note \cite{OMMY5}, we have treated the case $m{=}1$
and $K$ is given by 
$$
\begin{bmatrix}
i\rho& c'\\
c'&\rho
\end{bmatrix},\quad  
\rho{>}0,\,c'{=}c{+}i\theta,\, c{\in}{\mathbb R}, \quad \theta{\not=}0
\quad {\text{but small}}.
$$
In this expression, 
the polar element ${\e}_{00}$ splits into three cases:
$$
{:}(e_*^{\frac{\pi i}{i\h}\tilde{u}{\ctt}\tilde{v}})^2{:}_{_{K_{im}}}=-1,\quad 
{:}(e_*^{\frac{\pi i}{2i\h}(\tilde{u}^2{-}\tilde{v}^2)})^2{:}_{_{K_{im}}}=-1,\quad 
{:}(e_*^{\frac{\pi}{2i\h}(\tilde{u}^2+\tilde{v}^2)})^2{:}_{_{K_{im}}}{=}1.
$$

\noindent
{\bf Note}\,\,\,The replacement  
$(u,v)\to (e^{\frac{i\pi}{4}}u,e^{-\frac{i\pi}{4}}v)$ of generators
gives the exchange of  the second and third elements.

Here one may set
$e_*^{\frac{\pi i}{i\h}\tilde{u}{\ctt}\tilde{v}}{=}e_*^{-\frac{\pi i}{2i\h}(\tilde{u}^2{-}\tilde{v}^2)}$,
but $e_*^{\frac{\pi}{2i\h}(\tilde{u}^2+\tilde{v}^2)}$ does not relate
others. Although we have three square roots as above
$$
e_1{=}e_*^{\frac{\pi i}{2i\h}\tilde{u}{\ctt}\tilde{v}},
e_2{=}e_*^{\frac{\pi i}{2i\h}(\tilde{u}^2{-}\tilde{v}^2)},
e_3{=}e_*^{\frac{\pi}{2i\h}(\tilde{u}^2+\tilde{v}^2)},
$$
these do not generate an associative algebra. However, such phenomena
suggest only that one can not use $e_1, e_2, e_3$ as elements of 
binary operations. In the next note, we will show that certain 
compositions of elements form an associative algebra. It is
interesting that this has certain similarity to the so called 
``quark confinement''.

\bigskip
\noindent
{\bf Note}\,\, So far, $K$ was called an expression parameter and it was treated 
as a supplemental parameter to express the true nature of
elements. However, the observation for the polar elements in this section shows that
expression parameters represent certain essential nature of elements.
Here we note that
${\mathfrak{K}_s}\cap{\mathfrak{K}_{re}}{=}\emptyset$, 
${\mathfrak{K}_{im}}\cap{\mathfrak{K}_{re}}{=}\emptyset$. 
In the next section, we give another notion of elements which depend
essentially on the expression parameters.

\section{Vacuums, pseudo-vacuums}\label{VacVac}

As it is mentioned in the preface, there is no mathematical definition
of ``vacuum'', but this is a kind of target to which one makes actions 
to create the ``space{-}time'' or the ``configuration space''.

\medskip
Recall first several properties of the $*$-exponential
function $e_*^{zH_*}$, 
$H_*{=}\sum_{k=1}^m\frac{1}{i\h}{\tilde  u}_k{\ctt}{\tilde v}_k$. 

\bigskip
\noindent
({\bf a})\,\,In generic ordered expression, one may
suppose that there is no singular point on the real line.
$e_*^{zH_*}$ 
is $4\pi$ periodic w.r.t.$t$, i.e. 
${:}e_*^{(z{+}4\pi i)H_*}{:}_{_K}={:}e_*^{zH_*}{:}_{_K}$. More
precisely, the periodicity depends on the real part of $z$. 
In the case $m{=}1$, there is an interval $[a,b]$, called the 
{\it exchanging interval} in \cite{ommy6} such that   
$e_*^{(s{+}it)H_*}$ is $2\pi$ periodic if $a{<}s{<}b$, and   
alternating $2\pi$ periodic if $s{\not\in}[a,b]$.

\bigskip 
\noindent
({\bf b})\,\,$e_*^{tH_*}$ is rapidly decreasing on $\mathbb R$ of
$e^{-\frac{m}{2}|t|}$ order.  Hence the 
integral $\int_{\mathbb R}{:}e_*^{tH_*}{:}_{_K}dt$ is welldefined,
and it follows ${:}H_*{*}\int_{\mathbb R}e_*^{tH_*}dt{:}_{_K}{=}0$.

\bigskip
\noindent
({\bf c}) Double branched singular points are distributed $\pi i$
periodically along $2m$ lines parallel to the 
imaginary axis whose positions depend on  expression parameters.
In the case $m{=}1$, the real part of these lines are $a$ and $b$.

\bigskip
\noindent
({\bf d}) $\lim_{t\to{-}\infty}e^{\frac{m}{2}t}e_*^{tH_*}$ is a nontrivial
element denoted by $\widetilde{\varpi}(L)$. We have called this the (total) 
{\bf  vacuum} in \cite{ommy6}. The precise statement for the case
$m=1$ is as follows:

\begin{prop}\label{existvacuum}
In a generic ordered expression 
$e_*^{t\frac{1}{i\h}2\tilde{u}{\ctt}\tilde{v}}$ is rapidly decreasing with 
the growth order $e^{-|t|}$ along lines parallel to the real axis.
Noting $\tilde{v}{*}\tilde{u}{=}\tilde{u}{\ctt}\tilde{v}{+}\frac{1}{2}i\h$, we see the following:
In generic ordered expressions such that there is no 
singular point on the real axis, but the limit exist
$$
\lim_{t{\to}-\infty}e_*^{t\frac{1}{i\h}2\tilde{u}{*}\tilde{v}}
{=}{\varpi}_{00},\quad 
\lim_{t{\to}\infty}e_*^{t\frac{1}{i\h}2\tilde{v}{*}\tilde{u}}
{=}\overline{\varpi}_{00},\quad 
\lim_{t{\to}\infty}e_*^{t\frac{1}{i\h}2\tilde{u}{*}\tilde{v}}
{=}0,\quad 
\lim_{t{\to}-\infty}e_*^{t\frac{1}{i\h}2\tilde{v}{*}\tilde{u}}
{=}0. 
$$
\end{prop}
More precisely, in a fixed generic expression parameter 
$K{=}
\tiny{\begin{bmatrix}
\delta&c\\
c&\delta' 
\end{bmatrix}}$, 
${:}e_*^{t\frac{1}{i\h}2\tilde{u}{\ctt}\tilde{v}}{:}_{_{K}}$ is smooth   
rapidly decreasing in $\pm$ directions and \eqref{genericparam00} 
gives 
\begin{equation}
  \label{eq:vacuum123123}
  \begin{aligned}
{:}\varpi_{00}{:}_{_K}=
&\lim_{t{\to}-\infty}{:}e_*^{t\frac{1}{i\h}2\tilde{u}{*}\tilde{v}}{:}_{_K}
{=}\frac{2}{\sqrt{(1{+}c)^2{-}\delta\delta'}}
e^{-\frac{1}{i\h}\frac{1}{(1{+}c)^2{-}\delta\delta'}
(\delta'\tilde{u}^2{-}(1{+}c)\tilde{u}\tilde{v}{+}\delta\tilde{v}^2)},\\
{:}{\overline{\varpi}}_{00}{:}_{_K}
=&\lim_{t{\to}\infty}{:}e_*^{t\frac{1}{i\h}2\tilde{v}{*}\tilde{u}}{:}_{_K}
{=}\frac{2}{\sqrt{(1{-}c)^2{-}\delta\delta'}}
e^{\frac{1}{i\h}\frac{1}{(1{-}c)^2{-}\delta\delta'}
(\delta'\tilde{u}^2{+}(1{-}c)2\tilde{u}\tilde{v}{+}\delta\tilde{v}^2)},\\ 
&
\lim_{t{\to}\infty}{:}e_*^{t\frac{1}{i\h}2\tilde{u}{*}\tilde{v}}{:}_{_K}
{=}0,\quad 
\lim_{t{\to}{-}\infty}{:}e_*^{t\frac{1}{i\h}2\tilde{v}{*}\tilde{u}}{:}_{_K}
{=}0, 
 \end{aligned}
\end{equation}
without sign ambiguity by requesting that this is in the same sheet as
the starting point $t=0$. The idempotent properties of 
$\varpi_{00}$ and $\overline{\varpi}_{00}$ follows immediately, but 
the product $\varpi_{00}{*}\overline{\varpi}_{00}$ is not defined.  
We call 
$\varpi_{00}$ and $\overline{\varpi}_{00}$ {\bf vacuum} and 
{\bf bar-vacuum} respectively. 

Now compare \eqref{eq:vacuum123123} with the formula \eqref{formvac} in the case $m{=}1$. 
If $K{\in}{\mathfrak H}_+(V)$. Then 
$\delta^{(V)}_*(\tilde{\pmb v}){*}\delta^{(V)}_*(\tilde{\pmb u})$
is welldefined, but the $K$-ordered expression in \eqref{formvac}
has a sign ambiguity.

On the other hand, 
${:}\varpi_{00}{:}_{_K}$ is defined without sign ambiguity. 
We see that the $\pm$ sign of  
$\delta^{(V)}_*(\tilde{\pmb v}){*}\delta^{(V)}_*(\tilde{\pmb u})$ must
be chosen so that these equalize in $V$-class expressions. 

For every $k$,  
${\varpi}_{00}(k)=\lim_{s\to{-}\infty}e^{-\frac{1}{2}s}e_*^{s\frac{1}{i\h}{\tilde u}_k{\ctt}{\tilde v}_k}$
and 
$\overline{\varpi}_{00}(k)$ $=$
$\lim_{t\to\infty}e^{\frac{1}{2}t}e_*^{-t\frac{1}{i\h}{\tilde u}_k{\ctt}{\tilde v}_k}$
are called the {\bf partial vacuum} and  
the {\bf partial bar-vacuum} respectively.
As they are given by $\lim_{t}$, the exponential law gives the
idempotent property 
${\varpi}_{00}(k){*}{\varpi}_{00}(k){=}{\varpi}_{00}(k),\quad 
\overline{\varpi}_{00}(k){*}\overline{\varpi}_{00}(k){=}\overline{\varpi}_{00}(k)$.
In these definitions, the $*$-product ${\varpi}_{00}(k){*}\overline{\varpi}_{00}(k)$
is not fixed depending on $s{+}t$. We call 
$$
{\widetilde\varpi}(L){=}\varpi_{00}(1){*}\varpi_{00}(2){*}\cdots{*}\varpi_{00}(m)
$$
a standard vacuum, where $\varpi_{00}(k){=}\lim_{t\to{-}\infty}
e_*^{t\frac{1}{i\h}\tilde{u}_k{*}\tilde{v}_k}$, but to fix this
without sign ambiguity, we have the mention about the path to $t\to
-\infty$ so that the path does not change sheets, which is always
possible, but we can select the another sheet to obtain 
${-}{\widetilde\varpi}(L)$. Such a selection rule does not suffer the
definition of vacuums. As  
$$
\delta^{(V)}_*(\tilde{\pmb v}){*}\delta^{(V)}_*(\tilde{\pmb u})
{=}\prod_k \delta^{(V)}_*(\tilde{v}_k){*}\delta^{(V)}_*(\tilde{\pmb u})\quad {\text{and}}\quad 
\widetilde{\varpi}(L){=}\prod_k\lim_{t\to -\infty}
e^{-t\frac{1}{2}}e_*^{t\frac{1}{i\h}\tilde{u}_k{\ctt}\tilde{v}_k},
$$
the next one follows:
\begin{prop}
$\delta^{(V)}_*(\tilde{\pmb v}){*}\delta^{(V)}_*(\tilde{\pmb u})$ is what we called the 
(total) {\it vacuum} $\widetilde{\varpi}(L)$ in previous notes.  
Precisely, in every $V$-class expression we have 
${:}\delta^{(V)}_*(\tilde{\pmb v}){*}\delta^{(V)}_*(\tilde{\pmb u}){:}_{_K}{=}
{:}\widetilde{\varpi}(L){:}_{_K},$
but the r.h.s. is defined for generic ordered expressions. 
Similarly, we have 
${:}\delta^{(V)}_*(\tilde{\pmb u}){*}\delta^{(V)}_*(\tilde{\pmb v}){:}_{_K}{=}
{:}\widetilde{\varpi}(\overline{L}){:}_{_K}$.
\end{prop}

As $\tilde{v}_k{*}\widetilde{\varpi}(L){=}0$ we see in generic ordered 
expression 
\begin{equation}\label{polpol}
{\e}_{00}(L){*}\widetilde{\varpi}(L){=}i^m\widetilde{\varpi}(L).
\end{equation}

In the case $\widetilde{\varpi}(L)$, 
the regular representation space w.r.t. $\widetilde{\varpi}({L})$ 
is ${\mathbb C}[\tilde{\pmb u}]{*}\widetilde{\varpi}({L})$.
This is obviously ordinary commutative space of functions $f(\tilde{\pmb u})$.
The detail will be mentioned in \S\,\ref{VacRep}. 
Here we give only a several comments.
 
Extending $\mathbb C$-linearly the natural mapping $\iota(\tilde{\pmb u}){=}\tilde{\pmb v}$,
we obtain a nondegenerate bilinear form 
$$
\langle f(\tilde{\pmb u}), g(\tilde{\pmb u})\rangle{=}
\widetilde{\varpi}({L}){*}\iota(f(\tilde{\pmb u})){*}g(\tilde{\pmb u}){*}\widetilde{\varpi}({L})
$$
such that 
$$
\langle f(\tilde{\pmb u}), g(\tilde{\pmb u})\rangle{=}
\langle g(\tilde{\pmb u}), f(\tilde{\pmb u})\rangle,
\quad 
\langle f(\tilde{\pmb u}), g(\tilde{\pmb u}){*}h(\tilde{\pmb u})\rangle{=}
\langle f(\tilde{\pmb u}){*}g(\tilde{\pmb u}), h(\tilde{\pmb u})\rangle.
$$
We call this the {\it Frobenius algebra} w.r.t. $\widetilde{\varpi}({L})$.
(See Wikipedia for a quick view of Frobenius algebra.)

Although $K$-ordered expressions may have sign ambiguity,   
$*$-products of vacuums are defined without ambiguity.  
Namely under $V$-class expressions, the family of elements 
$\delta^{(V)}_*(\tilde{\pmb v}{-}\tilde{\pmb y}){*}
\delta^{(V)}_*(\tilde{\pmb u}{-}\tilde{\pmb x})$ for  
$(\tilde{\pmb x},\tilde{\pmb y})\in V$ generates a
 noncommutative associative algebra. 

\begin{prop}\label{prodvacassoc} 
The family of elements 
$\{\delta^{(V)}_*(\tilde{\pmb v}{-}\tilde{\pmb y}){*}
\delta^{(V)}_*(\tilde{\pmb u}{-}\tilde{\pmb x}); 
(\tilde{\pmb  x},\tilde{\pmb  y})\in V\}$
associatively closed under the $*$-product and every element
is an idempotent element.  
By Theorem\,\ref{Fundcal} together 
with changing variables gives the product formula: 
\begin{equation}\label{prodvac}
\big(\delta^{(V)}_*(\tilde{\pmb v}{-}\tilde{\pmb y}){*}
\delta^{(V)}_*(\tilde{\pmb u}{-}\tilde{\pmb x})\big){*}
\big(\delta^{(V)}_*(\tilde{\pmb v}{-}\tilde{\pmb y}'){*}
\delta^{(V)}_*(\tilde{\pmb u}{-}\tilde{\pmb x}')\big)
{=}
e^{{-}\frac{1}{i\h}\langle(\tilde{\pmb y}{-}\tilde{\pmb y}')J,\,\,
             \tilde{\pmb x}{-}\tilde{\pmb x}'\rangle}
\delta^{(V)}_*(\tilde{\pmb v}{-}\tilde{\pmb y}){*}
\delta^{(V)}_*(\tilde{\pmb u}{-}\tilde{\pmb x}')
\end{equation}
where 
$\langle(\tilde{\pmb  y}{-}\tilde{\pmb y}')J,\,\,\tilde{\pmb x}{-}\tilde{\pmb x}'\rangle
{=}\sum_{i=1}^m(\tilde{x}_i{-}\tilde{x}_i')(\tilde{y}_i{-}\tilde{y}_i')$. 
Denote by 
$\{\tilde{\pmb y}\tilde{\pmb x}\}_*{=}
\delta^{(V)}_*(\tilde{\pmb v}{-}\tilde{\pmb y}){*}
\delta^{(V)}_*(\tilde{\pmb u}{-}\tilde{\pmb x})$ for simplicity and
call it the {\bf field of vacuums}. Hence \eqref{prodvac}
is the product formula of fields of vacuums.
\end{prop}

\noindent
{\bf Proof}\,\, The associativity follows by the direct
calculation. It is enough to show the product formula. By
definition, we have 
$$
\begin{aligned}
&\delta^{(V)}_*(\tilde{\pmb v}{-}\tilde{\pmb y}){*}
\delta^{(V)}_*(\tilde{\pmb u}{-}\tilde{\pmb x}){*}
\delta^{(V)}_*(\tilde{\pmb v}{-}\tilde{\pmb y}'){*}
\delta^{(V)}_*(\tilde{\pmb u}{-}\tilde{\pmb x}')\\
&{=}
\!\iint\!\!\iint\! e_*^{\frac{1}{i\h}\langle\tilde{\pmb\eta},\,\tilde{\pmb v}{-}\tilde{\pmb y}\rangle}
{*}e_*^{\frac{1}{i\h}\langle\tilde{\pmb\xi},\,\tilde{\pmb u}{-}\tilde{\pmb x}\rangle}
{*}e_*^{\frac{1}{i\h}\langle\tilde{\pmb\eta}',\,\tilde{\pmb v}{-}\tilde{\pmb y}'\rangle}
{*}e_*^{\frac{1}{i\h}\langle\tilde{\pmb\xi}',\,\tilde{\pmb u}{-}\tilde{\pmb x}'\rangle}
\dbar{V_{\tilde{\pmb\xi}}}\dbar{V_{\tilde{\pmb\eta}}}
\dbar{V_{\tilde{\pmb\xi'}}}\dbar{V_{\tilde{\pmb\eta'}}}.
\end{aligned}
$$
Proposition\,\ref{Fundcal} gives 
$
e_*^{\frac{1}{i\h}\langle\tilde{\pmb\xi},\,\tilde{\pmb u}{-}\tilde{\pmb x}\rangle}
{*}e_*^{\frac{1}{i\h}\langle\tilde{\pmb\eta}',\,\tilde{\pmb v}{-}\tilde{\pmb y}'\rangle}{=}
e^{\frac{1}{i\h}\langle\tilde{\pmb\xi}J,\,\tilde{\pmb\eta}'\rangle}
e_*^{\frac{1}{i\h}\langle\tilde{\pmb\eta}',\,\tilde{\pmb v}{-}\tilde{\pmb y}'\rangle}
{*}e_*^{\frac{1}{i\h}\langle\tilde{\pmb\xi},\,\tilde{\pmb u}{-}\tilde{\pmb x}\rangle}$
where $\langle\tilde{\pmb\xi}J,\,\tilde{\pmb\eta}'\rangle{=}-\sum_i\tilde{\xi}_i\tilde{\eta}'_i$. 
Plugging this with the exponential law, we have 
$$
\iint\!\!\iint 
e^{\frac{1}{i\h}\langle\tilde{\pmb\xi}J,\,\tilde{\pmb\eta}'\rangle}
e^{-\frac{1}{i\h}(\langle\tilde{\pmb\xi},\,\tilde{\pmb x}\rangle{+}\langle\tilde{\pmb\xi}',\,\tilde{\pmb x}'\rangle)}
e^{-\frac{1}{i\h}(\langle\tilde{\pmb\eta},\,\tilde{\pmb y}\rangle{+}\langle\tilde{\pmb\eta}',\,\tilde{\pmb y}'\rangle)}
e_*^{\frac{1}{i\h}\langle\tilde{\pmb\eta}{+}\tilde{\pmb\eta}',\,\tilde{\pmb v}\rangle}
{*}e_*^{\frac{1}{i\h}\langle\tilde{\pmb\xi}{+}\tilde{\pmb\xi}',\,\tilde{\pmb u}\rangle}
\dbar{V_{\tilde{\pmb\xi}}}\dbar{V_{\tilde{\pmb\eta}}}\dbar{V_{\tilde{\pmb\xi'}}}
\dbar{V_{\tilde{\pmb\eta'}}}.
$$
This becomes  
$$
\iint\!\!\iint 
e^{\frac{1}{i\h}\langle\tilde{\pmb\xi}J,\,\tilde{\pmb\eta}'\rangle}
e^{\frac{1}{i\h}(\langle\tilde{\pmb\xi},\,\tilde{\pmb x}'\rangle{-}
\langle\tilde{\pmb\xi},\,\tilde{\pmb x}\rangle)}
e^{\frac{1}{i\h}
(\langle\tilde{\pmb\eta}',\tilde{\pmb y}\rangle{-}
\langle\tilde{\pmb\eta}',\,\tilde{\pmb y}'\rangle)}
e_*^{\frac{1}{i\h}\langle\tilde{\pmb\eta}'',\,\tilde{\pmb v}{-}\tilde{\pmb y}\rangle}
{*}e_*^{\frac{1}{i\h}\langle\tilde{\pmb\xi}'',\,\tilde{\pmb u}{-}\tilde{\pmb x}'\rangle}
\dbar{V_{\tilde{\pmb\xi}}}\dbar{V_{\tilde{\pmb\eta}'}}
\dbar{V_{\tilde{\pmb\xi}''}}\dbar{V_{\tilde{\pmb\eta}''}}
$$
$$
{=}\iint 
e^{\frac{1}{i\h}\langle\tilde{\pmb\xi}J,\,\,\tilde{\pmb\eta}'\rangle}
e^{\frac{1}{i\h}(\langle\tilde{\pmb\eta}',\,\tilde{\pmb y}{-}\tilde{\pmb y}'\rangle)}
e^{\frac{1}{i\h}(\langle\tilde{\pmb\xi},\,\tilde{\pmb x}'{-}\tilde{\pmb x}\rangle)}
\dbar{V_{\tilde{\pmb\xi}}}\dbar{V_{\tilde{\pmb\eta}'}}
\{\tilde{\pmb y}\tilde{\pmb x}'\}_*,
$$
where 
$\{\tilde{y}\tilde{x}'\}_*=
\delta_*^{(V)}(\tilde{v}{-}\tilde{y}){*} 
\delta_*^{(V)}(\tilde{u}{-}\tilde{x}').$  
Note that 
$\int_{V}e^{\frac{1}{i\h}
\langle\tilde{\pmb\eta},\,\tilde{\pmb y}{-}\tilde{\pmb y}'{+}\tilde{\pmb\xi}J\rangle}
\dbar{V_{\tilde{\pmb\eta}}}
{=}
\delta^{(V)}(\tilde{\pmb y}{-}\tilde{\pmb y}'{+}\tilde{\pmb\xi}J).
$
This is supported only at 
$\tilde{\pmb\xi}{=}(\tilde{\pmb y}{-}\tilde{\pmb y}')J$. 
Hence 
$$
\iint 
e^{\frac{1}{i\h}\langle\tilde{\pmb\xi}J,\,\,\tilde{\pmb\eta}'\rangle}
e^{\frac{1}{i\h}(\langle\tilde{\pmb\eta}',\,\tilde{\pmb y}{-}\tilde{\pmb y}'\rangle)}
e^{\frac{1}{i\h}(\langle\tilde{\pmb\xi},\,\tilde{\pmb x}'{-}\tilde{\pmb x}\rangle)}
\dbar{V_{\tilde{\pmb\xi}}}\dbar{V_{\tilde{\pmb\eta}'}}
{=}e^{\frac{1}{i\h}(\langle(\tilde{\pmb y}{-}\tilde{\pmb y}')J,\,\,\tilde{\pmb x}'{-}\tilde{\pmb x}\rangle)}.
$$
Plugging this into the integral, we obtain \eqref{prodvac}.
${ }$\hfill $\Box$

\bigskip

The product formulas \eqref{prodvac} are rewritten as 
$$
\{\tilde{\pmb y}\tilde{\pmb x}\}_*{*}\{\tilde{\pmb y}'\tilde{\pmb x}'\}_*
=C_{\tilde{\pmb x},\tilde{\pmb x}'}\{\tilde{\pmb y}\tilde{\pmb x}'\}_*,\quad 
C_{{\pmb x},{\pmb x}'}=e^{-\frac{1}{i\h}\langle(\tilde{\pmb y}{-}\tilde{\pmb y}')J,\,\,
        \tilde{\pmb x}{-}\tilde{\pmb x}'\rangle},
$$
$$
\{\tilde{\pmb y}\tilde{\pmb x}\}_*{*}\{\tilde{\pmb y}'\tilde{\pmb x}'\}_*
{*}
\{\tilde{\pmb y}''\tilde{\pmb x}\}_*
{=}C_{y, y',y''}\{\tilde{\pmb y}\tilde{\pmb x}\}_*,\quad C_{y, y',y''}\not=1.
$$
It is easy to see that 
$$
(\tilde{v}_i{-}\tilde{y}_i){*}\{\tilde{\pmb y}\tilde{\pmb x}\}_*{=}0{=}
\{\tilde{\pmb y}\tilde{\pmb x}\}_*(\tilde{u}_j{-}\tilde{x}_j),
$$
$$
\{\tilde{\pmb y}\tilde{\pmb x}\}_*{*}f^{(V)}_*(\tilde{\pmb u}){*}
\{\tilde{\pmb y}\tilde{\pmb x}\}_*{=}f(\tilde{\pmb x})\{\tilde{\pmb y}\tilde{\pmb x}\}_*.
$$
Setting $\tilde{\pmb y}{=}\tilde{\pmb\phi}(\tilde{\pmb x})$ for a smooth mapping, 
$\{\tilde{\pmb\phi}(\tilde{\pmb x})\tilde{\pmb x}\}_*$  will
  be called a field of vacuums on $V\!\cap{\mathbb C}^m$. 

Setting ${\mathbb C}[\tilde{\pmb u}]{*}\{\tilde{\pmb\phi}(\tilde{\pmb  x})\tilde{\pmb x}\}_*$ 
as the regular representation space,
$\tilde{u}_i{*}$ is represented as the multiplication operator, and  
$\frac{1}{i\h}\tilde{v}_i{*}$ does as the differential operator 
\begin{equation}\label{vacfieldonconfig}
\frac{1}{i\h}\tilde{v}_i{*}p_*(\tilde{\pmb u}){*}
\{\tilde{\pmb\phi}(\tilde{\pmb  x})\tilde{\pmb x}\}_*
{=}(\partial_{\tilde{u}_i}p_*(\tilde{\pmb u})
{+}\frac{1}{i\h}\tilde{\phi}_i(\tilde{\pmb  x})
p_*(\tilde{\pmb u})){*}\{\tilde{\pmb\phi}(\tilde{\pmb  x})\tilde{\pmb x}\}_*.
\end{equation}
This is often denoted by the notation of covariant differentiation 
$$
\frac{1}{i\h}\tilde{v}_i{*}p_*(\tilde{\pmb u}){*}\{\tilde{\pmb\phi}(\tilde{\pmb  x})\tilde{\pmb x}\}_*
{=}(\nabla_{\tilde{u}_i}p_*(\tilde{\pmb u}))
{*}\{\tilde{\pmb\phi}(\tilde{\pmb  x})\tilde{\pmb x}\}_*.
$$

\bigskip
Recall the definition of  $*$-delta functions \eqref{fulldelta}
of full variables.
The next formula gives a relation between $\delta^{(V)}_*(\pmb u{-}\pmb x)$ and ``half-variable''
$*$-delta functions.
\begin{equation}
\begin{aligned}
{:}\delta^{(V)}_*({\pmb u}{-}{\pmb x}'){:}_{_K} &{=} 
\int_{V}
{:}e_*^{\frac{1}{i\h}\langle{\pmb\xi},{\pmb u}{-}{\pmb x}'\rangle}
{:}_{_K}
\dbar V_{\pmb\xi}{=}
\int_{V}
e^{-\frac{1}{2i\h}\langle{\tilde{\pmb\xi}}J,\tilde{\pmb\eta}\rangle}
{:}e_*^{\frac{1}{i\h}\langle{\tilde{\pmb\xi}},\tilde{\pmb u}{-}\tilde{\pmb x}'\rangle}{*}
e_*^{\frac{1}{i\h}\langle{\tilde{\pmb\eta}},\tilde{\pmb v}{-}\tilde{\pmb y}'\rangle}{:}_{_K}
\dbar{V_{\tilde{\pmb\xi}}}\dbar{V_{\tilde{\pmb\eta}}} \\
&{=}
\int_{V}
{:}e_*^{\frac{1}{i\h}
\langle{\tilde{\pmb\xi}},\tilde{\pmb u}{-}\tilde{\pmb x}'{+}\frac{1}{2}\tilde{\pmb\eta}J\rangle}{*}
e_*^{\frac{1}{i\h}\langle{\tilde{\pmb\eta}},\tilde{\pmb v}{-}\tilde{\pmb y}'\rangle}{:}_{_K}
\dbar{V_{\tilde{\pmb\xi}}}\dbar{V_{\tilde{\pmb\eta}}}
=\int_{V}
{:}\delta^{(V)}_*(\tilde{\pmb u}{-}\tilde{\pmb x}'{+}\frac{1}{2}\tilde{\pmb\eta}J){*}
e_*^{\frac{1}{i\h}\langle{\tilde{\pmb\eta}},\tilde{\pmb v}{-}\tilde{\pmb y}'\rangle} {:}_{_K}
\dbar{V_{\tilde{\pmb\eta}}}.
\end{aligned}
\end{equation}
Replacing $\frac{1}{2}\tilde{\pmb\eta}J$ by $\tilde{\pmb\xi}'$, we see 
$$
\delta^{(V)}_* ({\pmb u}{-}\pmb x'){=}
\int_{V}
2^m\delta^{(V)}_*(\tilde{\pmb u}{-}\tilde{\pmb x}'{+}\tilde{\pmb\xi'}){*}
e_*^{-\frac{2}{i\h}\langle{\tilde{\pmb\xi}'\!J},\,\tilde{\pmb v}{-}\tilde{\pmb y}\rangle}
\dbar{V_{\tilde{\pmb\xi'}}}.
$$ 

\medskip
As $\langle{\tilde{\pmb\xi}'\!J},\,\tilde{\pmb v}{-}\tilde{\pmb y}'\rangle
{*}\delta_*(\tilde{\pmb v}{-}\tilde{\pmb y}'){=}0$ 
gives 
$e_*^{-\frac{2}{i\h}\langle{\tilde{\pmb\xi}'J},\,\tilde{\pmb v}{-}\tilde{\pmb y}'\rangle}
\delta_*(\tilde{\pmb v}{-}\tilde{\pmb y})=
e^{-\frac{2}{i\h}\langle{\tilde{\pmb\xi}'J},\,\tilde{\pmb y}{-}\tilde{\pmb y}'\rangle}
\delta_*(\tilde{\pmb v}{-}\tilde{\pmb y})$, we have  
$$
\delta^{(V)}_*({\pmb u}{-}\pmb x'){*}\{\tilde{\pmb y}\tilde{\pmb x}\}_*{=}
2^m\int_{V\cap{\mathbb C}^m}\!\!\!
e^{-\frac{2}{i\h}\langle{\tilde{\pmb\xi}'\!J},\,\tilde{\pmb y}{-}\tilde{\pmb y}'\rangle}
\delta^{(V)}_*(\tilde{\pmb u}{-}\tilde{\pmb x}'{+}\tilde{\pmb\xi'})
\dbar{V_{\tilde{\pmb\xi'}}}
{*}\{\tilde{\pmb y}\tilde{\pmb x}\}_*
{=}
2^me^{\frac{2}{i\h}\langle{\tilde{\pmb u}{-}\tilde{\pmb x}'\!J},\,\tilde{\pmb y}{-}\tilde{\pmb y}'\rangle}
\{\tilde{\pmb y}\tilde{\pmb x}\}_*.
$$
In particular, $\delta^{(V)}_*({\pmb u}){*}\{\tilde{0}\tilde{0}\}_*{=}2^m\{\tilde{0}\tilde{0}\}_*$. 
Several interesting properties will be given in the next section.

\bigskip
On the other hand, if $|{\rm{Re}}\,s|$ is sufficiently 
large, then 
$e_*^{(s{+}i\sigma)\frac{1}{i\h}\tilde{u}{*}\tilde{v}}$, 
$e_*^{(s{+}i\sigma)\frac{1}{i\h}\tilde{v}{*}\tilde{u}}$ are both 
$2\pi$-periodic w.r.t. $\sigma$. More precisely, 
these are $2\pi$-periodic w.r.t. $\sigma$, if $s$ is outside of 
the exchanging interval $[a,b]$.  
Thus, it is better to define vacuums as the limits of 
period integral: 
\begin{equation}\label{defvacuum}
2\pi\varpi_{00}=\lim_{t\to -\infty} 
\int_{-\pi}^{\pi}e_*^{(t{+}i\sigma)\frac{1}{i\h}\tilde{u}{*}\tilde{v}}d\sigma,
\quad 
2\pi\overline{\varpi}_{00}=\lim_{s\to\infty} 
\int_{-\pi}^{\pi}e_*^{(s{+}i\sigma)\frac{1}{i\h}\tilde{v}{*}\tilde{u}}d\sigma.
\end{equation}
In fact, we have no need to take the limit. Cauchy's
integral theorem  gives  
\begin{equation*}
\frac{1}{2\pi}\int_{-\pi}^{\pi}{:}e_*^{(s{+}i\sigma)\frac{1}{i\h}\tilde{u}{*}\tilde{v}}{:}_{_K}d\sigma
=\left\{
\begin{matrix}
{:}\varpi_{00}{:}_{_K},&  s{<}a\\
0,&    s{>}b
\end{matrix}
\right.,
\qquad
\frac{1}{2\pi}\int_{-\pi}^{\pi}{:}e_*^{(s{+}i\sigma)\frac{1}{i\h}\tilde{v}{*}\tilde{u}}{:}_{_K}d\sigma
=\left\{
\begin{matrix}
0,&  s{<}a\\
{:}\overline{\varpi}_{00}{:}_{_K},& s{>}b.
\end{matrix}
\right.
\end{equation*}

\medskip
The product $\varpi_{00}{*}\overline{\varpi}_{00}$ can not be 
defined directly by the definition, but the product 
$$
\int_{-\pi}^{\pi}e_*^{(s{+}i\sigma)\frac{1}{i\h}\tilde{u}{*}\tilde{v}}d\sigma{*}
\int_{-\pi}^{\pi}e_*^{(s'{+}i\sigma')\frac{1}{i\h}\tilde{v}{*}\tilde{u}}d\sigma'
=
\int_{-\pi}^{\pi}\int_{-\pi}^{\pi}
e_*^{(s{+}i\sigma)\frac{1}{i\h}\tilde{u}{*}\tilde{v}
{+}(s'{+}i\sigma')\frac{1}{i\h}\tilde{v}{*}\tilde{u}}
d\sigma d\sigma'
$$
can be defined always to give $0$, for by using 
$\frac{1}{i\h}\tilde{u}{*}\tilde{v}=\frac{1}{i\h}\tilde{u}{\ctt}\tilde{v}{-}\frac{1}{2}$,
and $\frac{1}{i\h}\tilde{v}{*}\tilde{u}
=\frac{1}{i\h}\tilde{u}{\ctt}\tilde{v}{+}\frac{1}{2}$, 
the change of variables gives
$$
\int_{-\pi}^{\pi}e^{+i\sigma}d\sigma\int_{-\pi}^{\pi}
e^{\frac{1}{2}(s'{-}s{-}i\tau)}e_*^{(s{+}s'{+}i\tau)\frac{1}{i\h}\tilde{u}{\ctt}\tilde{v}}d\tau=0.
$$
Thus, we have 
\begin{prop}\label{control}
For every polynomial $p(u,v)$, 
$\varpi_{00}{*}p_*(\tilde{u},\tilde{v}){*}\overline{\varpi}_{00}{=}0{=}
\overline{\varpi}_{00}{*}p(\tilde{u},\tilde{v}){*}{\varpi}_{00}$ in generic 
ordered expression. 
\end{prop}

\noindent
{\bf Note}\,\,
The next identities are easy to see 
$$
\varpi_{00}{*}\frac{1}{i\h}\tilde{u}{\ctt}\tilde{v}=
\frac{1}{2}\varpi_{00},\quad
\frac{1}{i\h}\tilde{u}{\ctt}\tilde{v}{*}\overline{\varpi}_{00}=
{-}\frac{1}{2}\overline{\varpi}_{00}.
$$ 
Hence in order to keep the associativity  
$(\varpi_{00}{*}\frac{1}{i\h}\tilde{u}{\ctt}\tilde{v})
{*}\overline{\varpi}_{00}=
\varpi_{00}{*}
(\frac{1}{i\h}\tilde{u}{\ctt}\tilde{v}{*}\overline{\varpi}_{00})$, 
we have to define 
$\frac{1}{2}\varpi_{00}{*}\overline{\varpi}_{00}=
{-}\frac{1}{2}\varpi_{00}{*}\overline{\varpi}_{00}=0.$ 
 To avoid such a strange phenomenon, 
we have to indicate how an element has been defined.

\subsection{Clifford vacuum}

Recall Propositions\,\ref{Cliffemarging} and \ref{GrassCliff}.
Note that as ${\e}_k^2{=}{-}1$ in ${\mathfrak K}_{s}$-ordered
expressions 
we have 
$$
\frac{1}{2}(1{+}i{\e}_k){+}\frac{1}{2}(1{-}i{\e}_k){=}1,\,\,  
\frac{1}{2}(1{+}i{\e}_k){*}\frac{1}{2}(1{-}i{\e}_k){=}0,\,\, 
(\frac{1}{2}(1{+}i{\e}_k))_*^2{=}\frac{1}{2}(1{+}i{\e}_k),\,\,
(\frac{1}{2}(1{-}i{\e}_k))_*^2{=}\frac{1}{2}(1{-}i{\e}_k).
$$
$$
(1{+}i{\e}_k){*}\widetilde{\varpi}_*(L){=}0{=}
\widetilde{\varpi}_*(L){*}(1{+}i{\e}_k), \quad 
\frac{1}{2}(1{-}i{\e}_k){*}\widetilde{\varpi}_*(L){=}\widetilde{\varpi}_*(L),
\quad 
\frac{1}{2}(1{+}i{\e}_k){*}\tilde{u}_k{=}\tilde{u}_k{*}\frac{1}{2}(1{-}i{\e}_k).
$$
Note next that 
\begin{prop}\label{invCliff}
In generic ordered expression, 
$\frac{1}{i\h}\int_{-\infty}^{0}e_*^{t\frac{1}{i\h}\tilde{v}_k{*}\tilde{u}_k}dt$
is welldefined to give the ${*}$-inverse
$(\tilde{v}_k{*}\tilde{u}_k)^{-1}$. Using these, 
$\tilde{u}_k^{\ctt}{=}(\tilde{v}_k{*}\tilde{u}_k)^{-1}{*}\tilde{v}_k$
satisfies $\tilde{u}_k^{\ctt}{*}\tilde{u}_k{=}1$,
$\tilde{u}_k{*}\tilde{u}_k^{\ctt}{=}
1{-}{\varpi}_{00}(k)$. $\tilde{u}_k^{\ctt}$ is called a (left) half-inverse of
$\tilde{u}_k$. 
As $(\tilde{v}_k{*}\tilde{u}_k)^{-1}$ commutes with ${\e}_k$ and 
${\e}_k{*}\tilde{v}_k{=}{-}\tilde{v}_k{*}{\e}_k$, $\tilde{u}_k^{\ctt}$
anti-commutes with ${\e}_k$. 
Moreover as $\tilde{v}_k{*}{\varpi}_{00}(k){=}0$, we see 
$\tilde{u}_k^{\ctt}{*}\widetilde{\varpi}_*(L){=}0$ for every $k$.
\end{prop}

We set now 
$$
\xi_k{=}\frac{1}{2}(1{+}i{\e}_k){*}\tilde{u}_k, \quad 
\eta_k{=}\frac{1}{2}(1{-}i{\e}_k){*}\tilde{u}^{\ctt}_k
{=}\frac{1}{2}(1{-}i{\e}_k){*}(\tilde{v}_k{*}\tilde{u}_k)^{-1}{*}\tilde{v}_k.
$$
The next formulas are easy to see 
$$
\xi_k^2{=}0{=}\eta^2_k,\quad \xi_k{*}\eta_k{+}\eta_k{*}\xi_k
{=}1{-}\frac{1}{2}(1{+}i{\e}_k){*}{\varpi}_{00}(k){=}1,\quad  
\eta_k{*}\widetilde{\varpi}_*(L){=}0{=}\widetilde{\varpi}_*(L){*}\xi_k. 
$$
Hence, $\{1, \xi_k, \eta_k\}$ generates $2{\times}2$ matrix algebra 
$M_2(\mathbb C)$. 
As ${\e}_k{*}{\e}_{\ell}{=}{-}{\e}_{\ell}{*}{\e}_{k}$ for $k{\not=}\ell$, we see in
general 
$$
\xi_k{*}\xi_{\ell}{+}\xi_{\ell}{*}\xi_k{=}0{=}
\eta_k{*}\eta_{\ell}{+}\eta_{\ell}{*}\eta_k, \quad 
\xi_k{*}\eta_{\ell}{+}\eta_{\ell}{*}\xi_k{=}\delta_{k\ell}.
$$
Hence $\xi_1,\cdots,\xi_m$ (resp. $\eta_1, \cdots, \eta_m$) form a 
Grassmann algebra ${\bigwedge}_m(\pmb{\xi})$ 
(resp. ${\bigwedge}_m(\pmb{\eta})$).
Grassmann algebra is called often a ``super commutative'' algebra.
Now setting 
${\widetilde\varpi}(\wedge){=}{\widetilde\varpi}(L)$, 
we call ${\widetilde\varpi}(\wedge)$ the {\bf Clifford vacuum}. 

The regular representation space is spanned by
the space of all differential forms 
$\sum f_{\alpha}(\tilde{\pmb u})\xi^{\alpha}{*}{\widetilde\varpi}(\wedge)$,
where
$\xi^{\alpha}{=}\xi_{i_1}{\wedge}\xi_{i_2}{\wedge}\cdots{\wedge}\xi_{i_{p}}$ 
using notations of Grassmann algebra. 
By this representation, the computations on Clifford algebra is 
translated into the calculus on the Grassmann algebra. 
Note that 
\begin{equation}
\tilde{v}_k{*}\xi^{\alpha}{*}{\widetilde\varpi}(\wedge){=}0, \quad
k=1{\sim}m.
\end{equation} 
It follows 
$$
\tilde{v}_k{*}f_{\alpha}(\tilde{\pmb u})\xi^{\alpha}{*}{\widetilde\varpi}(\wedge)
{=}i\h\big(\partial_{k}f_{\alpha}(\tilde{\pmb u})\big)\xi^{\alpha}{*}{\widetilde\varpi}(\wedge).
$$
It is easy to see  
\begin{equation}\label{bilinear}
{\widetilde\varpi}(\wedge){*}
{\mathbb C}[\tilde{\pmb u},\tilde{\pmb v}]
{*}\big({\bigwedge}_m(\pmb{\xi}){+}{\bigwedge}_m(\pmb{\eta})\big)
{\widetilde\varpi}(\wedge){=}
{\mathbb C}{\widetilde\varpi}(\wedge).
\end{equation}
Hence, the regular representation space has a natural nondegenerate
bilinear from over ${\mathbb C}$, by which we obtain a Frobenius
algebra structure. 
But the commutation relations with $\tilde{u}_j$ variables are a little
complicated. In a ${\mathfrak K}_{s}$-ordered expression  
$$ 
\xi_k{*}\tilde{u}_k^{2\ell}{=}\tilde{u}_k^{2\ell}{*}\xi_k,\quad 
\xi_k{*}\tilde{u}_k^{2\ell+1}{=}\tilde{u}_k^{2\ell{+}1}{*}\eta_k,\quad 
\xi_k{*}\tilde{u}_j^{2\ell}{=}\tilde{u}_j^{2\ell}{*}\xi_k, \,\,k{\not=}j.
$$
Hence, it is difficult to fix the formula of the exterior differentiation.

Note however that 
$$
\xi_k{*}(\tilde{u}^{\ctt}_k{*}\tilde{v}_k){=}(\tilde{u}^{\ctt}_k{*}\tilde{v}_k){*}\xi_k,\quad
(\tilde{u}^{\ctt}_k{*}\tilde{v}_k){*}\xi^{\alpha}{*}{\widetilde\varpi}(\wedge){=}0,\quad 
[\frac{1}{2}\tilde{u}^{2}_k, \tilde{u}^{\ctt}_k{*}\tilde{v}_k]{=}
-i\h{-}\tilde{u}_k{*}\varpi_{00}(k){*}\tilde{v}_k.
$$
Since
$\tilde{u}_k{*}\varpi_{00}(k){*}\tilde{v}_k{*}{\widetilde\varpi}(\wedge){=}0$,
these two are canonical conjugate in a sense. Then, one may use 
$$
\frac{1}{2}\tilde{u}^{2}_1,\frac{1}{2}\tilde{u}^{2}_2,\cdots,\frac{1}{2}\tilde{u}^{2}_m, \,\,
\tilde{u}^{\ctt}_1{*}\tilde{v}_1,\tilde{u}^{\ctt}_2{*}\tilde{v}_2,\cdots,\tilde{u}^{\ctt}_m{*}\tilde{v}_m
$$
instead of original generators 
$\tilde{u}_1,\cdots,\tilde{u}_m,
\tilde{v}_1,\cdots,\tilde{v}_m$.  
We denote 
$$
(x_1,\cdots,x_m,y_1,\cdots,y_m){=}
(\frac{1}{2}\tilde{u}^{2}_1,\frac{1}{2}\tilde{u}^{2}_2,\cdots,\frac{1}{2}\tilde{u}^{2}_m, \,\,
\tilde{u}^{\ctt}_1{*}\tilde{v}_1,\tilde{u}^{\ctt}_2{*}\tilde{v}_2,\cdots,\tilde{u}^{\ctt}_m{*}\tilde{v}_m).
$$
The regular representation space is spanned by elements such as 
$$
f_{\alpha}({\pmb x}){*}{\pmb\xi}^{\alpha}{*}{\widetilde\varpi}(\wedge).
$$
Thus, the exterior differential is defined by 
$$
d\Big(f_{\alpha}({\pmb x}){\pmb\xi}^{\alpha}{*}{\widetilde\varpi}(\wedge)\Big)
{=}\sum_k {\xi}_k{*}{y}_k{*}
f_{\alpha}({\pmb x})\xi^{\alpha}{*}{\widetilde\varpi}(\wedge),
\quad (\text{i.e.}\quad d{=}\sum_k {\xi}_k{*}\tilde{y}_k{*}\,\,).
$$
On this space, the computational rule is the same to those of differential forms.
As $x_k{=}\frac{1}{2}\tilde{u}^{2}_k$ and this looks ``positive'' in a
sense, the algebra generated by $(x_1,\cdots,x_m,y_1,\cdots,y_m)$ is not
isomorphic to the original Weyl algebra, but it looks to be the
algebra of $m$ copies of upper half planes. 

\subsection{Pseudo-vacuums}

Let $I_{\ctt}(K)=[a,b]$ be the exchanging interval (cf. \eqref{defvacuum}) of 
${:}e_*^{z\frac{1}{i\h}\tilde{u}{\ctt}\tilde{v}}{:}_{_K}$.  
If $a{<}s{<}b$, $e_*^{(s{+}it)\frac{1}{i\h}\tilde{u}{\ctt}\tilde{v}}$ is  
$2\pi i$-periodic in $t$, and if $s{<}a$ or
$b{<}s$ $e_*^{(s{+}it)\frac{1}{i\h}\tilde{u}{\ctt}\tilde{v}}$ 
is alternating $2\pi i$-periodic. For $a{<}s{<}b$, we set  
\begin{equation}\label{pseudoquasivac}
{:}\varpi_*(s){:}_{_K}{=}\frac{1}{2\pi}\int_0^{2\pi}
{:}e_*^{(s{+}it)(\frac{1}{i\h}\tilde{u}{\ctt}\tilde{v})}{:}_{_K}dt.
\end{equation}
This is independent of $s$ whenever $a{<}s{<}b$ and  
$(\frac{1}{i\h}\tilde{u}{\ctt}\tilde{v})
{*}\int_0^{2\pi}e_*^{(s{+}it)\frac{1}{i\h}\tilde{u}{\ctt}\tilde{v}}=0$.
We denote by ${\mathfrak K}_0$ the totality of expression parameter
$K$ such that $I_{\ctt}(K)$ contains the origin $0$. 

\begin{prop}\label{nonvanishing}
Suppose $K{\in}{\mathfrak K}_0$. Then 
$e_*^{it\frac{1}{i\h}\tilde{u}{\ctt}\tilde{v}}$ is $2\pi$-periodic  and 
$\frac{1}{2\pi}\int_0^{2\pi}{:}e_*^{it(\frac{1}{i\h}\tilde{u}{\ctt}\tilde{v})}{:}_{_K}dt$
has the idempotent property. 
This is called the {\bf pseudo-vacuum}. We denote this by
${:}\varpi_*(0){:}_{_K}$. This is a nontrivial element, 
but note that the pseudo-vacuum is expressed only by expression
parameter $K\in{\mathfrak K}_0$. 
\end{prop}
\noindent
{\bf Proof}\,\,It is enough to show that $\varpi_*(0){\not=}0$. To see
this, note that 
$\lim_{s\to -\infty}e_*^{(s{+}it)\frac{1}{i\h}\tilde{u}{\ctt}\tilde{v}{-}s\frac{1}{2}}
{=}\varpi_{00}{\not=}0$ for every fixed $t$. Hence 
$\lim_{s\to -\infty}e_*^{s\frac{1}{i\h}\tilde{u}{*}\tilde{v}}{*}
\int_0^{2\pi}e_*^{it\frac{1}{i\h}\tilde{u}{\ctt}\tilde{v}}{=}2\pi\varpi_{00}{\not=}0$.
${}$\hfill $\Box$

\bigskip
By Proposition\,\ref{nicenice22} and the definition of exchanging
interval, we see  
\begin{cor}
Obviously
${\mathfrak K}_{re}\subset{\mathfrak K}_{0}$. Hence one may treat the
pseudo-vacuum under $K_{re}$-expressions. 
\end{cor}
\bigskip
\noindent
{\bf Important remark}\,\,\,
${\e}_{00}$ satisfies ${\e}_{00}{*}{\e}_{00}{=}1$ in
$K_{re}$-expression, while ${\e}_{00}{*}\widetilde{\varpi}(L){=}i\widetilde{\varpi}(L)$. 
This may sound contradiction, but recall the definition of the product 
$e_*^{t\frac{1}{i\h}\tilde{u}{\ctt}\tilde{v}}{*}F$. This is defined 
as the solution of evolution equation 
$$
\frac{d}{dt}f_t={:}\frac{1}{i\h}\tilde{u}{\ctt}\tilde{v}{:}_{_K}{*_{_K}}f_t,\quad f_0={:}F{:}_{_K}.
$$
Note that the property ${:}{\e}_{00}^2{:}_{_{K_{re}}}{=}1$ is the
property of the solution with the initial data $1$.  
$e_*^{t\frac{1}{i\h}\tilde{u}{\ctt}\tilde{v}}{*}\widetilde{\varpi}(L)$ 
is alternating $2\pi i$-periodic in generic ordered expressions, for
the vacuum is very far from origin $1$. 

It should be noted that the identity
$e_*^{\frac{z}{i\h}\tilde{u}{\ctt}\tilde{v}}{*}
\int_0^{2\pi}e_*^{(s{+}it)\frac{1}{i\h}\tilde{u}{\ctt}\tilde{v}}{=}
\int_0^{2\pi}e_*^{(s{+}it)\frac{1}{i\h}\tilde{u}{\ctt}\tilde{v}}$ 
 holds only for $z$ is pure imaginary or for a small real part. 

\bigskip
In the case $m{>}1$ we define 
$$
\widetilde{\varpi}_*(0){=}
\frac{1}{(2\pi)^m}\int_0^{2\pi}\!{\cdots}\!
\int_0^{2\pi}{:}e_*^{it_1(\frac{1}{i\h}\tilde{u}_1{\ctt}\tilde{v}_1)}
{*}\cdots{*}e_*^{it_m(\frac{1}{i\h}\tilde{u}_m{\ctt}\tilde{v}_m)}
{:}_{_K}dt_1\cdots dt_m.
$$
We see then
$$
(\frac{1}{i\h}\tilde{u}_k{\ctt}\tilde{v}_k){*}\widetilde{\varpi}_*(0){=}0,\quad 
e_*^{it\frac{1}{i\h}\tilde{u}_k{\ctt}\tilde{v}_k}{*}\widetilde{\varpi}_*(0)
{=}\widetilde{\varpi}_*(0),\quad t{\in}{\mathbb R},\quad 
k{=}1{\sim} m.
$$
Hence the regular representation space
w.r.t. $\widetilde{\varpi}_*(0)$  is 
$\big({\mathbb C}[\tilde{\pmb u}]{+}{\mathbb C}[\tilde{\pmb v}]\big)
{*}\widetilde{\varpi}_*(0)$, on which $\frac{1}{i\h}u_k{\ctt}v_k$, 
$(\frac{1}{i\h}u_k{\ctt}v_k)^2$ act as operators 
$u_k\partial_{u_k}{-}v_k\partial_{v_k}$,
$(u_k\partial_{u_k})^2{+}(v_k\partial_{v_k})^2$ e.t.c.. 

Moreover, repeated use of the bumping
identity gives the following:
\begin{prop}
$\widetilde{\varpi}_*(0){*}
{\mathbb C}[\tilde{\pmb u}, \tilde{\pmb v}]{*}\widetilde{\varpi}_*(0){=}
{\mathbb C}\widetilde{\varpi}_*(0)$.
\end{prop}

\noindent
{\bf Proof}\,\, It is enough to show that 
${\varpi}_*(0){*}\tilde{u}{*}{\varpi}_*(0){=}
0{=}{\varpi}_*(0){*}\tilde{v}{*}{\varpi}_*(0)$ in the case $m=1$. 
Note that the bumping identity and the exponential law give  
$$
\begin{aligned}
\int_0^{2\pi}&\!e_*^{is\frac{1}{i\h}\tilde{u}{\ctt}\tilde{v}}ds{*}\,\tilde{u}{*}\!
\int_0^{2\pi}\!e_*^{it\frac{1}{i\h}\tilde{u}{\ctt}\tilde{v}}dt{=}
\tilde{u}{*}\!\int_0^{2\pi} e_*^{is(\frac{1}{i\h}\tilde{u}{\ctt}\tilde{v}{+}1)}ds{*}
\int_0^{2\pi}e_*^{it\frac{1}{i\h}\tilde{u}{\ctt}\tilde{v}}dt\\
&{=}
\tilde{u}{*}\!\int_0^{2\pi}\int_0^{2\pi}
e_*^{i(s{+}t)(\frac{1}{i\h}\tilde{u}{\ctt}\tilde{v})}{*}
e_*^{is}dsdt{=}
\tilde{u}{*}\!
\int_0^{2\pi}e_*^{i\tau(\frac{1}{i\h}\tilde{u}{\ctt}\tilde{v})}d\tau
\int_0^{2\pi}e_*^{is}ds{=}0.
\end{aligned}
$$
The similar computation gives the second one. \hfill $\Box$ 

In particular, we have 
\begin{equation}\label{Frobprod}
\widetilde{\varpi}_*(0){*}\tilde{v}_k^p{*}\tilde{u}_k^p{*}\widetilde{\varpi}_*(0){=}
(i\h)^p(1/2)_p\widetilde{\varpi}_*(0),\quad
\widetilde{\varpi}_*(0){*}\tilde{u}_k^p{*}\tilde{v}_k^p{*}\widetilde{\varpi}_*(0){=}
(-i\h)^p(1/2)_p\widetilde{\varpi}_*(0),
\end{equation}
where $(a)_p{=}a(a{+}1)\cdots(a{+}p{-}1)$, all others are $0$. 
Hence by defining an involution  
$$
\iota: {\mathbb C}[\tilde{\pmb u}]+{\mathbb C}[\tilde{\pmb v}]\to
{\mathbb C}[\tilde{\pmb v}]+{\mathbb C}[\tilde{\pmb u}],\quad  
\iota(\tilde{u}_k^p){=}\tilde{v}_k^p,\quad  
\iota(\tilde{v}_k^p){=}{-}\tilde{u}_k^p,\quad \iota(1){=}1,  
$$
the regular representation space $\big({\mathbb C}[\tilde{\pmb u}]{+}{\mathbb  C}[\tilde{\pmb v}]\big)
{*}\widetilde{\varpi}_*(0)$ 
has a natural nondegenerate bilinear product 
$\widetilde{\varpi}_*(0){*}f{*}g^{\iota}{*}\widetilde{\varpi}_*(0)$. 

On the other hand by the property ({\bf b}) mentioned in
\S\,\ref{VacVac}, integrals 
$\int_{-\infty}^{0}e_*^{s\frac{1}{i\h}\tilde{u}{\ctt}\tilde{v}}ds$, 
$-\int_0^{\infty}e_*^{s\frac{1}{i\h}\tilde{u}{\ctt}\tilde{v}}ds$  
converge to give a $*$-inverse of
$\frac{1}{i\h}\tilde{u}{\ctt}\tilde{v}$. Similarly, integrals 
$\int_{-\infty}^{0}e_*^{s\frac{1}{i\h}\tilde{v}{*}\tilde{u}}ds,
\,\, -\int_0^{\infty}e_*^{s\frac{1}{i\h}\tilde{u}{*}\tilde{v}}ds$  
converge to give $*$-inverses of
$\frac{1}{i\h}\tilde{v}{*}\tilde{u}$,
$\frac{1}{i\h}\tilde{u}{*}\tilde{v}$ respectively. 
By setting 
$\tilde{u}^{\btt}{=}\tilde{v}{*}(\frac{1}{i\h}\tilde{u}{*}\tilde{v})_*^{-1}$,
we have   
$$
\tilde{u}{*}\tilde{u}^{\btt}{=}1,\quad 
\tilde{u}^{\btt}{*}\tilde{u}{=}1{-}\overline{\varpi}_{00}.
$$ 
Note also that
$$
\overline{\varpi}_{00}{*}\overline{\varpi}_{00}{=}\overline{\varpi}_{00},\quad
\tilde{u}{*}\overline{\varpi}_{00}{=}0{=}\overline{\varpi}_{00}{*}\tilde{v},
\quad 
(\tilde{u}^{\btt})^k{*}\overline{\varpi}_{00}{*}\tilde{u}^{\ell}
\,\,{\text{is the $(k,\ell)$-matrix element}}.
$$
Since 
$(\tilde{u}{*}\tilde{v}){*}\varpi_*(0){=}{-}\frac{i\h}{2}\varpi_*(0)$,
if the associativity is expected, then
one may set 
\begin{equation}\label{strginv}
\tilde{u}^{\btt}{*}\varpi_*(0){=}
\tilde{v}{*}(\frac{1}{i\h}\tilde{u}{*}\tilde{v})^{-1}{*}\varpi_*(0)
{=}-{2}\tilde{v}{*}\varpi_*(0).
\end{equation}
However note that the double integral 
$$
\int_0^{\infty}\int_0^{2\pi}
e_*^{s\frac{1}{i\h}\tilde{u}{*}\tilde{v}}{*}e_*^{it\frac{1}{i\h}\tilde{u}{\ctt}\tilde{v}}
dsdt
$$
does not converge suffered by a singular point in the domain. However,
if we take the integral by $dt$ first, then this gives an interesting 
result. Note that $\varpi_*(0)$ is given also by  
$\varpi_*(0){=}\frac{1}{4\pi}\int_0^{4\pi}e_*^{it\frac{1}{i\h}\tilde{u}{\ctt}\tilde{v}}dt$ 
and
$\int_0^{4\pi}e_*^{(s{+}it)\frac{1}{i\h}\tilde{u}{\ctt}\tilde{v}}dt{=}0$
for $s{<}a$ or $s{>}b$, where $[a,b]$ is the exchanging interval of 
${:}e_*^{z\frac{1}{i\h}\tilde{u}{\ctt}\tilde{v}}{:}_{_K}$.
\begin{lem}\label{delicate}
Under the expression $K\in {\mathfrak K}_0$, it holds 
$e_*^{(s{+}it)\frac{1}{i\h}\tilde{u}{\ctt}\tilde{v}}{*}\varpi_*(0){=}\varpi_*(0)$
for $a{<}s{<}b$ and vanishes outside. Hence
$$
\int_0^{\infty}
e_*^{s\frac{1}{i\h}\tilde{u}{*}\tilde{v}}{*}
\varpi_*(0)ds{=}2(1{-}e^{-\frac{b}{2}})\varpi_*(0),\quad 
\tilde{u}^{\btt}{*}\varpi_*(0){=}-2(1{-}e^{-\frac{b}{2}})\tilde{v}{*}\varpi_*(0)
$$   
which is \eqref{strginv} only in the case that 
the exchanging interval is $[a,\infty)$.   
\end{lem} 
We think this might be the true nature of the formula of the product 
$\tilde{u}^{\btt}{*}\varpi_*(0)$ depending on the expression parameter.  

\medskip
Anyhow, by setting  
$\tilde{u}^{\btt}{*}\varpi_*(0){=}\alpha_0\tilde{v}{*}\varpi_*(0)$, we
compute 
$$
(\tilde{u}^{\btt})^2{*}\varpi_*(0){=}
\alpha_0\tilde{v}{*}(\frac{1}{i\h}\tilde{u}{*}\tilde{v})_*^{-1}{*}v{*}\varpi_*(0)
{=}
{-}\alpha_0\tilde{v}{*}
\int_0^{\infty}e_*^{it\frac{1}{i\h}\tilde{u}{*}\tilde{v}}{*}v{*}\varpi_*(0)dt.
$$
The bumping identity and the argument about the exchanging interval
give 
$$
(\tilde{u}^{\btt})^2{*}\varpi_*(0){=}
{-}\alpha_0\tilde{v}^2{*}
\int_0^{b}e_*^{it(\frac{1}{i\h}\tilde{u}{*}\tilde{v}{-}1)}{*}\varpi_*(0)dt
{=}
{-}\alpha_0\tilde{v}^2{*}\int_0^{b}e^{-t\frac{3}{2}}dt.
$$ 
Repeating these we have the next result:

\begin{lem}\label{lem}
Depending on the expression parameter $K\in {\mathfrak K}_0$, there
are constants $\alpha_n{\not=}0$ such that 
$(\tilde{u}^{\btt})^n{*}\varpi_*(0){=}\alpha_n v^n{*}\varpi_*(0)$. 
\end{lem}
Thus the regular representation space is the linear space 
${\mathbb C}[\tilde{u}^{\btt}, \tilde{u}]$, which is linearly
isomorphic to the Laurent polynomial space ${\mathbb C}[z^{-1},z]$.
However, the space ${\mathbb C}[\tilde{u}^{\btt},\tilde{u}]$ is not 
closed under the $*$-product. This generates an algebra ${\mathcal A}$ 
which is the direct sum of the Laurent polynomial space
and the space of matrices:  
$$
{\mathcal A}{=}{\mathbb C}[\tilde{u}^{\btt},\tilde{u}]\oplus {\mathcal M}, 
\quad 
{\text{where}}\quad 
{\mathcal M}{=}
\{(\tilde{u}^{\btt})^k{*}\overline{\varpi}_{00}{*}\tilde{u}^{\ell};
k,\ell\in {\mathbb N}\}.
$$
${\mathcal A}$ is  a noncommutative
associative algebra containing the two-sided ideal 
${\mathcal M}$ of matrices of finite rank such that the quotient algebra
is the Laurent polynomial ring ${\mathbb C}[z^{-1},z]$.
$$
0\to {\mathcal M}\to {\mathcal A}\to {\mathbb C}[z^{-1},z]\to 0.
$$  
Hence ${\mathcal A}$ may be regarded as a nontrivial extension of 
${\mathbb C}[z^{-1},z]$ by ${\mathcal M}$ such that 
$$
z{*}z^{-1}{=}1,\quad z^{-1}{*}z{=}1-\overline{\varpi}_{00},\quad 
z^{-2}{*}z{=}z^{-1}{-}\tilde{u}^{\btt}{*}\overline{\varpi}_{00},\quad 
z^{-2}{*}z^2{=} 
1{-}\overline{\varpi}_{00}{-}\tilde{u}^{\btt}{*}\overline{\varpi}_{00}{*}\tilde{u},\,e.t.c.
$$


\subsection{$SU(2)$-vacuum}\label{SU(2)vacuum}

Note that the pseudo-vacuum $\widetilde{\varpi}_*(0)$ is given by the
period integral $\int_{S^1}dt$ of a $*$-exponential function. 
By recalling the argument in \S\,\ref{anothersp}, it is natural 
to think such a period integral may extend to higher dimensions, e.g.
$\int_{S^3}dg$. In this section, we restrict our attention to the case $m=1$.

By using $x, y, \rho$ in \eqref{x,y,rho}, $*$-exponential functions 
$e_*^{it\frac{1}{i\h}H_*}$ for 
$H_*{=}\langle \tilde{\pmb u}g, \tilde{\pmb u}g\rangle{\in}{\mathfrak{su}_1}(2)J$ 
is written in the form 
$$
{:}\gamma_*(t,g){:}_{_K}{=}{:}e_*^{it\frac{1}{i\h}x(\tilde{u}^2{+}\tilde{v}^2)
{+}iy(\tilde{u}^2{-}\tilde{v}^2){+}2i\rho\tilde{u}{\ctt}\tilde{v}}{:}{_{_K}}.
$$
Recalling Proposition\,{\ref{twocover}}, we apply the formula 
\eqref{tildeKKK} for the cases  
$$
K{=}
\begin{bmatrix}
1&ic\\
ic&1
\end{bmatrix}\in{\mathfrak K}_{re},\quad
K'{=}
\begin{bmatrix}
1&ic'\\
ic'&1
\end{bmatrix}
\,\,\quad c'{=}c{+}i\theta, \,\,\,\theta{\not=}0\quad {\text{but small {for simplicity}}}.
$$
The $K'$-ordered expression is given as 
\begin{equation}\label{holomorphic}
{:}\gamma_*(t,g){:}_{_{K'}}{=}\frac{1}{\sqrt{\det(\cos t I{-}(\sin t){}^t\!gK'g)}}
e^{\frac{1}{i\h}\langle{\pmb u}g 
\frac{\sin t}{\cos tI-\sin t\,{}^t\!gK'g},{\pmb u}g\rangle}.
\end{equation}
Note that ${:}\gamma_*(t,g){:}_{_{K'}}$ is holomorphic in 
$(t, g)\in {V{\times}S'}$ where $V$ is a neighborhood of $[0, 2\pi]$ in $\mathbb C$.   

\medskip 
Note also that $\det(\cos t I{-}(\sin t){}^t\!gK'g){=}\det(\cos t I{-}(\sin t)g{}^t\!gK')$ 
and set 
$$
g\,{}^t\!g{=}
\begin{bmatrix}
x{+}iy&i\rho\\
i\rho&x{-}iy
\end{bmatrix},\quad  g{\in}S\!L(2,{\mathbb C}).
$$

Then, $x^2{+}y^2{+}\rho^2{=}1$, and the amplitude part is  
$$
\begin{aligned}
\Big(
\det&
\begin{bmatrix}
\cos t{-}(x{-}\rho c'{+}iy)\sin t& -i(\rho{+}c'(x{+}iy))\sin t\\
-i(\rho{+}c'(x{-}iy))\sin t &\cos t{-}(x{-}\rho c'{-}iy)\sin t
\end{bmatrix}\Big)^{-1/2}\\
&\qquad{=}
\Big(\big(\cos t{-}(x{-}\rho c')\sin t\big)^2{+}y^2\sin^2t{+}
  \big((\rho{+}c'x)^2{+}(c')^2y^2\big)\sin^2t\Big)^{-1/2}.
\end{aligned}
$$
This is singular at a point $y{=}0$, $\rho{+}c'x{=}0$, 
$\cot t{=}x(1{+}(c')^2)$. Hence under the $K'$-ordered expression, 
there is only one singular point 
$(\cot t, x, y,\rho){=}(0,0,0,0)$. 
As we use only $K'$-ordered expression, we need
not care about cocycle conditions, and this expressions 
determine a point set $D^4$. 
We easily see that the following: 
\begin{lem}\label{zeropt}
There is only one singular point at $(\cos t, x, y,\rho){=}(0,0,0,0)$
in the unit $4$-disk 
$$
D^4{=}\{(\cos t, x, y,\rho);\,\,
\cos^2t{+}\sin^2t(x^2{+}y^2{+}\rho^2)\leq 1\}.
$$
\end{lem}
 
\bigskip
Note that $\gamma{=}(\cos t, x, y,\rho)$ such that 
$\cos^2t{+}\sin^2t(x^2{+}y^2{+}\rho^2)=1$  
corresponds to the element  
$$
\gamma_*{=}
e_*^{it(\frac{1}{\h}x(\tilde{u}^2{+}\tilde{v}^2)
{+}iy\frac{1}{i\h}(\tilde{u}^2{-}\tilde{v}^2){+}2i\rho\frac{1}{i\h}\tilde{u}{\ctt}\tilde{v})}
\,\, {\in} \,\,SU(2)
$$
and ${:}SU(2){:}_{_{K'}}$ forms a group under the $*_{_{K'}}$-product. 
Thus, one may identify the boundary $\partial D^4$ with the group
$SU(2)$. Letting $dm_\gamma$ be the standard invariant measure on
$SU(2){=}S^3$, we set 
$$
{:}\Omega_*{:}_{_{K'}}{=}\int_{\partial D^4}{:}\gamma_*{:}_{_{K'}}dm_\gamma$$ 
and call $\Omega_*$ the $SU(2)$-{\bf vacuum}, although this is a 
collection of expressed elements.  
Obviously,  
$$
{:}SU(2){:}_{_{K'}}{*_{_{K'}}}{:}\Omega_*{:}_{_{K'}}{=}{:}\Omega_*{:}_{_{K'}}{=}
{:}\Omega_*{:}_{_{K'}}{*_{_{K'}}}{:}SU(2){:}_{_{K'}}, \,\,\text{ and hence }\,\, 
{:}(\mathfrak{su}(2)J){*}\Omega_*{:}_{_{K'}}{=}{:}\Omega_*{*}(\mathfrak{su}(2)J){:}_{_{K'}}{=}\{0\}.
$$ 

\begin{prop}
The $SU(2)$-vacuum does not vanish, i.e.
$\Omega_*\not=0$.
\end{prop}

\noindent
{\bf Proof}\,\,Recall that \eqref{holomorphic} is holomorphic in
particular on 
$(t, g)\in V{\times}V_{SU(2)}$ where  
$V$ is a neighborhood of $[0,\pi]$ in $\mathbb C$ and $V_{SU(2)}$ is
a neighborhood of $SU(2)$ in $S\!L(2,{\mathbb C})$. 
We use the projection $\iota: {\mathcal A}\to\frac{1}{2}(1{+}{\e}_{00}){\mathcal A}$
defined in Theorem\,\ref{surprise0} where $\iota({\e}_{00})$ may be
treated as $1$. It is enough to prove $\iota(\Omega_*)\not=0$. 

\setlength{\unitlength}{1mm}
\begin{picture}(30,40)(0,10)
\put(0,0){\line(1,0){30}}
\put(0,0){\line(0,1){40}}
\put(-2,20){$\iota\gamma_*(t,g)$}
\put(30,0){\line(0,1){40}}
\put(28,20){$\iota\gamma_*(t,h)$}
\put(0,40){\line(1,0){30}}
\put(10,-3){$t{=}0$}
\put(10,2){$\iota(1){=}1$}
\put(10,42){$t{=}\pi$}
\put(10,38){$\iota({\e}_{00}){=}1$}
\put(15,25){$D$}
\put(5,10){\tiny{no singular point}}
\end{picture}
\hfill\parbox[b]{.7\linewidth}
{For every fixed $g, h{\in}S'$, consider a mapping  
$C: \partial D\to V{\times}V_{SU(2)}$ as in the l.h.s.figure. 
{Note that $C$ extends to a holomorphic mapping 
$\tilde{C}: D\to  V{\times}V_{SU(2)}$.  
By Cauchy's integral theorem we see 
$\iota(\tilde{\gamma}(g)){=}\int_0^{\pi}\iota(\gamma_*(t,g)$
does not depend on $g$. Setting $g{=}1$, we have 
${:}\iota(\tilde{\gamma}(g)){:}_{_{K'}}{=}
\int_0^{\pi}{:}e_*^{t\frac{1}{i\h}(\tilde{u}^2{+}\tilde{v}^2)}{:}_{_{K'}}dt$.} 
This is non vanishing by the same reasoning as in 
Proposition\,\ref{nonvanishing}. 
Now, note that $SU(2)/S^1{=}S^2$, then integrating by the standard
volume form $dm(S^2)$ on $S^2$ gives }
$$
{}\qquad \int_{\partial D^4}{:}\iota(\gamma_*){:}_{_{K'}}dm_\gamma{=}
\int_{S^2}{:}\iota(\tilde{\gamma_*}){:}_{_{K'}}dm(S^2){=}
4\pi\int_0^{\pi}{:}e_*^{t\frac{1}{i\h}(\tilde{u}^2{+}\tilde{v}^2)}{:}_{_{K'}}dt
$$
In a sense, $\Omega_*$ is the group $SU(2)$ itself, just
as the invariant 3-form on $SU(2)$.\hfill $\Box$

Note that $e_*^{tQ_*(\tilde{u},\tilde{v})}{*}\Omega_*{=}\Omega_*$ holds only
for $t\in{\mathbb R}$. 
Note also that the space of quadratic forms is given as
${\mathfrak{sl}}(2,{\mathbb C})J$ and 
${\mathfrak{sl}}(2,{\mathbb C)}{=}{\mathfrak{su}}(2)\oplus{\mathfrak{h}}(2)$
where ${\mathfrak{h}}(2)$ is the space of all traceless hermitian matrices. 
The space ${\mathfrak{h}}(2)J$ forms a Lie algebra  
under a new bracket product $[|X,Y|]{=}i[iX,iY]_*$. But this is
isomorphic to ${\mathfrak{su}}(2)J$ and hence  
${\mathfrak{h}}(2)J$ may be regarded as the Lie algebra of $SU(2)$.

\medskip
Now, it is a problem how the regular representation space
w.r.t. $\Omega_*$ should be defined. 
One may think that this must contain the enveloping algebra of ${\mathfrak{h}}(2)J$ under 
the $*$-product, that is, $({\mathfrak{h}}(2)J)_*^n{*}\Omega_*$ are
contained. But this refuses $i$ in the constant term, 
for $i({\mathfrak{h}}(2)J){=}{\mathfrak{su}}(2)J$. 
Hence it is difficult to make an infinite dimensional algebra under 
the computation of modulo ${\mathfrak{su}}(2)J$.  

\medskip
On the other hand 
$\{1, i, \tilde{u}, i\tilde{u}, \tilde{v}, i\tilde{v}, {\mathfrak{h}}(2)J\}$
is a Lie algebra over ${\mathbb R}$ under the new bracket product
$[|X,Y|]$, and there are Lie subalgebras   
$$
\begin{aligned}
E_0{=}&\{1, \tilde{u}, \tilde{v}, \frac{1}{2\h}(\tilde{u}^2{+}\tilde{v}^2)\},\quad 
E_{+}{=}\{1, e^{\frac{\pi i}{4}}\tilde{u},\,e^{-\frac{\pi i}{4}}\tilde{v},\,
\frac{i}{2\h}(\tilde{u}^2{-}\tilde{v}^2)\}\\
&E_{-}{=}\{1, \,e^{\frac{\pi i}{4}}\frac{1}{2}(\tilde{u}{+}\tilde{v}),\, 
e^{-\frac{\pi i}{4}}\frac{1}{2}(\tilde{u}{-}\tilde{v}), \,\frac{i}{\h}\tilde{u}{\ctt}\tilde{v}\}.
\end{aligned}
$$ 
These are mutually isomorphic under suitable linear change of
generators, and their enveloping algebras are infinite dimensional. 
In general, every closed one-parameter subgroup $T_s$ of $SU(2)$ 
causes a rotation on a 2-dimensional $\mathbb R$ linear subspace 
$V_2$ of ${\mathbb C}\tilde{u}{\oplus}{\mathbb C}\tilde{v}$ such that 
$[|V_2, V_2|]\subset {\mathbb R}$. 

The $\mathbb R$-linear subspaces where the rotations occurs by 
$\frac{1}{2\h}(\tilde{u}^2{+}\tilde{v}^2)$, 
$\frac{i}{2\h}(\tilde{u}^2{-}\tilde{v}^2)$ and
$\frac{i}{\h}\tilde{u}{\ctt}\tilde{v}$ span a 
$3$-dimensional subspace $E^3$ over ${\mathbb R}$:
$$
E^3{=}
\mathbb R(\tilde{u}, \tilde{v}){\oplus}
\mathbb R(e^{\frac{\pi i}{4}}\tilde{u},\,e^{-\frac{\pi i}{4}}\tilde{v})\oplus
\mathbb R(\frac{e^{\frac{\pi i}{4}}}{2}(\tilde{u}{+}\tilde{v}),
\frac{e^{-\frac{\pi i}{4}}}{2}(\tilde{u}{-}\tilde{v}))
$$
with complementary subspace ${\mathbb R}e_0{=}{\mathbb R}((1{-}i)\tilde{u}{-}(1{+}i)\tilde{v})$.

Setting 
$$
(s,\xi_1,\xi_2, t){=}s{+}\xi_1\tilde{u}{+}\xi_2\tilde{v}{+}
t\frac{1}{\h}(\tilde{u}^2{+}\tilde{v}^2),
$$ 
the Lie bracket product on $E_0$ is 
$$
[|(s,\xi_1,\xi_2, t),\,(s',\eta_1,\eta_2, t')|]=
(\xi_1\eta_2-\xi_2\eta_1, t\eta_2{-}t'\xi_2, {-}t\eta_1{+}t'\xi_1, 0).
$$
The next one is proved by direct calculations: 
\begin{prop}\label{smallerLie}
$E_0$ forms a Lie subalgebra
under the bracket product $[|X,Y|]$. Furthermore,  
$E_0$ has a Lorentz bilinear form 
$$
\langle (s,\xi_1,\xi_2,t), (s',\eta_1,\eta_2,t')\rangle{=} 
st'{+}ts'{-}\xi_1\eta_1{-}\xi_2\eta_2,
$$
which is adjoint invariant:
$$
\langle[|(s,\xi_1,\xi_2,t), (s',\eta_1,\eta_2,t')|],
(s'',\eta'_1,\eta'_2,t'')\rangle{+}
\langle(s',\eta_1,\eta_2,t'), [|(s,\xi_1,\xi_2,t),
(s'',\eta'_1,\eta'_2,t''|])\rangle{=}0.
$$
In particular, $E_0$ is a Minkowski space.
\end{prop}
It is not hard to see that the Lie group $G$ with the Lie algebra 
$(E_0, [|\,\,,\,\,|])$ is obtained by giving a group structure on the 
space $E_0$.
Note that $E_0$ is not invariant under the Lorentz group $SO(1,3)$.

As it is naturally expected, $E_{\pm}$ are also Minkowski spaces, 
and these three $\{E_0, E_+, E_-\}$ jointly define the standard 
Minkowski metric on ${\mathbb R}^{1{+}3}$ such that the positive 
cone $V_+$ in ${\mathbb R}^{1{+}3}$ is the subset 
$$
V_+{=}\{AA^*; A{\in}S\!L(2,{\mathbb C})\}
$$ 
of all hermitian matrices ${\mathbb R}{\oplus}{\mathfrak{h}}(2)$. 
The Lorentz group $SO(1,3)$ acts only on such a joint object.
This gives a big change of the role of the Lorentz group $SO(1,3)$ 
and gives an entrance key to the relativistic quantum theory in physics.
These will be discussed in the next note.

\section{Pseudo-differential operators as vacuum representations }\label{VacRep}

Before entering into the field theory, we have to make it clear what
the infinite dimensional regular representation space by a vacuum 
makes a configuration space. 
In this subsection we set $V{=}{\mathbb R}^{2m}$ for simplicity, and we 
show that ordinary pseudo-differential operators are obtained by 
the vacuum representation of a certain $\mu$-regulated algebra. 

Let $C^{\infty}({\mathbb R}^m)$, $C_0^{\infty}({\mathbb R}^m)$ 
be the space of smooth functions on ${\mathbb R}^m$ and the space of 
all compactly supported smooth functions on ${\mathbb R}^m$ respectively.
First for every $f(\tilde{\pmb x}){\in}C_0^{\infty}({\mathbb R}^m)$
consider the correspondence  
$$
f(\tilde{\pmb x})\to f_*(\tilde{\pmb x}{+}\tilde{\pmb u})
{=}\int_{\mathbb R^m}f(\tilde{\pmb x}{+}\tilde{\pmb x}'){*}
\delta_*(\tilde{\pmb u}{-}\tilde{\pmb x}')\dbar{\tilde{\pmb x}'}.
$$  
This may be regarded as the $*$-function field defined by 
$f\in C_0^{\infty}({\mathbb R}^m)$.  
The inverse correspondence is obtained by taking the
Weyl ordered expression of the r.h.s. and replacing $\tilde{\pmb u}$ by $0$.

\medskip
Let $h(\tilde{\pmb x},\tilde{\pmb y}')$ be a smooth function 
w.r.t.$\tilde{\pmb x}$ and a tempered distribution 
w.r.t. $\tilde{\pmb y}'$ 
on ${\mathbb R}^{2m}$, we define hybrid $*$-function, 
$h_*(\tilde{\pmb x}, \tilde{\pmb v})$ by 
$$
h_*(\tilde{\pmb x}, \tilde{\pmb v}){=}
\iint
h(\tilde{\pmb x},\tilde{\pmb y}')
\delta_*(\tilde{\pmb v}{-}\tilde{\pmb y}')
\dbar{\tilde{\pmb y}'}.
$$
Consider the operator 
$$ 
f_*(\tilde{\pmb u}{+}\tilde{\pmb x}){*}\{\tilde{\pmb\phi}(\tilde{\pmb  x})\tilde{\pmb x}\}_*
\to 
h_*(\tilde{\pmb x},\tilde{\pmb v}){*}
f_*(\tilde{\pmb u}{+}\tilde{\pmb x}){*}\{\tilde{\pmb\phi}(\tilde{\pmb x})\tilde{\pmb x}\}_*.
$$
Note that 
$$
h_*(\tilde{\pmb x},\tilde{\pmb v})
{*}f_*(\tilde{\pmb u}{+}\tilde{\pmb x}){*}\{\tilde{\pmb\phi}(\tilde{\pmb  x})\tilde{\pmb x}\}_*
{=}\!\iint\!\! h(\tilde{\pmb x},\tilde{\pmb y}')f(\tilde{\pmb x}{+}\tilde{\pmb x}')
\{\tilde{\pmb y}'\tilde{\pmb x}'\}_*{*}
\{\tilde{\pmb\phi}(\tilde{\pmb  x})\tilde{\pmb x}\}_*
\dbar{\tilde{\pmb y}'}
\dbar{\tilde{\pmb x}'}.
$$
Following Proposition\,\ref{$*$-funcWeyl}, we would like to write the above in the form  
$$
h_*(\tilde{\pmb x},\tilde{\pmb v})
{*}f_*(\tilde{\pmb u}{+}\tilde{\pmb x}){*}\{\tilde{\pmb\phi}(\tilde{\pmb  x})\tilde{\pmb x}\}_*
{=}\int_{\mathbb R^m}P(h,f)(x{+}x')
\delta_*(\tilde{\pmb u}{-}\tilde{\pmb  x}')\dbar{\tilde{\pmb x}'}{*}
\{\tilde{\pmb\phi}(\tilde{\pmb  x})\tilde{\pmb x}\}_* 
$$
and to take out $P(h,f)(x{+}x')$-part. For that purpose, rewrite   
$\{\tilde{\pmb y}'\tilde{\pmb x}'\}_*{*}
\{\tilde{\pmb\phi}(\tilde{\pmb  x})\tilde{\pmb x}\}_*$ as follows
$$
\begin{aligned}
\delta_*(\tilde{\pmb v}{-}\tilde{\pmb y}'){*}
\delta_*(\tilde{\pmb u}{-}\tilde{\pmb x}'){*}
\{\tilde{\pmb\phi}(\tilde{\pmb x})\tilde{\pmb x}\}_*
&{=}
\iint e^{\frac{1}{i\h}\langle\tilde{\pmb\eta}J,\,\tilde{\pmb\xi}\rangle} 
e^{\frac{1}{i\h}\langle\tilde{\pmb\eta},\,\tilde{\pmb\phi}(\tilde{\pmb x}){-}\tilde{\pmb y}'\rangle}            
e_*^{\frac{1}{i\h}\langle\tilde{\pmb\xi},\,\tilde{\pmb u}{-}\tilde{\pmb x}'\rangle}
\{\tilde{\pmb\phi}(\tilde{\pmb x})\tilde{\pmb x}\}_*
\dbar{\tilde{\pmb\eta}}\dbar{\tilde{\pmb\xi}}\\
&{=}
\iint e^{\frac{1}{i\h}\langle\tilde{\pmb\eta},\,\tilde{\pmb\phi}(\tilde{\pmb x}){-}\tilde{\pmb y}'\rangle}            
e_*^{\frac{1}{i\h}\langle\tilde{\pmb\xi},\,\tilde{\pmb u}{-}\tilde{\pmb x}'{+}\tilde{\pmb\eta}J\rangle}
\{\tilde{\pmb\phi}(\tilde{\pmb x})\tilde{\pmb x}\}_*
\dbar{\tilde{\pmb\eta}}\dbar{\tilde{\pmb\xi}}
\end{aligned}
$$
by using  
$\langle\tilde{\pmb\eta}J,\,\tilde{\pmb\xi}\rangle{=}\langle\,\tilde{\pmb\xi},\,\tilde{\pmb\eta}J\rangle$.
Integrating by $\tilde{\pmb\xi}$, we have 
$\int e_*^{\frac{1}{i\h}\langle\tilde{\pmb\xi},\,\tilde{\pmb u}{-}\tilde{\pmb x}'{+}\tilde{\pmb\eta}J\rangle}
\dbar{\tilde{\pmb\xi}}
{=}\delta_*(\tilde{\pmb u}{-}\tilde{\pmb x}'{+}\tilde{\pmb\eta}J)$,
and 
$$
\begin{aligned}
h_*(\tilde{\pmb x},\tilde{\pmb v})&
{*}f_*(\tilde{\pmb u}{+}\tilde{\pmb x}){*}\{\tilde{\pmb\phi}(\tilde{\pmb  x})\tilde{\pmb x}\}_*\\
&{=}
\!\iint\!\!\int h(\tilde{\pmb x},\tilde{\pmb y}')f(\tilde{\pmb x}{+}\tilde{\pmb x}')
e^{\frac{1}{i\h}\langle\tilde{\pmb\eta},\,\tilde{\pmb\phi}(\tilde{\pmb x}){-}\tilde{\pmb y}'\rangle}
\delta_*(\tilde{\pmb u}{-}\tilde{\pmb x}'{+}\tilde{\pmb\eta}J)
\dbar{\tilde{\pmb\eta}}\dbar{\tilde{\pmb x}'}\dbar{\tilde{\pmb y}'}
{*}\{\tilde{\pmb\phi}(\tilde{\pmb x})\tilde{\pmb x}\}_*.
\end{aligned}
$$
Now, note that the $*$-variable $\tilde{\pmb v}$ does not contained in
the integral. By Proposition\,\ref{$*$-funcWeyl}, we take 
its Weyl ordered expression. Since 
${:}\delta_*(\tilde{\pmb u}{-}\tilde{\pmb x}'{+}\tilde{\pmb\eta}J){:}_0
{=}\delta(\tilde{\pmb u}{-}\tilde{\pmb x}'{+}\tilde{\pmb\eta}J)$ 
and this is supported only at $\tilde{\pmb\eta}{=}(\tilde{\pmb u}{-}\tilde{\pmb x}')J$,
we replace $\tilde{\pmb\eta}$ in the integral by $(\tilde{\pmb u}{-}\tilde{\pmb x}')J$.
Then putting $\tilde{\pmb u}{=}0$, we have 
$$
\begin{aligned}
h_*(\tilde{\pmb x},\tilde{\pmb v})
{*}f_*(\tilde{\pmb u}{+}\tilde{\pmb x}){*}\{\tilde{\pmb\phi}(\tilde{\pmb  x})\tilde{\pmb x}\}_*
{=}
\Big(\iint 
h(\tilde{\pmb x},\tilde{\pmb y}')f(\tilde{\pmb x}{+}\tilde{\pmb x}')
e^{-\frac{1}{i\h}\langle\tilde{\pmb x}'J,\,
\tilde{\pmb\phi}(\tilde{\pmb x}){-}\tilde{\pmb y}'\rangle}
\dbar{\tilde{\pmb x}'}\dbar{\tilde{\pmb y}'}\Big)
\{\tilde{\pmb\phi}(\tilde{\pmb x})\tilde{\pmb x}\}_*.
\end{aligned}
$$
If $\tilde{\pmb\phi}(\tilde{\pmb  x}){=}0$, 
the operator in the big parentheses may be regarded 
as the pseudo-differential operator of H{\"o}rmander. The product
formula is given by $\Psi$DO-product formula.
In general,   
$\tilde{\pmb\phi}(\tilde{\pmb  x})$ is regarded as the electromagnetic potential.

\bigskip
\noindent
{\bf Note}\,\, 
$\iint h(\tilde{\pmb x}',\tilde{\pmb y}')
f(\tilde{\pmb x}'')
e^{\frac{1}{i\h}\langle\tilde{\pmb y}'{-}\tilde{\pmb y},\,\tilde{\pmb x}'{-}\tilde{\pmb x}''\rangle}
\dbar{V_{\tilde{\pmb x}''}}\dbar{V_{\tilde{\pmb y}'}}$
does {\it not} converge in general,  
even if $f(\tilde{\pmb x}'')\in C_0^{\infty}(V)$, 
but if 
$h(\tilde{\pmb x}',\tilde{\pmb y}'){=}(\tilde{\pmb y}'{-}\tilde{\pmb y})^{\alpha}$,
then the integration by parts gives that 
$$
\int(\tilde{\pmb y}'{-}\tilde{\pmb y})^{\alpha}
f(\tilde{\pmb x}''){*}
e_*^{\frac{1}{i\h}\langle\tilde{\pmb y}'{-}\tilde{\pmb y},\,\tilde{\pmb x}'{-}\tilde{\pmb x}''\rangle}
\dbar{V_{\tilde{\pmb x}''}}
\dbar{V_{\tilde{\pmb y}'}}{=}
(i\h\partial_{\tilde{\pmb x}})^{\alpha}f(\tilde{\pmb x}').
$$
By this reason, what is essential for the convergence is the remainder term of the expansion 
$$
h(\tilde{\pmb x}',\tilde{\pmb y}'){=}\sum_{|\alpha|\leq k}\frac{1}{\alpha !}
\partial_{\tilde{\pmb y}'}^{\alpha}
h(\tilde{\pmb x}',\tilde{\pmb y})(\tilde{\pmb y}'{-}\tilde{\pmb y})^{\alpha}
{+}h_{(k)}(\tilde{\pmb x}',\tilde{\pmb y}').
$$
The integral converges if $h_{(k)}(\tilde{\pmb x}', \tilde{\pmb y}')$ belongs to
$L_1(V\cap{\mathbb C}^m)$ for some $k$. To make the required condition
clear, we have to define the notion of 
``symbol class'' of functions. 
Such a method of calculation of integrals is often called 
oscillatory integrals and it is denoted by 
$$
{os}\text{-}\!\!\iint h(\tilde{\pmb x}',\tilde{\pmb y}')
f(\tilde{\pmb x}'')
e^{\frac{1}{i\h}\langle\tilde{\pmb y}'{-}\tilde{\pmb y},\,\tilde{\pmb x}'{-}\tilde{\pmb x}''\rangle}
\dbar{\tilde{\pmb x}''}\dbar{\tilde{\pmb y}'}.
$$

\medskip
Note also that the variables of configuration space play only
parameters. In contrast, these play essential role 
in pseudo-differential operators of Weyl type mentioned below.

\bigskip
Setting $\pmb u{=}(\tilde{\pmb u},\tilde{\pmb v})$, 
$\pmb x{=}(\tilde{\pmb x},\tilde{\pmb y})$, and using the correspondence
$$
f_*(\tilde{\pmb u}){=}\int f(\tilde{\pmb x})\delta_*(\tilde{\pmb
  u}{-}\tilde{\pmb x}),\quad
h_*(\pmb u){=}\iint h({\pmb x})\delta_*(\pmb u{-}\pmb x)\dbar{\pmb x},
$$ 
we consider the operator 
$$
h_*(\pmb u){*}f_*(\tilde{\pmb u}){*}\{\tilde{0}\tilde{0}\}_*
{=}\iint\!\!\int
h({\pmb x})f(\tilde{\pmb x}')
\delta_*(\pmb u{-}\pmb x){*}\delta_*(\tilde{\pmb u}{-}\tilde{\pmb x}')
\dbar{\tilde{\pmb x}}
\dbar{\tilde{\pmb y}}\dbar{\tilde{\pmb x}'}
{*}\{\tilde{0}\tilde{0}\}_*.
$$
But we want to rewrite this in the form 
$\int P(h,f)(\tilde{\pmb x}")\delta_{*}(\tilde{\pmb u}{-}\tilde{\pmb x}"){*}\{\tilde{0}\tilde{0}\}_*$. 
Note that by Proposition\,\ref{Fundcal}  we have 
$$
\begin{aligned}
\delta_*({\pmb u}{-}{\pmb x}){*}\delta_*(\tilde{\pmb u}{-}\tilde{\pmb x}')&{=}
\iint\!\int
e^{-\frac{1}{2i\h}\langle{\tilde{\pmb\xi}}J,\,\tilde{\pmb\eta}\rangle}
e_*^{\frac{1}{i\h}\langle{\tilde{\pmb\xi}},\tilde{\pmb u}{-}\tilde{\pmb x}\rangle}{*}
e_*^{\frac{1}{i\h}\langle{\tilde{\pmb\eta}},\tilde{\pmb v}{-}\tilde{\pmb y}\rangle}{*}
e_*^{\frac{1}{i\h}\langle{\tilde{\pmb\xi}'},\tilde{\pmb u}{-}\tilde{\pmb x}'\rangle}
\dbar{\tilde{\pmb\xi}}\dbar{\tilde{\pmb\eta}}\dbar{\tilde{\pmb\xi}'}\\
&{=}
\iint\!\int
e^{-\frac{1}{2i\h}\langle{\tilde{\pmb\xi}}J,\,\tilde{\pmb\eta}\rangle}
e^{\frac{1}{i\h}\langle{\tilde{\pmb\eta}}J,\,\tilde{\pmb\xi}'\rangle}
e_*^{\frac{1}{i\h}\langle{\tilde{\pmb\xi}},\tilde{\pmb u}{-}\tilde{\pmb x}\rangle}{*}
e_*^{\frac{1}{i\h}\langle{\tilde{\pmb\xi}'},\tilde{\pmb u}{-}\tilde{\pmb x}'\rangle}{*}
e_*^{\frac{1}{i\h}\langle{\tilde{\pmb\eta}},\tilde{\pmb v}{-}\tilde{\pmb y}\rangle}{*}
\dbar{\tilde{\pmb\xi}}\dbar{\tilde{\pmb\eta}}\dbar{\tilde{\pmb\xi}'}.
\end{aligned}
$$
Since $\tilde{v}_i{*}\{\tilde{0}\tilde{0}\}_*{=}0$, we see 
$
e_*^{\frac{1}{i\h}\langle\tilde{\pmb\eta}, \tilde{\pmb v}{-}\tilde{\pmb y}\rangle}
{*}\{\tilde{0}\tilde{0}\}_*{=}
e^{-\frac{1}{i\h}\langle{\tilde{\pmb\eta}},\tilde{\pmb y}\rangle}
{*}\{\tilde{0}\tilde{0}\}_*$ and 
$$
\begin{aligned}
&\delta_*({\pmb u}{-}{\pmb x}){*}\delta_*(\tilde{\pmb u}{-}\tilde{\pmb x}'){*}
\{\tilde{0}\tilde{0}\}_*\\
&{=}
\iint\!\!\int
e^{-\frac{1}{2i\h}\langle{\tilde{\pmb\xi}}J,\,\tilde{\pmb\eta}\rangle}
e^{\frac{1}{i\h}\langle{\tilde{\pmb\eta}}J,\,\tilde{\pmb\xi}'\rangle}
e^{-\frac{1}{i\h}\langle{\tilde{\pmb\eta}},\,\tilde{\pmb y}\rangle}
e_*^{\frac{1}{i\h}\langle{\tilde{\pmb\xi}},\,\tilde{\pmb u}{-}\tilde{\pmb x}\rangle
{+}\frac{1}{i\h}\langle{\tilde{\pmb\xi}'},\,\tilde{\pmb u}{-}\tilde{\pmb x}'\rangle}
\dbar{\tilde{\pmb\xi}}\dbar{\tilde{\pmb\eta}}\dbar{\tilde{\pmb\xi}'}
\{\tilde{0}\tilde{0}\}_*\\
&{=}
\iint\!\!\int
e^{\frac{1}{2i\h}\langle{\tilde{\pmb\xi}}J,\,\tilde{\pmb\eta}\rangle}
e^{-\frac{1}{i\h}\langle{\tilde{\pmb\eta}},\,\tilde{\pmb y}\rangle}
e^{\frac{1}{i\h}\langle{\tilde{\pmb\xi}},\,\tilde{\pmb x}'{-}\tilde{\pmb x}\rangle}
e_*^{\frac{1}{i\h}\langle{\tilde{\pmb\xi}"},\,
\tilde{\pmb u}{-}\tilde{\pmb x}'{+}\tilde{\pmb\eta}J\rangle}
\dbar{\tilde{\pmb\xi}}\dbar{\tilde{\pmb\eta}}\dbar{\tilde{\pmb\xi}"}
\{\tilde{0}\tilde{0}\}_*\\
&{=}
\iint\!\!\int
e^{-\frac{1}{i\h}\langle{\tilde{\pmb\eta}},\,\tilde{\pmb y}\rangle}
e^{\frac{1}{i\h}
\langle{\tilde{\pmb\xi}},\,\tilde{\pmb x}'{-}\tilde{\pmb x}{-}\frac{1}{2}\tilde{\pmb\eta}J\rangle}
e_*^{\frac{1}{i\h}\langle{\tilde{\pmb\xi}"},\,
\tilde{\pmb u}{-}\tilde{\pmb x}'{+}\tilde{\pmb\eta}J\rangle}
\dbar{\tilde{\pmb\xi}}\dbar{\tilde{\pmb\eta}}\dbar{\tilde{\pmb\xi}"}
\{\tilde{0}\tilde{0}\}_*.\\
\end{aligned}
$$
Integrating by ${\tilde{\pmb\xi}}$ and ${\tilde{\pmb\xi}"}$, we have 
$$
\begin{aligned}
\iint\!\!&\int h(\tilde{\pmb x},\tilde{\pmb y})f(\tilde{\pmb x}')
\delta_*({\pmb u}{-}{\pmb x}){*}\delta_*(\tilde{\pmb u}{-}\tilde{\pmb x}')
\dbar\tilde{\pmb x}\dbar\tilde{\pmb y}\dbar\tilde{\pmb x}'{*}
\{\tilde{0}\tilde{0}\}_*\\
&{=}
\iint\!\!\iint  h(\tilde{\pmb x},\tilde{\pmb y})f(\tilde{\pmb x}')
e^{-\frac{1}{i\h}\langle{\tilde{\pmb\eta}},\,\tilde{\pmb y}\rangle}
\delta(\tilde{\pmb x}'{-}\tilde{\pmb  x}{-}\frac{1}{2}\tilde{\pmb\eta}J)
\delta_*(\tilde{\pmb u}{-}\tilde{\pmb x}'{+}\tilde{\pmb\eta}J)\dbar{\tilde{\pmb\eta}}
\dbar\tilde{\pmb x}\dbar\tilde{\pmb y}\dbar\tilde{\pmb x}'
\{\tilde{0}\tilde{0}\}_*.
\end{aligned}
$$
Now, recall that the Weyl ordered expression of 
$\delta_*(\tilde{\pmb u}{-}\tilde{\pmb x}'{+}\tilde{\pmb\eta}J)$
is $\delta(\tilde{\pmb u}{-}\tilde{\pmb x}'{+}\tilde{\pmb\eta}J)$. 
Hence setting $\tilde{\pmb u}{=}\tilde{\pmb x}"$, there must be the 
relation 
$$
2(\tilde{\pmb x}'{-}\tilde{\pmb x})=\tilde{\pmb\eta}J=\tilde{\pmb x}'{-}\tilde{\pmb x}".
$$
It follows 
$$
\tilde{\pmb x}{=}\frac{1}{2}(\tilde{\pmb x}'{+}\tilde{\pmb x}").
$$
Changing variables gives  
\begin{equation}
\begin{aligned}
h_*(\pmb u){*}f_*(\tilde{\pmb u}){*}\{\tilde{\pmb y}\tilde{\pmb x}\}_*{=}
\int\Big(\!\iint h(\frac{1}{2}(\tilde{\pmb x}'{+}\tilde{\pmb x}"),\tilde{\pmb y})
f(\tilde{\pmb x}')
e^{\frac{1}{i\h}\langle\tilde{\pmb y}',\,\,\tilde{\pmb x}'{-}\tilde{\pmb x}"\rangle}
\dbar{\tilde{\pmb x}'}\dbar{\tilde{\pmb y}'}\Big)
\delta_*(\tilde{\pmb u}{-}\tilde{\pmb x}")\dbar{\tilde{\pmb x}"}\{\tilde{0}\tilde{0}\}_*.
\end{aligned}
\end{equation}
The operator in the big parentheses is known as the
pseudo-differential operator of Weyl-type. The product formula is
given by the Moyal product formula.

However a certain care about the convergence as in the case of
H{\"o}rmander type is requested. In precise, we have to use the oscillatory integrals:
\begin{equation}
\begin{aligned}
{=}
\int\Big(os{\text{-}}\!\!\iint h(\frac{1}{2}(\tilde{\pmb x}'{+}\tilde{\pmb x}"),\tilde{\pmb y})
f(\tilde{\pmb x}')
e^{\frac{1}{i\h}\langle\tilde{\pmb y}',\,\,\tilde{\pmb x}'{-}\tilde{\pmb x}"\rangle}
\dbar{\tilde{\pmb x}'}\dbar{\tilde{\pmb y}'}\Big)
\delta_*(\tilde{\pmb u}{-}\tilde{\pmb x}'')\dbar{\tilde{\pmb x}"}
\{\tilde{0}\tilde{0}\}_*.
\end{aligned}
\end{equation}

\subsection{Fourier integral operators as vacuum representations}

For the vacuum representation of a $*$-exponential function 
$e_*^{\frac{1}{i\h}\langle{\pmb a},{\pmb u}\rangle}$, 
we first apply the decomposition \eqref{Fundcal2} or Proposition\,\ref{Fundcal} and note that 
$e_*^{\frac{1}{i\h}\langle\tilde{\pmb b},\tilde{\pmb v}\rangle}
{*}\{\tilde{\pmb\phi}(\tilde{\pmb x})\tilde{\pmb x}\}_*{=}
e^{\frac{1}{i\h}\langle\tilde{\pmb b},\tilde{\pmb\phi}(\tilde{\pmb x})\rangle}
\{\tilde{\pmb\phi}(\tilde{\pmb x})\tilde{\pmb x}\}_*$.
We have then 
$$
e_*^{\frac{1}{i\h}\langle\tilde{\pmb b},{\pmb v}\rangle}
{*}f(\tilde{\pmb u}){*}
\{\tilde{\pmb\phi}(\tilde{\pmb x})\tilde{\pmb x}\}_*{=}
e_*^{\frac{1}{i\h}\langle\tilde{\pmb b},{\pmb v}\rangle}
{*}f(\tilde{\pmb u}){*}e_*^{-\frac{1}{i\h}\langle\tilde{\pmb b},{\pmb v}\rangle}
{*}e_*^{\frac{1}{i\h}\langle\tilde{\pmb b},{\pmb v}\rangle}
{*}\{\tilde{\pmb\phi}(\tilde{\pmb x})\tilde{\pmb x}\}_*
{=}f(\tilde{\pmb u}{+}\tilde{\pmb b}){*}
e^{\frac{1}{i\h}\langle\tilde{\pmb b},\tilde{\pmb\phi}(\tilde{\pmb x})\rangle}
\{\tilde{\pmb\phi}(\tilde{\pmb x})\tilde{\pmb x}\}_*.
$$
We can apply this formula to obtain a vacuum representation of certain 
diffeomorphisms of the configuration space. Let 
$\tilde{\pmb x}\to\psi(\tilde{\pmb x})$ be a diffeomorphism on
${\mathbb R}^m$ which is the identity except on a compact subset. 
Then, we see the following vacuum representation is the desired one:
$$
e_*^{\frac{1}{i\h}\langle\psi(\tilde{\pmb x}),{\pmb v}\rangle}
{*}f(\tilde{\pmb u}){*}
\{\tilde{\pmb\phi}(\tilde{\pmb x})\tilde{\pmb x}\}_*{=}
f(\tilde{\pmb u}{+}\psi(\tilde{\pmb x})){*}
e^{\frac{1}{i\h}\langle\psi(\tilde{\pmb x}),\tilde{\pmb\phi}(\tilde{\pmb x})\rangle}
\{\tilde{\pmb\phi}(\tilde{\pmb x})\tilde{\pmb x}\}_*.
$$
One may extend this to make the vacuum representations of 
symplectic transformations of positively homogeneous of degree 1
which is near the identity. This is called a Fourier integral operator.
Here we only mention what is symplectic transformations of positively homogeneous of degree 1.
Regard ${\mathbb R}^{m}{\times}({\mathbb R}^{m}{\setminus}\{0\})$ as
the cotangent bundle $T^*{\mathbb R}^{m}{\setminus}\{0\}$ with removed $0$-section. 
A symplectic diffeomorphism 
$\varphi{=}(\varphi_1,\varphi_2): 
T^*{\mathbb R}^{m}{\setminus}\{0\}\to 
T^*{\mathbb R}^{m}{\setminus}\{0\}$ 
is of positively homogeneous of degree 1, if it satisfies 
$$ 
{\varphi}_1(\tilde{\pmb x},r\tilde{\pmb y}){=}{\varphi}_1(\tilde{\pmb x},\tilde{\pmb y}),  
\quad 
{\varphi}_2(\tilde{\pmb x},r\tilde{\pmb y}){=}r{\varphi}_2(\tilde{\pmb
  x},r\tilde{\pmb y}),\quad r>0.
$$
Therefore, this method cannot be applied to the vacuum representations of 
$*$-exponential functions of quadratic forms.
As it will be seen in the next section, such an equation appears in 
Schr{\"o}dinger equation of harmonic oscillators. 

However, as it is wellknown, Schr{\"o}dinger equations are not
invariant under the Lorentz group. To obtain Lorentz invariance, we
have to use the ``square root'' of the Hamiltonian by loosing the 
locality of the operator, but such a non-local operator of degree 1
can be treated by Fourier integral operators.

\section{Vacuum representations of $*$-exponential functions \\
of  quadratic forms}\label{Two-comp}

The vacuum representations of $*$-exponential functions of quadratic
forms are little strange. This is because the vacuum
representations give a kind of {\bf double cover} of the adjoint representations.

Setting 
$e_*^{itH_*}{*}f(\tilde{\pmb u}){*}\widetilde{\varpi}(L){=}
f_t(\tilde{\pmb u}){*}\widetilde{\varpi}(L)$,
we have to solve the initial value problem of    
$$
\frac{d}{dt}f_t(\tilde{\pmb u}){*}\widetilde{\varpi}(L)
{=}iH_*{*}f_t(\tilde{\pmb  u}){*}\widetilde{\varpi}(L),\quad  
 f_t(\tilde{\pmb u}){*}\widetilde{\varpi}(L)|_{t=0}
=f(\tilde{\pmb u}){*}\widetilde{\varpi}(L).
$$ 
Since this is a very difficult problem in general, we restrict our
attention to the case $H_*$ is a quadratic form 
$\frac{1}{i\h}\langle\tilde{\pmb u}g, \tilde{\pmb u}g\rangle$, $g\in Sp(m,{\mathbb C})$   
so that $e_*^{itH_*}$ has a periodical property. In this case, we
consider the eigenvalue problem first 
$$
iH_*{*}f_n(\tilde{\pmb u}){*}\widetilde{\varpi}(L){=}
\lambda_n f_n(\tilde{\pmb  u}){*}\widetilde{\varpi}(L).
$$
If $H_*$ is fixed, then this belongs to the ordinary representation theory of a compact
group $S^1$, the eigenvectors $\{f_n(\tilde{\pmb u}),n{=}0,1,2,\cdots\}$ 
form an orthonormal system to make a Hilbert space. The initial value
problem is solved by making the initial function by a linear
combination of eigenvectors $\{f_n(\tilde{\pmb u})\}$. 
Here we have to care about the two possible cases where 
$e_*^{itH_*}$ is alternating $\pi$-periodic and $\pi$-periodic. 

Set $e_*^{itH_*}{*}\widetilde{\varpi}(L){=}
\phi_t(\tilde{\pmb u}){*}\widetilde{\varpi}(L)$, and 
according to the periodicity, we define elements 
$$
\Phi_{+}(\tilde{\pmb u}){*}\widetilde{\varpi}(L){=}
\frac{1}{2\pi}\int_{0}^{2\pi}e^{it}\phi_t(\tilde{\pmb u}){*}\widetilde{\varpi}(L)dt,\quad 
\Phi_{0}(\tilde{\pmb u}){*}\widetilde{\varpi}(L){=}
\frac{1}{2\pi}\int_{0}^{2\pi}\phi_t(\tilde{\pmb u}){*}\widetilde{\varpi}(L)dt.
$$
We easily see that 
$$
H_*{*}\Phi_{+}(\tilde{\pmb u}){*}\widetilde{\varpi}(L){=}
i\Phi_{+}(\tilde{\pmb u}){*}\widetilde{\varpi}(L),\quad 
H_*{*}\Phi_{0}(\tilde{\pmb u}){*}\widetilde{\varpi}(L){=}
\Phi_{0}(\tilde{\pmb u}){*}\widetilde{\varpi}(L).
$$
For the initial value problem we set the initial function as 
$f(\tilde{\pmb u}){=}\phi(\tilde{\pmb u}){*}\Phi_{+}(\tilde{\pmb u})$
or $\phi(\tilde{\pmb u}){*}\Phi_{0}(\tilde{\pmb u})$. Then 
$$
e_*^{itH_*}{*}\phi(\tilde{\pmb u}){*}\Phi_{+}(\tilde{\pmb u}){*}\widetilde{\varpi}(L)
{=}
\Big({\rm{Ad}}(e_*^{itH_*})\phi(\tilde{\pmb u})\Big){*}e^{it}
\Phi_{+}(\tilde{\pmb u}){*}\widetilde{\varpi}(L),
$$
$$
e_*^{itH_*}{*}\phi(\tilde{\pmb u}){*}\Phi_{0}(\tilde{\pmb u}){*}\widetilde{\varpi}(L)
{=}
\Big({\rm{Ad}}(e_*^{itH_*})\phi(\tilde{\pmb u})\Big){*}
\Phi_{0}(\tilde{\pmb u}){*}\widetilde{\varpi}(L)
$$
give the solutions. 
Hence, $\Phi_{+}(\tilde{\pmb u})$ is {\it not} a strict state vector of Heisenberg,
but the combination with the ground state vibration, which makes the periodicity change
from the periodicity of adjoint representation. 

\noindent
{\bf Note}\,\,If ${\rm{Ad}}(e_*^{itH_*})\phi(\tilde{\pmb u})$ is
$\pi$-periodic (resp. alternating $\pi$-periodic), then $e^{it}{\rm{Ad}}(e_*^{itH_*})\phi(\tilde{\pmb u})$
is alternating $\pi$-periodic (resp. $\pi$-periodic).

\bigskip 
Now, we want to see the above observation more concretely 
for the case $m{=}1$. Setting for each $k$ 
$$
e_*^{t\frac{1}{i\h}(a\tilde{u}^2_k{+}b\tilde{v}^2_k
{+}2c\tilde{u}_k{\ctt}\tilde{v}_k)}{*}f(\tilde{\pmb u}){*}\widetilde{\varpi}(L)
{=}f_t(\tilde{\pmb u}){*}\widetilde{\varpi}(L),
$$
$f_t(\tilde{\pmb u})$ 
must satisfy the equation involving Schr{\"o}dinger equation in a
special case 
\begin{equation}\label{Schrodinger00}
\begin{aligned}
i\frac{d}{dt}f_t(\tilde{u}_k)&{*}\widetilde{\varpi}(L){=}
\frac{1}{\h}\Big(a\tilde{u}^2_k{+}b\tilde{v}^2_k{+}2c\tilde{u}_k{\ctt}\tilde{v}_k\Big)
{*}f_t(\tilde{u}_k){*}\widetilde{\varpi}(L)\\
&=\Big(\frac{1}{\h}a\tilde{u}^2_k{*}f_t(\tilde{u}_k)
{-}\h b\partial_{\tilde{u}_k}^2f_t(\tilde{u}_k){+}
2ci(\tilde{u}_k\partial_{\tilde{u}_k}{+}\frac{1}{2})
    f_t(\tilde{u}_k)\Big){*}\widetilde{\varpi}(L).\\
\end{aligned}
\end{equation} 
As we are interested in the periodical solutions, we assume $ab{-}c^2{=}1$ in what follows. 
The feature of this equation is that the term 
$2ci\frac{1}{2}f_t(\tilde{u}_k)$ of (trivial) central
extension term (Schwinger term) appears. If $c{=}0$, a big difference appears
between the cases $a, b$ are reals and pure imaginaries. (
See the last part of this section.)

\bigskip
If $b{=}0$, then setting $c{=}\pm i$, the initial value problem is 
\begin{equation}\label{singorigin}
i\big(\partial_t{\mp}2i\tilde{u}_k\partial_{\tilde{u}_k}\big)f_t{=}
\big(\frac{1}{\h}a\tilde{u}^2_k{\mp}1\big)f_t,\quad f_0{=}1.
\end{equation}
Changing variables $(t,\tilde{u}_k){=}(t',e^{\mp 2it'}x)$ gives 
$$
i\partial_{t'}f_{t'}(e^{\mp 2it'}x){=}
\big(\frac{1}{\h}ae^{\mp 4it'}x^2{\mp}1\big){*}f_{t'}(e^{\mp 2it'}x),
\quad f_0(x){=}1.
$$
Hence, $f_{t'}(e^{\mp 2it'}x){=}C(x)e^{\mp it}e^{\frac{a}{4i\h}e^{\mp 4it'}x^2}$
and adjusting the initial condition gives 
$$
e_*^{t\frac{1}{i\h}(a\tilde{u}_k^2{\pm}2i\tilde{u}_k{\ctt}\tilde{v}_k)}
{*}\widetilde{\varpi}(L)   
{=}f_t(\tilde{u}_k){*}\widetilde{\varpi}(L)
{=}e^{\mp it}e^{\frac{a}{4i\h}(1{-}e^{\mp 4it})\tilde{u}^2_k}{*}\widetilde{\varpi}(L).
$$
This is alternating $\pi$-periodic and 
$$
{\e}_{00}(k){*}\widetilde{\varpi}(L){=}
e_*^{\pi\frac{1}{2i\h}(a\tilde{u}_k^2{\pm}2i\tilde{u}_k{\ctt}\tilde{v}_k)}{*}\widetilde{\varpi}(L)
{=}\mp i\widetilde{\varpi}(L),\quad \forall a.
$$

On the other hand, the Fourier expansion solves the eigenvalue problem:
$$
e^{-it}e^{\frac{a}{4i\h}(1{-}e^{-4it})\tilde{u}^2_k}{=}
e^{-it}\sum_{n{=}-\infty}^{\infty}a_n(\tilde{u}_k)e^{int}  
$$
However, the eigenvalue problem is solved by itself. 
$\big(\frac{1}{\h}a\tilde{u}^2_k{\mp}2\tilde{u}_k\partial_{\tilde{u}_k}{\mp}1\big)f_{\lambda}(\tilde{u}_k)
{=}i\lambda{*}f_{\lambda}(\tilde{u}_k).$
Setting    
$f_{\lambda}(\tilde{u}_k){=}g_{\lambda}e^{\pm\frac{a}{4\h}\tilde{u}^2_k}$. We
easily see that $f_{\lambda}(\tilde{u}_k){=}
\tilde{u}_k^{\frac{\mp i\lambda{+}1}{2}}
e^{\frac{a}{4\h}\tilde{u}^2_k}$. 
This is smooth only if $i\lambda{=}\mp(2n{+}1)$. It follows 
$$
e_*^{t\frac{1}{i\h}(a\tilde{u}^2_k{\pm}2i\tilde{u}_k{\ctt}\tilde{v}_k)}{*}
{u}^n_k{*}e^{\pm\frac{a}{4\h}\tilde{u}^2_k}{*}\widetilde{\varpi}(L){=}
e^{it(\mp 2n{+}1)}{u}^n_k{*}e^{\pm\frac{a}{4\h}\tilde{u}^2_k}{*}\widetilde{\varpi}(L)
$$
for every integer $n$ and these are alternating $\pi$-periodic and 
$$
{\e}_{00}(k){*}{u}^n_k{*}e^{\pm\frac{a}{4\h}\tilde{u}^2_k}{*}\widetilde{\varpi}(L)
{=}
e^{i\frac{\pi}{2}(\mp 2n{+}1)}{u}^n_k{*}e^{\pm\frac{a}{4\h}\tilde{u}^2_k}{*}\widetilde{\varpi}(L)
{=}i(-u_k)^n{*}e^{\pm\frac{a}{4\h}\tilde{u}^2_k}{*}\widetilde{\varpi}(L)
$$
for every $a$. In particular  
${\e}_{00}(k)^2{*}{u}^n_k{*}
e^{\pm\frac{a}{4\h}\tilde{u}^2_k}{*}\widetilde{\varpi}(L)
{=}-{u}^n_k{*}
e^{\pm\frac{a}{4\h}\tilde{u}^2_k}{*}\widetilde{\varpi}(L)$
for every $a$ and $n$. 

To make a Hilbert space by using 
$\{e^{\frac{a}{4\h}\tilde{u}^2_k},\,\,
\tilde{u}_k\,e^{\frac{a}{4\h}\tilde{u}^2_k},\,\,
\tilde{u}^2_k\,e^{\frac{a}{4\h}\tilde{u}^2_k},\cdots \}$,  
we have to restrict the variable $\tilde{u}_k$ so that 
${\rm{Re}}(\frac{a}{4\h}\tilde{u}^2_k)<0$.

\noindent
{\bf Note}\,\, If there is no constant term, then the eigenvalues
are $2n$, $n=1,2,3,\cdots$. Hence  
the solutions of the initial value problems are 
 $\pi$-periodic. 

\bigskip
Suppose next that $b{\not=}0$. We set first
$f_t(\tilde{u}_k){=}h(t,\tilde{u}_k)e_*^{\alpha\tilde{u}_k^2}$, 
$h(0,\tilde{u}_k){=}e_*^{{-}\alpha\tilde{u}_k^2}$ and 
$$
\begin{aligned}
ic\,f_t(\tilde{u}_k)&{=}ic\,h(t,\tilde{u}_k)e_*^{\alpha\tilde{u}_k^2},\\
\frac{1}{\h}a\,\tilde{u}^2_kf_t(\tilde{u}_k)&{=}
\frac{1}{\h}a\,\tilde{u}^2_k h(t,\tilde{u}_k)e_*^{\alpha\tilde{u}_k^2},\\
2ic\,\tilde{u}_k{*}\partial_{\tilde{u}_k}f_t(\tilde{u}_k)&{=}
2ic\tilde{u}_k\partial_{\tilde{u}_k}\,h(t,\tilde{u}_k)e_*^{\alpha\tilde{u}_k^2}
{+}4ic\alpha\tilde{u}^2_k{*}h(t,\tilde{u}_k)e_*^{\alpha\tilde{u}_k^2},\\
-\h\,b\,\partial^2_{\tilde{u}_k}f_t(\tilde{u}_k)&{=}
-\h\,b\,\Big(\partial^2_{\tilde{u}_k}\,h(t,\tilde{u}_k){+}4\alpha\tilde{u}_k
\partial_{\tilde{u}_k}\,h(t,\tilde{u}_k){+}2\alpha\,h(t,\tilde{u}_k){+}
4\alpha^2\tilde{u}^2_k{*}\,h(t,\tilde{u}_k)\Big){*}e_*^{\alpha\tilde{u}_k^2}.
\end{aligned}
$$
Plugging these into \eqref{Schrodinger00} we have
\begin{equation}\label{kihon}
\begin{aligned}
i\partial_th(t,\tilde{u}_k){=}
&-\h\,b\,\partial^2_{\tilde{u}_k}\,h(t,\tilde{u}_k)
{+}2(ic{-}2\h b\alpha)\tilde{u}_k{*}\partial_{\tilde{u}_k}h(t,\tilde{u}_k)\\
&\quad {+}(ic{-}2\h b\alpha)\,h(t,\tilde{u}_k)
{+}(\frac{1}{\h}a{+}4ic\alpha{-}4\h b\alpha^2)\tilde{u}^2_k{*}h(t,\tilde{u}_k).
\end{aligned}
\end{equation}
Setting $\alpha{=}\frac{ic}{2\h b}$ to
eliminate the second and third terms and recalling $ab{-}c^2{=}1$, we have 
$$
i\partial_th(t,\tilde{u}_k){=}
-\h\,b\,\partial^2_{\tilde{u}_k}\,h(t,\tilde{u}_k)
{+}\frac{1}{\h b}\tilde{u}^2_k{*}h(t,\tilde{u}_k).
$$
Putting $\tilde{u}_k{=}\sqrt{\h b}\,x$ 
changes the above equation to 
$$
x^2h_t(x){-}\partial_x^2h_t(x){=}i\partial_th_t(x), \quad 
h_0(x){=}e^{-\frac{ci}{2}x^2}{=}e^{-\frac{ci}{2\h b}\tilde{u}_k^2}.
$$
This is the Schr{\"o}dinger equation of standard harmonic oscillator.
The eigenvalue problem is wellknown. The eigenvectors are given by
using Hermite polynomials $H_n(x)$
$$
(x^2{-}\partial_x^2)H_n(x)e^{-\frac{1}{2}x^2}e^{int}
{=}(2n{+}1)H_n(x)e^{-\frac{1}{2}x^2}e^{int},\quad n{=}1,2,\cdots.
$$
By the original variables these are 
$$
H_n(x)e^{-\frac{1}{2}x^2}e^{-\frac{ci}{2}x^2}
{=}H_n(\sqrt{\h b}\,\tilde{u}_k)\,e^{-\frac{1}{2\h b}\tilde{u}_k^2}\,e^{-\frac{ci}{2\h b}\tilde{u}_k^2}.
$$
It follows 
$$
e_*^{t\frac{1}{i\h}(a\tilde{u}^2_k{+}b\tilde{v}^2_k{+}2c\tilde{u}_k{\ctt}\tilde{v}_k)}{*}
H_n(\sqrt{\h b}\,\tilde{u}_k)\,e^{-(\frac{1}{2\h b}{+}\frac{ci}{2\h b})\tilde{u}_k^2}
{*}\widetilde{\varpi}(L){=}
e^{it(2n{+}1)}H_n(\sqrt{\h b}\,\tilde{u}_k)\,
e^{-(\frac{1}{2\h b}{+}\frac{ci}{2\h b})\tilde{u}_k^2}
{*}\widetilde{\varpi}(L)
$$ 
and 
$$
\begin{aligned}
{\e}_{00}(k){*}&H_n(\sqrt{\h b}\,\tilde{u}_k)\,
e^{-(\frac{1}{2\h b}{+}\frac{ci}{2\h b})\tilde{u}_k^2}
{*}\widetilde{\varpi}(L)\\
&{=}
e_*^{\pi\frac{1}{2i\h}(a\tilde{u}^2_k{+}b\tilde{v}^2_k{+}2c\tilde{u}_k{\ctt}\tilde{v}_k)}{*}
H_n(\sqrt{\h b}\,\tilde{u}_k)\,
e^{-(\frac{1}{2\h b}{+}\frac{ci}{2\h b})\tilde{u}_k^2}{*}
\widetilde{\varpi}(L)\\
&{=}
i^{2n{+}1}H_n(\sqrt{\h b}\,\tilde{u}_k)\,
e^{-(\frac{1}{2\h b}{+}\frac{ci}{2\h b})\tilde{u}_k^2}
{*}\widetilde{\varpi}(L).
\end{aligned}
$$

To make a Hilbert space we have to restrict the variable to an
$\mathbb R$ linear subspace so that 
$$
{\rm{Re}}\frac{1{+}ci}{2\h b}\tilde{u}_k^2 <0.
$$
But here the $\sqrt{\h b}$ causes a delicate sign change, as we have
to use $\sqrt{-\h b}$ in the opposite quadratic form. 

\bigskip
For simplicity, we investigate the case $ab{=}1$, but $a, b$ are reals
and the case $a, b$ are pure imaginary. If $a, b$ are reals, then 
\eqref{kihon} is nothing but the equation of standard harmonic
oscillators, but if $a, b$ are pure imaginary, then by setting
$\alpha{=}\pm\frac{i}{2\h b}$ the equation 
\eqref{kihon} turns out 
$$
\partial_th_{it}(\tilde{u}_k){=}
\Big(i\h\partial^2_{\tilde{u}_k}\pm \tilde{u}_k\partial_{\tilde{u}_k}{\pm}1\Big)h_{it}(\tilde{u}_k),
\quad 
h_0(\tilde{u}_k){=}e_*^{i\frac{1}{2\h |b|}\tilde{u}^2_k}.
$$
Hence  setting $h_{it}(\tilde{u}_k){=}e^{\pm it}g_{it}(\tilde{u}_k)$, we have
to solve 
$$
\partial_tg_{it}(\tilde{u}_k){=}
\Big(i\h\partial^2_{\tilde{u}_k}\pm \tilde{u}_k\partial_{\tilde{u}_k}\Big)g_{it}(\tilde{u}_k),
\quad 
g_0(\tilde{u}_k){=}e_*^{i\frac{1}{2\h |b|}\tilde{u}^2_k}.
$$
These looks similar to the equation \eqref{singorigin}, but the point
here is that we have to use the Fourier transform. This procedure makes
the periodicity change. 

By setting 
$g_t(\tilde{u}_k){=}\int\hat{g}_t(\tilde{u}_k)e^{\frac{1}{i\h}\xi\tilde{u}_k}\dbar\xi$
we have 
$\hat{g}_t(\xi){=}e^{\frac{1}{2i\h}\xi^2}\hat{g}_0(e^t\xi)$,
but adjusting the initial condition gives 
$$
\hat{g}_{it}(\xi){=}\int
e^{\frac{1}{2i\h}\xi^2(e^{2it}{-}1)}
e^{\frac{i}{2\h|b|}u^2}e^{\frac{1}{i\h}e^{it}\xi u}\dbar u.
$$
Hence change variables $u$ to
$e^{it}u$, we see that $\hat{g}_{it}(\xi)$ is alternating
$\pi$-periodic. As the result, $h_{it}(\tilde{u}_k)$
is $\pi$-periodic. 

\bigskip
\noindent
{\bf Note}\,\,As it was seen, solutions of Schr{\"o}dinger equations
of harmonic oscillators have no singular point on the real line, as we choose a generic
ordered expressions where the $*$-exponential functions has no
singular point on the real line. However in computations after 
\eqref{Schrodinger00} are done without specifying the expression
parameter. As these are computations only for $\tilde{u}_k$ variable, 
these are in fact same to the computations under the normal ordered
expressions.

On the other hand, there is another set up of Schr{\"o}dinger equations
of harmonic oscillators from classical harmonic oscillators. This is
set up so that the Weyl ordered expression coincide to the classical 
one. Those two approaches to quantum harmonic oscillators give
slightly different results. The later contains singular points on the 
real line. The reason is that the Weyl ordered expression 
$$ 
{:}e_*^{t\frac{1}{i\h}(au^2{+}b^2{+}2cu{\ctt}v)}{:}_{0}, \,\, ab{-}c^2{=}1,
$$
has singular points on real line corresponding to polar elements 
(cf.\eqref{tildeKKK}). Here, Maslov's theory can be applied to control 
the discontinuous jump. 
However, we need not to use Maslov's theory by 
selecting a suitable expression parameters. 
Thus, we are thinking that Maslov's theory may be applied to 
understand what does the next 
Theorem \ref{nongo} implies.

Recall the argument in \S\,\ref{subsecpolar},\eqref{sp-inverse} and take the 
opposite quadratic form. The next Theorem shows that 
it is difficult to make a group of Fourier integral operators on ${\mathbb R}^{4}$ 
which extends the group $SU(2)$.
\begin{thm}\label{nongo}
In any fixed generic ordered expression,  
there is no common Hilbert space on which 
$$
e_*^{t\frac{1}{i\h}(a\tilde{u}^2_k{+}b\tilde{v}^2_k
{+}2c\tilde{u}_k{\ctt}\tilde{v}_k)}, \quad ab{-}c^2{=}1,
$$ 
is regularly represented w.r.t. the vacuum $\widetilde{\varpi}(L)$. 
\end{thm}
\noindent
{\bf Proof}\,\,Suppose there is a Hilbert space on which  
$e_*^{t\frac{1}{i\h}(a\tilde{u}^2_k{+}b\tilde{v}^2_k{+}2c\tilde{u}_k{\ctt}\tilde{v}_k)}$
is represented for all $(a,b,c)$ such that $ab{-}c^2{=}1$. Then, the
argument above shows that ${\e}_{00}(k)^2$ is represented by 
$i^{\ell}$ by some integer $\ell$. 
As the space $ab{-}c^2{=}1$ is connected this must be fixed throughout the space
$ab{-}c^2{=}1$. Since there is a case that ${\e}_{00}(k)^{2}{=}-1$,
$i^{\ell}$ must be $-1$. 
However, the existence of 
opposite quadratic forms gives a contradiction by the same argument as
in \S\,\ref{subsecpolar}.\hfill $\Box$

\bigskip
\noindent
{\bf Note}\,\,As it is mentioned in \cite{OMMY5}, 
$e_*^{t\frac{1}{i\h}(a\tilde{u}^2_k{+}b\tilde{v}^2_k{+}2c\tilde{u}_k{\ctt}\tilde{v}_k)}$,
$ab{-}c^2{=}1$, generate a group-like object which looks a ``double
cover'' of $S\!L(2,{\mathbb C})$ and this contains the  ``double cover'' of
$SU(2)$. Therefore, if an expression parameter $K$ is fixed, these
objects cannot form genuine groups and they must contain some singular
points. Theorem\,\ref{nongo} shows such singular points are not
removable by the vacuum representation. 

To overcome the difficulty, we have to use the $SU(2)$-vacuum, but we 
have to restrict the expression parameters.

\end{document}